\documentclass[twocolumn,prd,superscriptaddress,preprintnumbers,nofootinbib,floatfix]{revtex4}
\usepackage[utf8]{inputenc}
\usepackage{amsmath}
\usepackage{appendix}
\usepackage{natbib}
\usepackage{graphicx}
\usepackage{bm}        
\usepackage{rviewport}
\usepackage{ragged2e}
\usepackage{amsfonts,amssymb}
\usepackage{array}
\usepackage{mathpazo}
\usepackage{amsmath}
\usepackage{amssymb}
\usepackage{bm}
\usepackage{graphicx}
\usepackage{url}
\usepackage{xcolor}
\usepackage{booktabs}

\definecolor{rossoCP3}{cmyk}{0,.88,.77,.40}
\definecolor{darkBlue}{rgb}{0, 0, 0.8}
\newcommand{\xmax}{X_{\rm max}}

\newcolumntype{C}[1]{>{\centering\let\newline\\\arraybackslash\hspace{0pt}}m{#1}}
\usepackage[backref,breaklinks,colorlinks]{hyperref}

\newcommand{\lsim}{\mathrel{\hbox{\rlap{\lower.75ex \hbox{$\sim$}} \kern-.3em \raise.4ex \hbox{$<$}}}}
\newcommand{\gsim}{\mathrel{\hbox{\rlap{\lower.75ex \hbox{$\sim$}} \kern-.3em \raise.4ex \hbox{$>$}}}}

\definecolor{hookgreen}{rgb}{0.0,0.44,0.0}

\usepackage{color}     
\usepackage{bm}        
\usepackage{tocloft}   

\usepackage{comment}   
\usepackage[nottoc]{tocbibind}      

\usepackage{multirow}

\begin{document}

\title{Performance and science reach of POEMMA for ultrahigh-energy particles}

\author{Luis A. Anchordoqui}

\affiliation{Department of Physics and Astronomy,  Lehman College, City University of
  New York, NY 10468, USA
}

\affiliation{Department of Physics,
 Graduate Center, City University
  of New York,  NY 10016, USA
}

\affiliation{Department of Astrophysics,
 American Museum of Natural History, NY
 10024, USA
}

\author{Douglas R. Bergman}
\affiliation{High Energy Astrophysics Institute and Department of Physics and Astronomy, University of Utah, Salt Lake City, Utah, USA}

\author{Mario E. Bertaina}
\affiliation{Dipartimento di Fisica, Universit\'a di Torino, Torino 10125, Italy}

\author{Francesco~\nolinebreak~Fenu}
\affiliation{Dipartimento di Fisica, Universit\'a di Torino, Torino 10125, Italy}

\author{John~\nolinebreak~F.~\nolinebreak~Krizmanic}
\affiliation{CRESST/NASA Goddard Space Flight Center, Greenbelt, MD 20771, USA \\
University of Maryland, Baltimore County, Baltimore, MD 21250, USA}

\author{Alessandro Liberatore}
\affiliation{Dipartimento di Fisica, Universit\'a di Torino, Torino 10125, Italy}

\author{Angela V. Olinto}
\affiliation{Department of Astronomy \& Astrophysics, KICP, EFI, The University of Chicago, Chicago, IL 60637, USA}

\author{Mary Hall Reno}
\address{Department of Physics and Astronomy,
University of Iowa, Iowa City, IA 52242, USA}

\author{Fred Sarazin}
\address{Department of Physics,
Colorado School of Mines, Golden, CO 80401, USA}

\author{Kenji~\nolinebreak~Shinozaki}
\affiliation{Dipartimento di Fisica, Universit\'a di Torino, Torino 10125, Italy}

\author{Jorge F. Soriano}
\affiliation{Department of Physics and Astronomy,  Lehman College, City University of
  New York, NY 10468, USA
}

\affiliation{Department of Physics,
 Graduate Center, City University
  of New York,  NY 10016, USA
}

\author{Ralf Ulrich}
\affiliation{Institut f\"ur Kernphysik, Karlsruher Institut f\"ur Technologie, 76021 Karlsruhe, Germany}

\author{Michael Unger}
\affiliation{Institut f\"ur Kernphysik, Karlsruher Institut f\"ur Technologie, 76021 Karlsruhe, Germany}

\author{Tonia M. Venters}
\affiliation{Astrophysics Science Division, NASA Goddard Space Flight Center, Greenbelt, MD 20771, USA}

\author{Lawrence Wiencke}
\address{Department of Physics,
Colorado School of Mines, Golden, CO 80401, USA}

\date{\today}

\begin{abstract}
\noindent The Probe Of Extreme Multi-Messenger Astrophysics (POEMMA) is a potential NASA Astrophysics Probe-class mission designed to observe ultra-high energy cosmic rays (UHECRs) and cosmic neutrinos from space. POEMMA will monitor colossal volumes of the Earth's atmosphere to detect extensive air showers (EASs) produced by extremely energetic cosmic messengers:
UHECRs above 20~EeV over the full sky and cosmic neutrinos above 20~PeV. We focus most of this study on the impact of POEMMA for UHECR science by simulating  the detector response and mission performance for EAS from UHECRs.  We show that POEMMA will provide a significant increase in the statistics of observed UHECRs at the highest energies over the entire sky. POEMMA will be the first UHECR fluorescence detector deployed in space that will provide high-quality stereoscopic observations of the longitudinal development of air showers. Therefore it will be able to provide
event-by-event estimates of the calorimetric energy and nuclear mass of 
UHECRs.  The particle physics in the interactions limits the interpretation of the {\it shower maximum} on an event by event basis.  In contrast, the calorimetric energy measurement is significantly less sensitive to the different possible final states in the early interactions. We study the prospects to discover the origin and nature of UHECRs using expectations  for measurements of the energy spectrum, the distribution of arrival direction, and the atmospheric column depth at which the EAS longitudinal development reaches maximum. We also explore supplementary science capabilities of POEMMA through its sensitivity to particle interactions at extreme energies and its ability to detect ultra-high energy neutrinos and photons produced by top-down models including cosmic strings and super-heavy dark matter particle decay in the halo of the Milky Way.
\end{abstract}
\maketitle
\section{Introduction}

After over 80 years of the first measurement of extensive air showers
(EASs) by Pierre Auger~\cite{Auger:1939sh}, the astrophysical sources of these extremely
energetic cosmic rays remain unknown. Ultra-high energy (UHE) cosmic rays
(CRs) have been observed with energies \mbox{$E \agt 10^{20}~{\rm eV} \equiv 100~{\rm EeV}$,}
which is more than 7 orders of magnitude higher than what the LHC beam
can currently achieve.  The nature of the astrophysical sources and
their acceleration mechanism(s) remains a
mystery~\cite{Kotera:2011cp,Anchordoqui:2018qom,AlvesBatista:2019tlv}.  The understanding
is further muddled by the uncertainty in the nuclear composition of
UHECRs above $50~{\rm EeV}$.  

A succession of increasingly sized ground-based experiments has led to the Pierre Auger
Observatory (Auger)~\cite{Abraham:2010zz,Abraham:2009pm}, with an exposure \mbox{${\cal E} \sim 65,000~{\rm km^2 \ sr \ yr}$} collected in 15
years of operation~\cite{Aab:2017njo}, and the Telescope
Array (TA)~\cite{AbuZayyad:2012kk,Tokuno:2012mi}, with  
\mbox{${\cal E} \sim 10,000~{\rm km}^2 \, {\rm sr} \, {\rm yr}$} collected in 10
years. Both of these experiments have precisely measured key features
in the UHECR spectrum: a pronounced hardening around 5~EeV (the so-called “ankle” feature) and a suppression
of the flux above about
40~EeV~\cite{Abbasi:2007sv,Abraham:2008ru,Abraham:2010mj,AbuZayyad:2012ru}.
The differential energy spectra measured by TA and Auger agree within systematic errors below $10~{\rm EeV}$. However, even after energy re-scaling, a large difference remains at and 
beyond the flux suppression~\cite{TheTelescopeArray:2018dje}.

The EAS longitudinal development is characterized by the number of particles as a function of the atmospheric column depth $X$ in ${\rm g /cm}^{2}$. A well-defined peak of the longitudinal profile is observed when the number of $e^\pm$ in the electromagnetic shower is at its maximum. This shower maximum or $X_{\rm max}$ becomes a powerful observable for studying the UHECR nuclear composition. This is because breaks in the elongation rate -- the rate of change of $\langle X_{\rm max} \rangle$ per decade of energy --  can be related to changes in the nuclear composition~\cite{Linsley:1978bm,Linsley:1981gh}, even when uncertainties in the UHE particle physics limit the accuracy of mapping between $X_{\rm max}$ and the nucleus baryon number  $A$. 

At around the ankle, the measurements of $X_{\rm max}$ by TA~\cite{Abbasi:2014sfa,Abbasi:2018nun}
and Auger~\cite{Abraham:2010yv,Aab:2014kda,Aab:2014aea,Aab:2017cgk}
are both consistent with a predominantly light composition. For $E \gtrsim 10~{\rm EeV}$, Auger data show both a
significant decrease in the elongation rate and a decrease of the shower-to-shower fluctuations of
$X_{\rm max}$ with energy.  These two effects indicate a gradual increase of $A$  with rising energy. Indeed, at $E \approx 30~{\rm EeV}$ the interpretation of Auger
data with LHC-tuned hadronic interaction models leads to $A \approx 14 - 20$.  The 
Auger and TA collaborations have also conducted a thorough joint analysis concluding that at the current level of statistics and understanding of systematics, both datasets are compatible with being drawn from the same parent distribution~\cite{Abbasi:2015xga,Hanlon:2018dhz}. Above about 10~EeV TA data are compatible with a pure protonic composition, but also with the mixed composition determined by the Auger Collaboration~\cite{Abbasi:2015xga,Hanlon:2018dhz}; Auger data are more constraining and not compatible with the pure protonic option available with TA alone. Moreover, the Auger Collaboration has reported additional model-independent evidence for a mixed nuclear composition around the ankle in the correlations between $X_{\rm max}$ and the shower size at ground~\cite{Aab:2016htd}. All in all, while there remain some differences in the details of composition measurements between Auger and TA, 
Auger has provided evidence of heavier composition with increasing energies above $10~{\rm EeV}$ and TA measurements are not in contradiction with that interpretation.  

UHECR deflections by intervening magnetic fields constitute the main
challenge for source identification.  The typical deflection of UHECRs
in the extragalactic magnetic field,
$B \sim 1~{\rm nG}$~\cite{Pshirkov:2015tua}, can be estimated to be
\begin{equation}
\theta_{\rm eg}
\approx 1.5^\circ Z  \sqrt{\frac{d}{3.8~{\rm Mpc}} \ \frac{\lambda}{0.1~{\rm
  Mpc}}} \ \frac{B}{{\rm nG}} \ \left(\frac{E}{10~{\rm EeV}}\right)^{-1},
\end{equation}
where $d$ is the source distance, $\lambda$ the magnetic field
coherence length, and  $Z$ is the charge of the UHECR in units of the proton
charge~\cite{Waxman:1996zn,Farrar:2012gm}. Typical values of
the deflections of UHECRs crossing the Galaxy are somewhat
larger~\cite{Farrar:2017lhm}
\begin{equation}
\theta_{\rm G} \sim 10^\circ \ Z \ \left(\frac{E}{10~{\rm
  EeV}}\right)^{-1} \,,
  \label{eq:BG}
\end{equation}
preventing small-scale clustering with directional pointing to the
sources. However, individual sources could still be isolated in the
sky if the UHECR flux is dominated by the contribution of a limited
number of sources. Indeed, the reduction of the UHECR horizon,
because of the so-called ``Greisen-Zatsepin-Kuzmin (GZK) interactions'' on the cosmic microwave background (CMB)~\cite{Greisen:1966jv,Zatsepin:1966jv}, implies that fewer and fewer sources contribute to the flux at higher and higher energy.

Assuredly, the most recent results concerning the origin of UHECRs
has been the discovery (statistical significance $>5\sigma$) of a
large scale hemispherical asymmetry in the arrival direction
distribution of events recorded by Auger~\cite{Aab:2017tyv,Aab:2018mmi}. The data
above $8~{\rm EeV}$ are well-represented by a dipole with an
amplitude $A = (6.5^{+1.3}_{-0.9})\%$ pointing in the direction $(l,
b) = (233^\circ, -13^\circ) \pm 10^\circ$ in Galactic coordinates,
favoring an extragalactic origin for UHECRs. However, Auger and TA data have yielded
only a few clues to the precise location of the sources. For instance, the TA Collaboration has reported an excess above the isotropic background-only expectation in cosmic rays with energies above $57~{\rm EeV}$~\cite{Abbasi:2014lda}. In addition, searches in Auger data revealed a possible correlation with nearby starburst galaxies, with a (post-trial) $4\sigma$ significance, for events above $39~{\rm EeV}$~\cite{Aab:2018chp}. The smearing angle and the anisotropic fraction
corresponding to the best-fit parameters are ${13^{+4}_{-3}}^\circ$ and $(10\pm 4)\%$,
respectively. The energy threshold coincides with the observed
suppression in the spectrum, implying that when we properly account
for the barriers to UHECR propagation in the form of energy loss
mechanisms~\cite{Greisen:1966jv,Zatsepin:1966jv} we obtain a self
consistent picture for the observed UHECR horizon. With current
statistics the TA Collaboration cannot make a statistically
significant corroboration or refutation of the starburst hypothesis~\cite{Abbasi:2018tqo}. A slightly weaker association
($2.7\sigma$) with active galactic nuclei emitting $\gamma$-rays
($\gamma$AGN) is also found in Auger events above $60~{\rm EeV}$~\cite{Aab:2018chp}. 
For $\gamma$AGN, the maximum deviation from isotropy is found at an intermediate angular scale of
${7^{+4}_{-2}}^\circ$ with an anisotropic fraction of $(7 \pm 4)\%$.

Extremely fast spinning young
pulsars~\cite{Blasi:2000xm,Fang:2012rx,Fang:2013cba},
active galactic nuclei
(AGNs)~\cite{Biermann:1987ep,Rachen:1992pg,Romero:1995tn,Matthews:2018laz},
starburst galaxies (SBGs)~\cite{Anchordoqui:1999cu,Anchordoqui:2018vji,Anchordoqui:2019mfb}, and
gamma-ray bursts
(GRBs)~\cite{Waxman:1995vg,Vietri:1995hs,Dermer:2006bb,Wang:2007xj,Murase:2008mr,Baerwald:2013pu, Globus:2014fka,Zhang:2017moz} can partially accommodate Auger and TA observations, but a convincing
unified explanation of all data is yet to 
be realized. What is clearly needed is a
more dramatic increase in UHECR exposure. The Probe of Extreme
Multi-Messenger Astrophysics (POEMMA) will accomplish this by using
space-based UHECR observations with excellent angular, energy, and
nuclear composition resolution~\cite{Olinto:2017xbi}. In addition, both Auger and TA are undergoing upgrades. TA$\times$4 is designed to cover the equivalent of Auger's aperture~\cite{Kido:2019enj}. Auger's upgrade (“AugerPrime”~\cite{Aab:2016vlz}) focuses on more detailed measurements of each shower observed. This will enable event-by-event probabilistic composition assignment (hence selection of low-$Z$ events). 

In this paper we investigate the sensitivity of POEMMA to address challenges of UHECR astrophysics and explore the potential of this experiment to probe fundamental physics. The layout of the paper is as follows. In Sec.~\ref{sec:2} we provide an overview of the POEMMA design and the mission specifications. Aspects of the simulated detector response are discussed in Sec.~\ref{sec:3} and the POEMMA UHECR performance is studied in Sec.~\ref{sec:4}. After that, in Sec.~\ref{sec:5} we examine the POEMMA science reach. 
We also evaluate whether the UHECR capabilities of POEMMA for baryonic cosmic rays are also applicable to fluorescence detection of extreme energy photons and ultrahigh-energy neutrino interactions deep in the atmosphere. Finally, we summarize our results and draw our conclusions in Sec.~\ref{sec:6}.

\section{The POEMMA Experiment}
\label{sec:2}

The POEMMA instruments and mission leveraged from previous
work developed for the Orbiting Wide-field Light-collectors (OWL) mission~\cite{Stecker:2004wt}, together with the experience on the fluorescence detection camera for the Joint Experimental Missions of the Extreme Universe Space Observatory (JEM-EUSO)~\cite{Adams:2013vea}, and the recently proposed CHerenkov from Astrophysical Neutrinos Telescope (CHANT) concept~\cite{Neronov:2016zou} to form a multi-messenger probe of the most extreme environments in the universe. 

POEMMA was selected by NASA as one of the several concept study proposals to provide science community input for a new class of NASA missions, called Astrophysics Probes. Such Astrophysics Probes will be examined by the 2020 Astronomy and Astrophysics Decadal Survey in support of the development of a recommended portfolio of future astrophysics missions.\footnote{NASA Research Announcement Astrophysics Probe Mission Concept Studies, Solicitation: NNH16ZDA001N-APROBES.} The Astrophysics Probe mission concepts were funded for 18 month studies, including week-long dedicated engineering study runs. The POEMMA study was performed at the Instrument Design Lab (IDL) and the Mission Design Lab (MDL) in the Integrated Design Center (IDC) at NASA/GSFC.\footnote{For details, see {\color{blue} \tt https://idc.nasa.gov/idc/}} The probe studies were developed with specific instructions to define this unique NASA Class B mission, including Phase A start date (1-Oct-2023), launch date (1-Nov-2029), and launch vehicle guidance,  with a total lifecycle (NASA Phases A through E) costs between  $400{\rm M}$ and $1{\rm B}$ in FY2018 dollars. In this context, POEMMA is considered as a potential probe mission in terms of the 2020 astrophysics decadal review assessment of the probe-class concept. In this section we provide a summary of the POEMMA mission specifications developed under the probe study.

\subsection{Instrument Design}

\begin{figure}[tb]
\centering
\includegraphics[width=0.97\columnwidth]{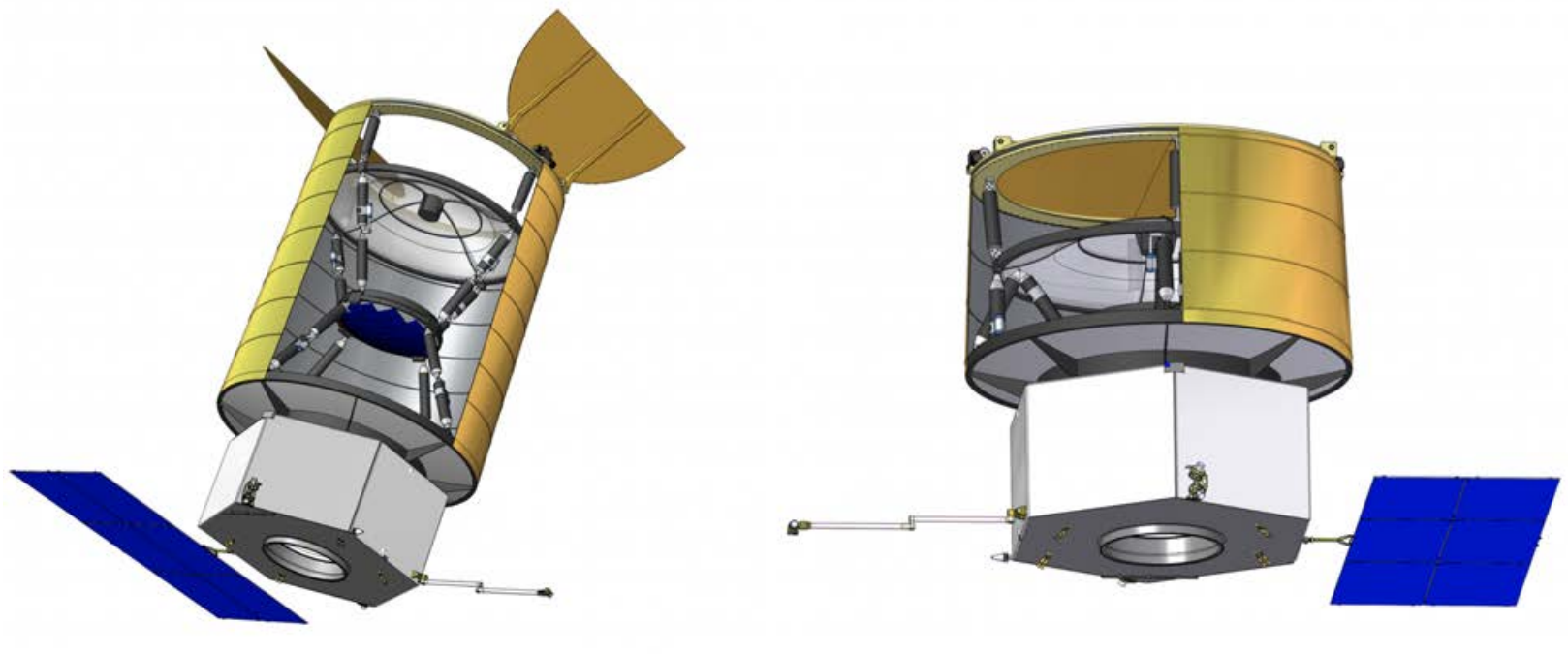}
\caption{Schematic representation of POEMMA (photometer and spacecraft) deployed with open shutter doors (left) and in stowed position for launch (right). Cutaways in the light shield display the internal structure of corrector plate and focal surface in the middle of the payload (blue). Spacecraft bus shown with solar panel (blue) and communications antenna deployed in both images.}
\label{fig:poemma_spacecrft}
\end{figure}

Building on the OWL study~\cite{Stecker:2004wt}, POEMMA is composed of two identical satellites flying in formation with the ability to observe overlapping atmospheric volumes during nearly moonless nights in configurations ranging from nadir viewing to that just above the limb of the Earth. The satellites will fly at an altitude of about 525~km with separations ranging from 300~km for stereo fluorescence UHECR observations to 25~km when pointing at the Earth’s limb for both fluorescence and Cherenkov observations of UHECRs and cosmic neutrinos. Each POEMMA satellite consists of a 4-meter diameter Schmidt telescope with a fast (f/0.64) optical design. The optical effective area ranges from 6 to 2~m$^2$ depending on the angle of incidence of the signal.  The visible portion of the EAS disk is a few hundreds of meters wide and determines the
required pixel angular resolution in the Schmidt telescope focal
plane: a spatial size of 1~km from 525~km leads to pixel angular
resolution of $\sim 0.1^\circ$ to accurately view the EAS
development. Each POEMMA telescope monitors a substantial $45^\circ$ field of view (FoV) with fine pixel angular resolution of
$0.084^\circ$ and a refractive aspheric aberration corrector plate. A lens-cap lid (or shutter door) and a cylindrical light shield shroud protect the mirror from stray light and micrometeoroids. The mirrors act as large light collectors with modest imaging requirements, i.e., the POEMMA optics imaging requirements are $\sim 10^4$ away from the diffraction limit. The primary mirror is 4~m diameter, whereas the corrector lens 3.3~m diameter. The concept of the POEMMA photometer and spacecraft is shown in Fig.~\ref{fig:poemma_spacecrft}.

\begin{figure}[tb]
  \centering
\includegraphics[width=0.99\linewidth]{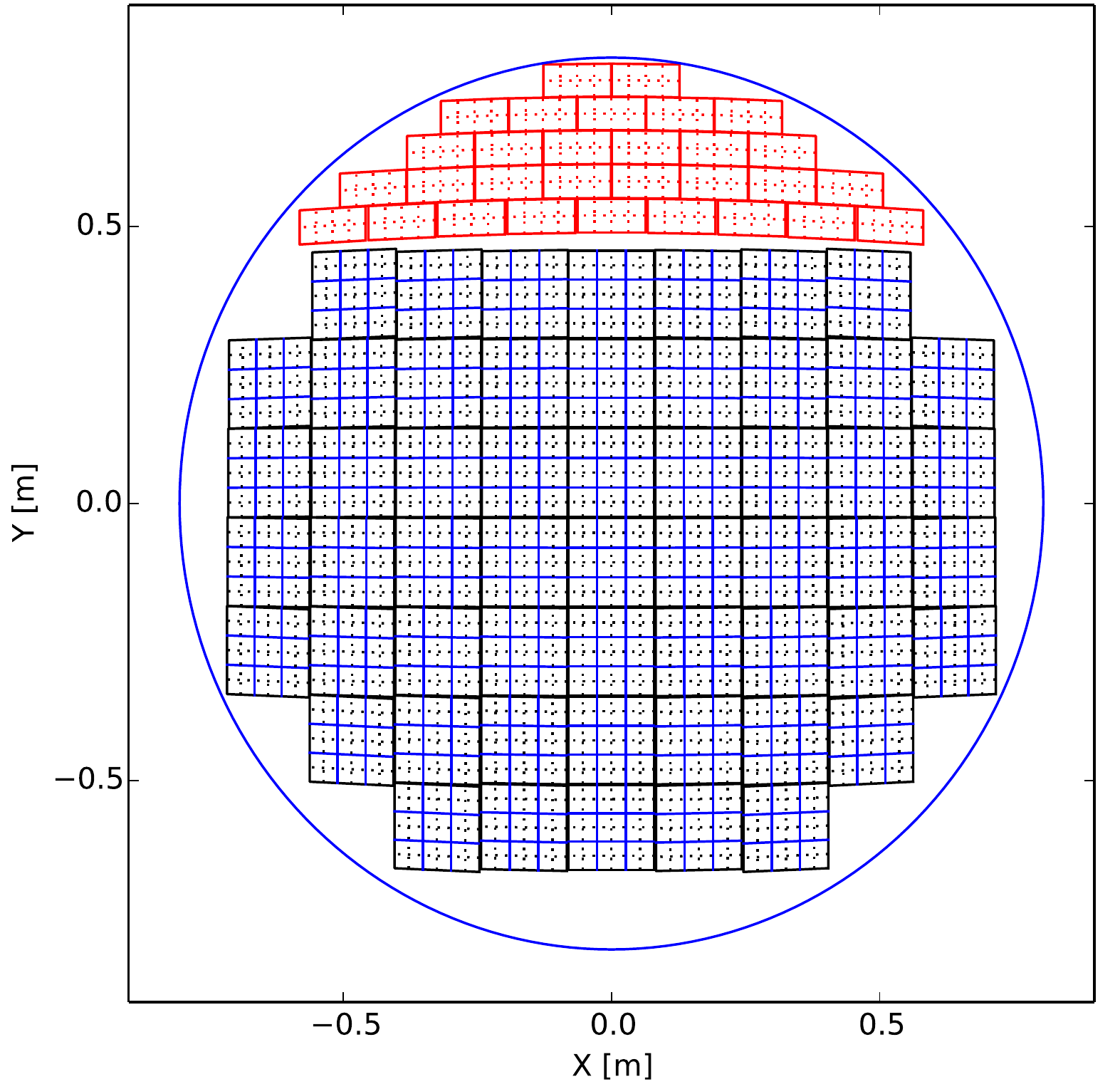}  
\caption{POEMMA's hybrid focal surface of 1.6~m diameter. The PFC (blue), composed of 55 PDMs (total 126,720 MAPMT pixels) with 1~$\mu{\rm s }$ time gates, and the PCC (red) with 28 SiPM focal surface units. The PCC observes a solid angle of $9^\circ$ by $30^\circ$ to monitor the Earth's limb for up-going EASs.}
\label{fig:fs}
\end{figure}

The POEMMA focal surface (1.6~m diameter) is composed of a hybrid of two types of cameras: over 85\% of the focal surface is dedicated to the POEMMA fluorescence camera (PFC), while the POEMMA Cherenkov camera (PCC) occupies the crescent moon shaped edge of the focal surface, which images the limb of the Earth. The PFC is composed of the EUSO designed Photo Detector Modules (PDM) based on multi-anode photomultiplier tubes (MAPMTs) as flown in sub-orbital missions in EUSO-Balloon~\cite{Abdellaoui:2019qmg}, EUSO- SPB1~\cite{Wiencke:2017cfi}, and EUSO-SPB2~\cite{Wiencke:2019vke}, and in the mini-EUSO~\cite{Belov:2017ksp} and CALET~\cite{2015JPhCS.632a2023A} experiments onboard the International Space Station. The sampling time between images for the PFC is $1~\mu{\rm s}$. The
much faster POEMMA Cherenkov camera (PCC) is composed of Silicon photo-multipliers (SiPMs) designed to detect the 10~ns to 100~ns Cherenkov flashes. The PFC registers UHECR tracks from near nadir when in stereo mode extended to just below the Earth’s limb when in tilted neutrino mode, where the PCC registers light within the Cherenkov emission cone of up-going showers around the limb of the Earth and also from UHECRs above the limb of the Earth.
An schematic representation of the POEMMA's hybrid focal surface is shown in Fig.~\ref{fig:fs}.

\subsection{Mission Synopsis}

The POEMMA mission involves two satellites flying in formation in a relatively low-altitude (525~km), near-equatorial orbit (28.5$^\circ$ inclination). Each satellite operates independently and telemeters data to the ground for combined analysis. Both satellites will be launched in a stowed configuration on a single launch vehicle. Once on orbit, the telescopes will be deployed along with the solar array, light shield, and the communications antenna. The mirror and data module are attached to the satellite bus.

Both satellites will be launched as a dual-manifest in an Atlas V using the long payload fairing. The satellites will be inserted into a circular orbit at an inclination of about 28.5$^\circ$ and an initial altitude of 525~km and a separation of 300~km. The most common flight configuration will be the UHECR stereo observation mode. The target-of-opportunity (ToO) observing mode will involve the instruments slewing to view the celestial source location as well as a maneuver to closer distance such that both satellites observe the same Cherenkov signal from Earth-skimming neutrinos from the transient source~\cite{Venters:2019xwi} for long-duration ($>$ 1 day) bursts. Once extreme transient event alerts are received, for example, from the gravitational wave signature of a binary neutron-star merger, the satellites will maneuver to a closer separation distance of about 25~km and an appropriate attitude and slew to follow the ToO of the transient source as it rises and sets over the Earth’s limb. While the PCC is searching for neutrinos from the ToO, each PFC continues to observe UHECRs in a common volume, providing two correlated monocular views of EASs. A sequence of observing formation stages, varying between stereo and ToO modes, will be planned to address each science goal for the minimum 3 year mission with a 5 year mission goal.

\section{Simulated Detector Response}
\label{sec:3}

POEMMA's two identical satellites fly in formation to observe EASs in two different modes: precision UHECR stereo mode and a tilted configuration, denoted as neutrino mode, when also viewing $\tau$-lepton EASs sourced near the Earth's limb. In stereo mode, the UHECR measurement performance is optimized by separating the POEMMA satellites by 300 km and tilted slightly away from nadir for each to view a common area and stereoscopically reconstruct the EASs from UHECRs. In neutrino mode, the satellites slew to view the source and are separated by 25~km to put both in the Cherenkov light pool from the upward-moving $\tau$-lepton EASs, using the PCCs, and also measure UHECRs using the PFCs with a higher geometry factor, but reduced performance due the more monocular-like UHECR reconstruction response. Note that both instruments also view a common atmospheric volume in limb-viewing mode. Here we detail the two separate simulations used to determine the stereo and monocular UHECR performance.

\begin{figure*}[tb]
\centering
\includegraphics[width=0.97\columnwidth]{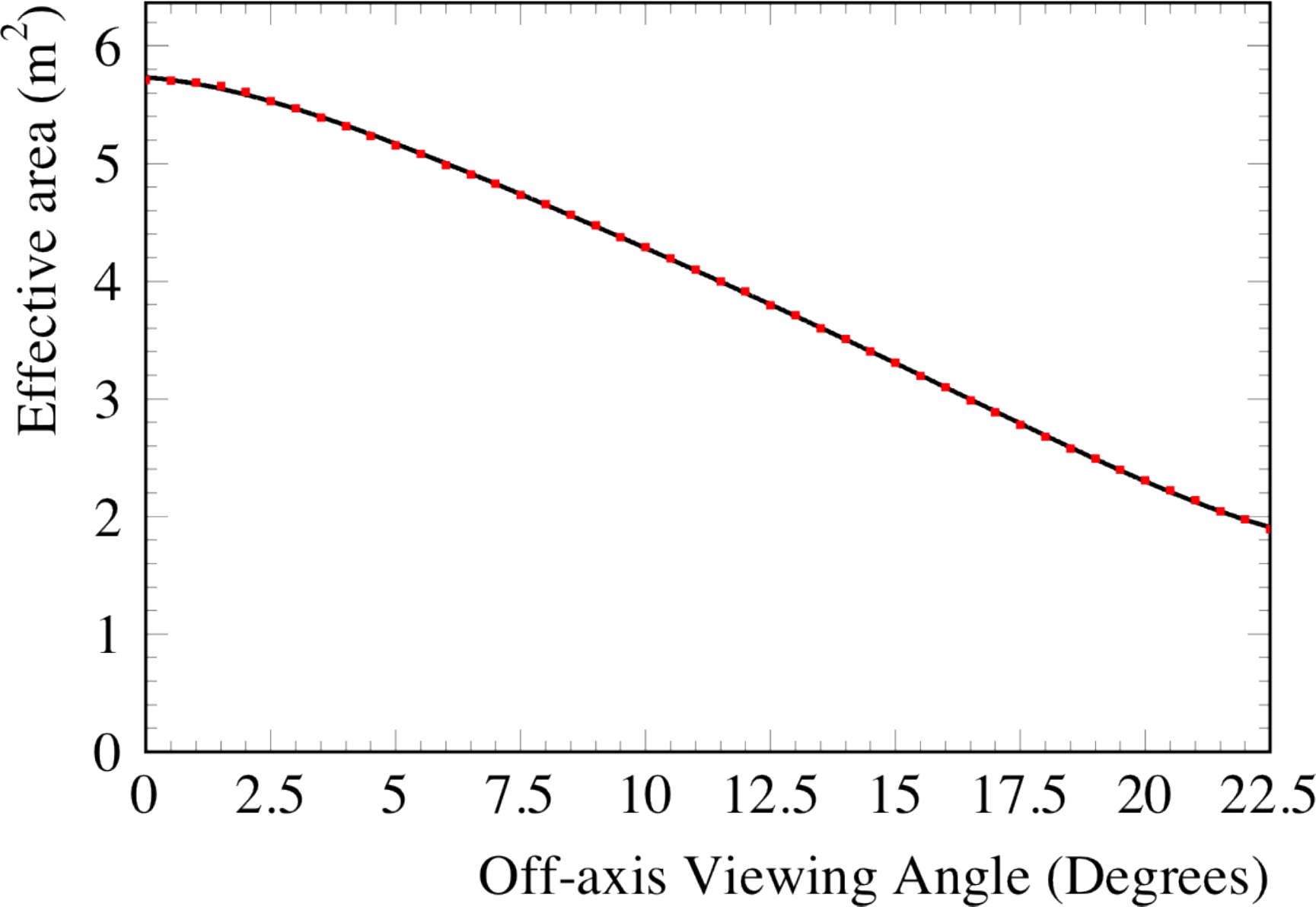}
\hspace{0.1 cm}
\includegraphics[width=0.99\columnwidth]{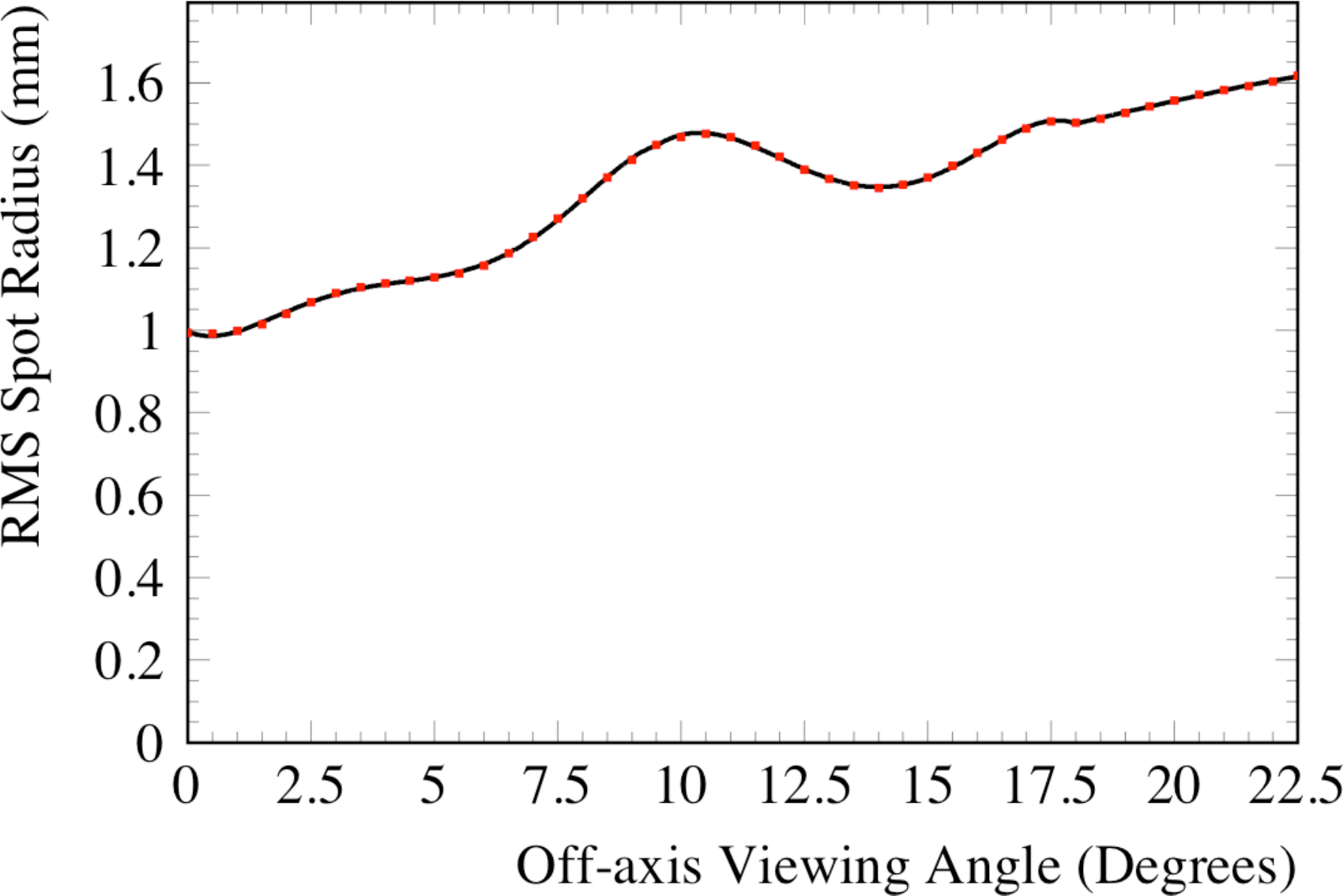}
    \caption{{\small The effective area (left) and the RMS spot radius (right) as a function of viewing angle for a POEMMA Schmidt telescope.}}
\label{ParamOptics}
\end{figure*}

Both simulations use the same parametric optical models derived from ray tracing optical design software (Zemax). Each satellite carries a Schmidt telescope with a 3.3 meter diameter optical aperture, defined by the corrector lens, and $45^\circ$ full FoV. In Fig.~\ref{ParamOptics} we show the effective area and RMS spot radius as a function of the viewing angle.  The effective area includes the effects of the transmission through the Corrector Lens (94\% transmission) and mirror reflectivity (95\%).

\subsection{Stereo Simulations}

The UHECR simulated stereo response was performed using an end-to-end simulation originally developed for the OWL~\cite{Stecker:2004wt} study but updated for POEMMA. The simulation assumes an isotropic UHECR flux impinging the Earth and uses a fast EAS generator~\cite{Mikulski} that provides the 1-dimensional EAS profiles as Gaisser-Hillas functions~\cite{Gaisser-Hillas-1977-ICRC-15-8-353}. POEMMA's 0.084$^\circ$ focal plane pixel FoV translates to a spatial distance of $\sim 0.8~{\rm km}$ at sea level for nadir viewing from an altitude of 525~km, indicating that the 1-dimensional EAS modeling is a good approximation. Starting point fluctuations of the EAS are included and the EAS generator can model any nuclei. A comparison of proton primaries to that produced by the CONEX~\cite{CONEX} 1-dimensional EAS simulation using QGSJETII-04 are shown in Table~\ref{CONEXcomp}.

\begin{table*}[tb]
\caption{Comparison of the POEMMA EAS simulation based on the distributions from 1000 modeled proton UHECRs with 30$^\circ$ zenith angle to that from CONEX.}
\centering
\begin{tabular}{c c c c c}
\hline
\hline
~~EAS  Energy (EeV)~~ & ~~POEMMA $N_{\rm max}$~~ & CONEX $N_{\rm max}$  & ~~POEMMA $X_{\rm max}$ $({\rm g/cm}^2)$~~ & ~~CONEX $X_{\rm max}$ $({\rm g/cm}^2)$~~ \\ \hline 
1 & $6.30 \pm 0.27 \times 10^8$ & ~~$6.15 \pm  0.22 \times 10^8$~~ & $750 \pm 62$ & $739 \pm 70$ \\ 
10 &  $6.12 \pm 0.21 \times 10^9$ & $6.10 \pm  0.19 \times 10^9$ & $815 \pm 58$ & $793 \pm 62$ \\ 
100 &  $5.91 \pm 0.18 \times 10^{10}$ & $5.94 \pm  0.20 \times 10^{10}$ & $868 \pm 53$ & $849 \pm 60$ \\ \hline \hline
\end{tabular}
\label{CONEXcomp}
\end{table*}

A detailed atmospheric model is required to define the EAS development, the fluorescence light generation, the generated and scattered Cherenkov light emission, and the fluorescence and Cherenkov light attenuation based on an optical depth between the EAS and POEMMA observations.
We employ a static, baseline model for the definition of the atmosphere profile using the model of Shibata~\cite{Shibata} to define the overburden and density. The temperature profile, needed to account for the altitude dependence of the fluorescence yield, is taken from the 1976 standard atmosphere while the index of refraction of air uses a parametric model of Hillas~\cite{Hillas1}.  The fluorescence light is generated in the wavelength band from 282 nm to 523 nm based on the measured relative yields from each specific line~\cite{Bunner1966,Davidson}, with the total fluorescence yield defined by recent measurements compiled by the Particle Data Group~\cite{Tanabashi:2018oca}.  The pressure and temperature dependence of the fluorescence emission uses the model of Kakimoto {\it et al.}~\cite{Kakimoto}. For simulating the PFC response, Cherenkov light is generated in the wavelength band 200 - 600~nm in bins of 25~nm based on a standard prescription~\cite{Fernow1986}.

The fluorescence and Cherenkov light attenuation includes the effects of Rayleigh scattering~\cite{Sokolsky1989} and ozone absorption.   The Earth's ozone layer efficiently attenuates optical signals at shorter wavelengths ($\lambda \lsim 330$~nm). An ozone attenuation model~\cite{Krizmanic1999} is used with an altitude dependent profile derived from Total Ozone Mapping Spectrometer (TOMS) measurements~\cite{McPeters}. The optical fluorescence and (scattered) Cherenkov wavelength dependent signals delivered to the POEMMA satellites at 525~km altitude are then convolved with a model of performance of each POEMMA Schmidt telescope.

\begin{figure}[tb]
\begin{center}
\includegraphics[width=0.37\textwidth]{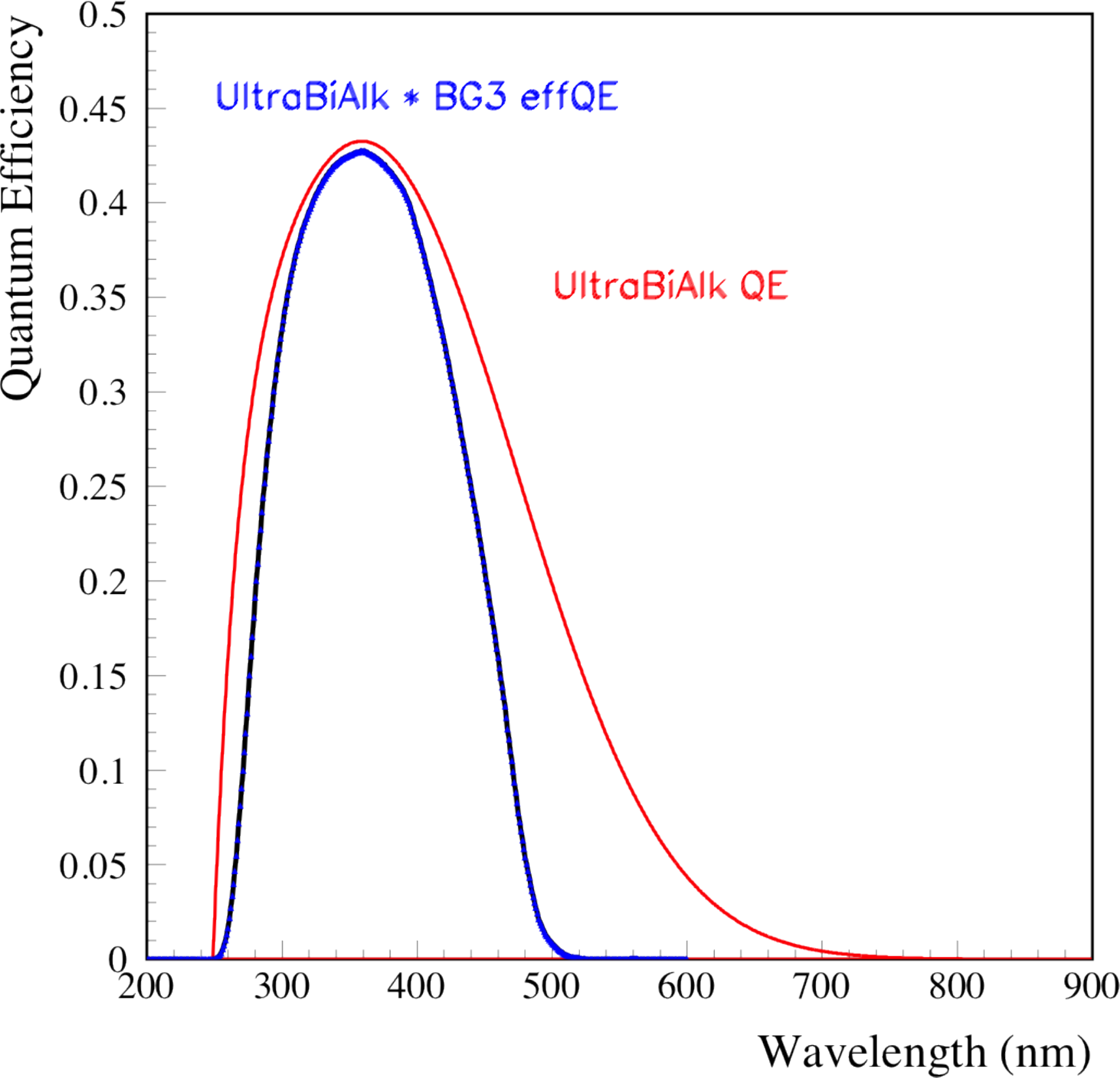}
\end{center}
\caption{The QE as a function of wavelength used to model the PFC response as well as the effective QE after taking into account the transmission of the BG3 UV filter.}
\label{QE}
\end{figure}

The effective area and RMS point-spread-function (PSF) are modeled using a parametric description based upon the optical design that relies on ray tracing.  The wavelength-dependent signals are attenuated by a UV filter and then mapped onto a POEMMA focal plane assuming 3 mm spatial pixel size.  The quantum efficiency (QE) of the photo-detector is modeled using the wavelength dependence based on that reported by the manufacturer (Hamamatsu).  The incident angle of the EAS optical signal determines the effective collecting area and the PSF of the optics, these are used to then generate photo-electrons (PE) by using Poisson statistics after accounting for the bandpass of the BG3 UV filter and QE of the photo-detectors (see Fig.~\ref{QE}).  This process uses 1~$\mu$s sampling time to record the EAS signals while also providing the integrated EAS profile in the POEMMA focal planes.

\begin{figure*}[tbp]
\centering
\includegraphics[width=0.98\columnwidth]{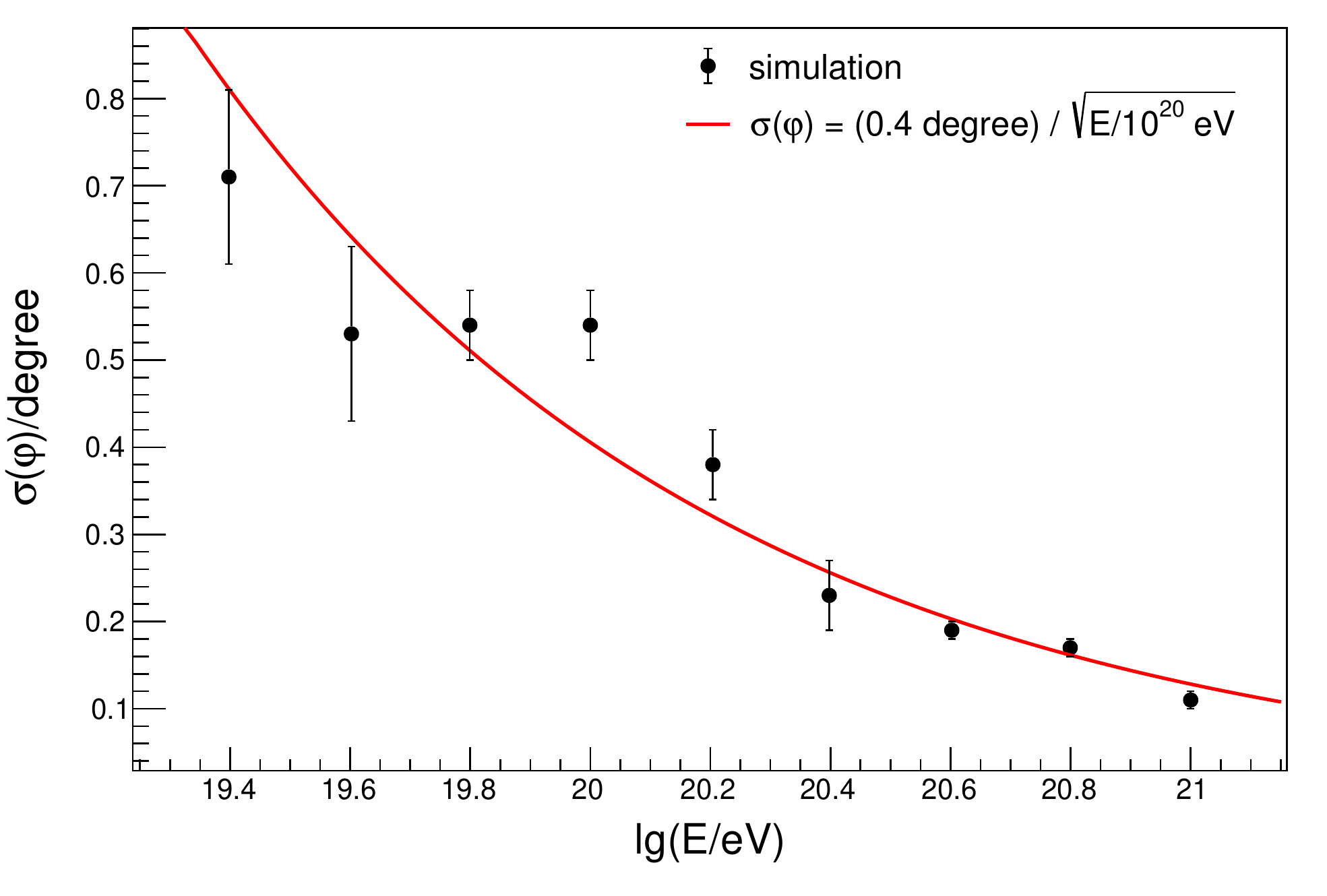}
\includegraphics[width=0.98\columnwidth]{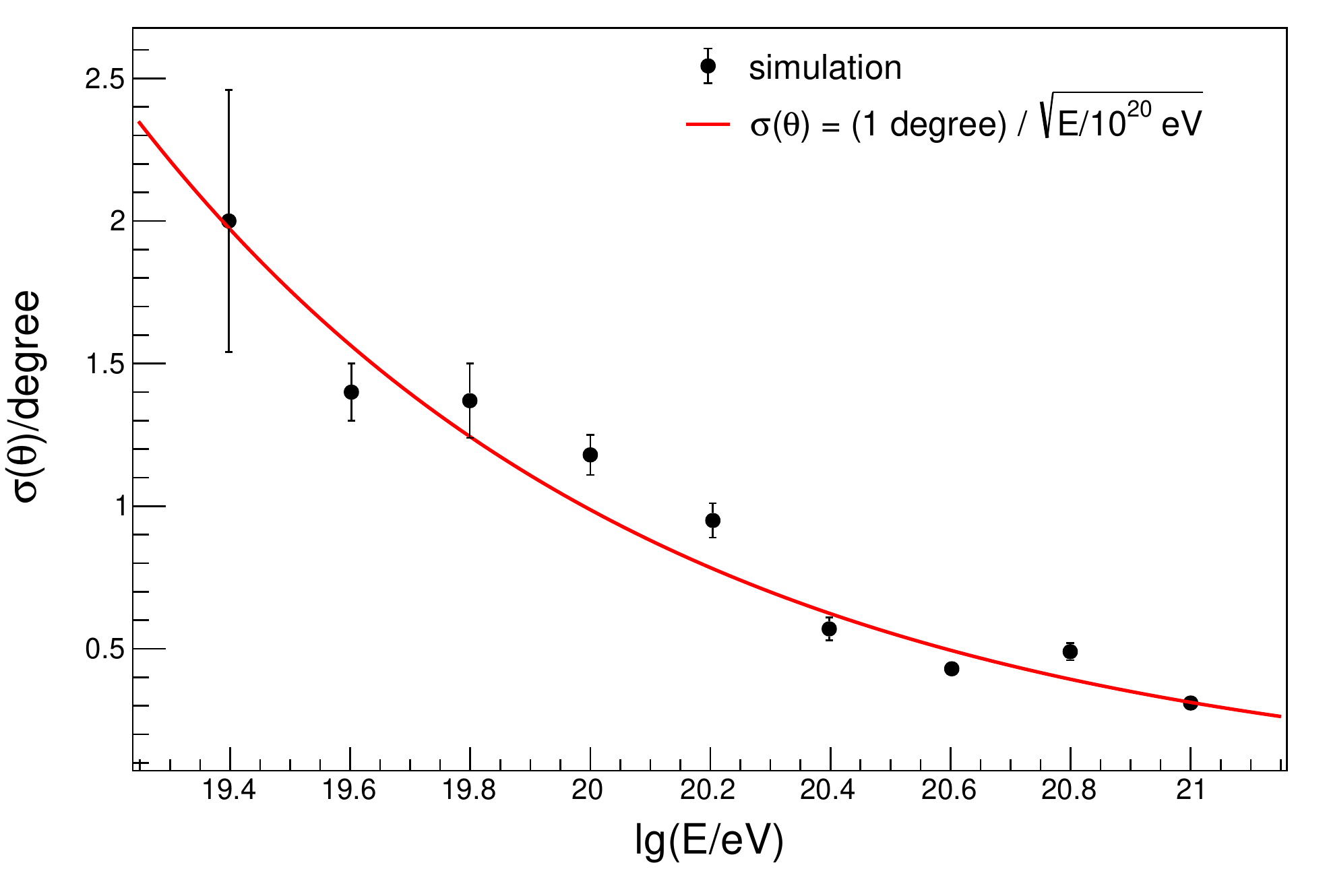}
    \caption{POEMMA's simulated stereo-reconstructed angular resolution versus
UHECR energy. Left: Azimuth angle, right: zenith angle. \label{AngRes}}
\end{figure*}

\begin{figure}[tb]
  \centering
\includegraphics[width=0.99\linewidth]{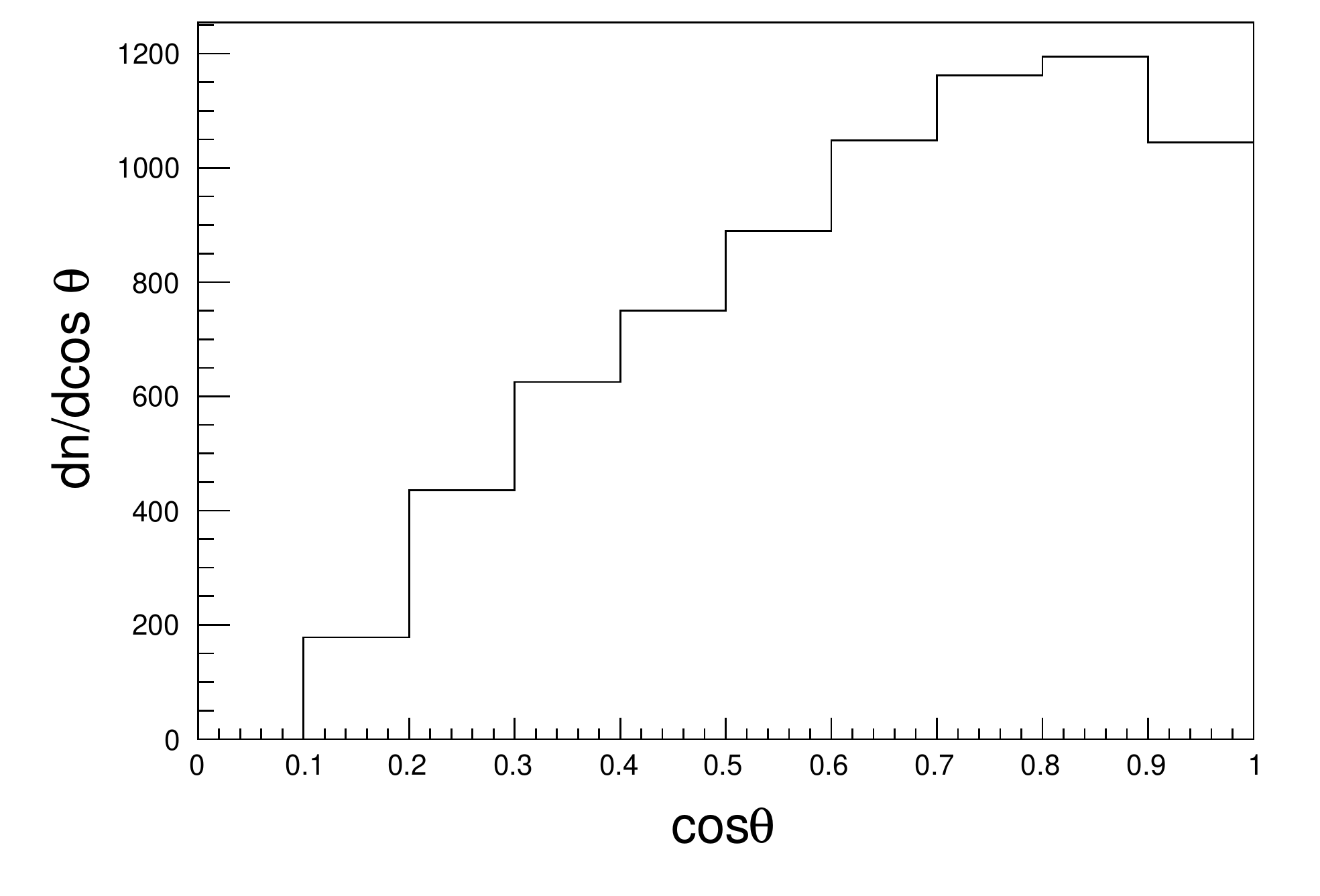}  \caption{Distribution of zenith angles
  of triggered events above 50~EeV.}
\label{fig:dndcostheta}
\end{figure}

\begin{figure*}[tbp]
\centering
\includegraphics[width=0.98\columnwidth]{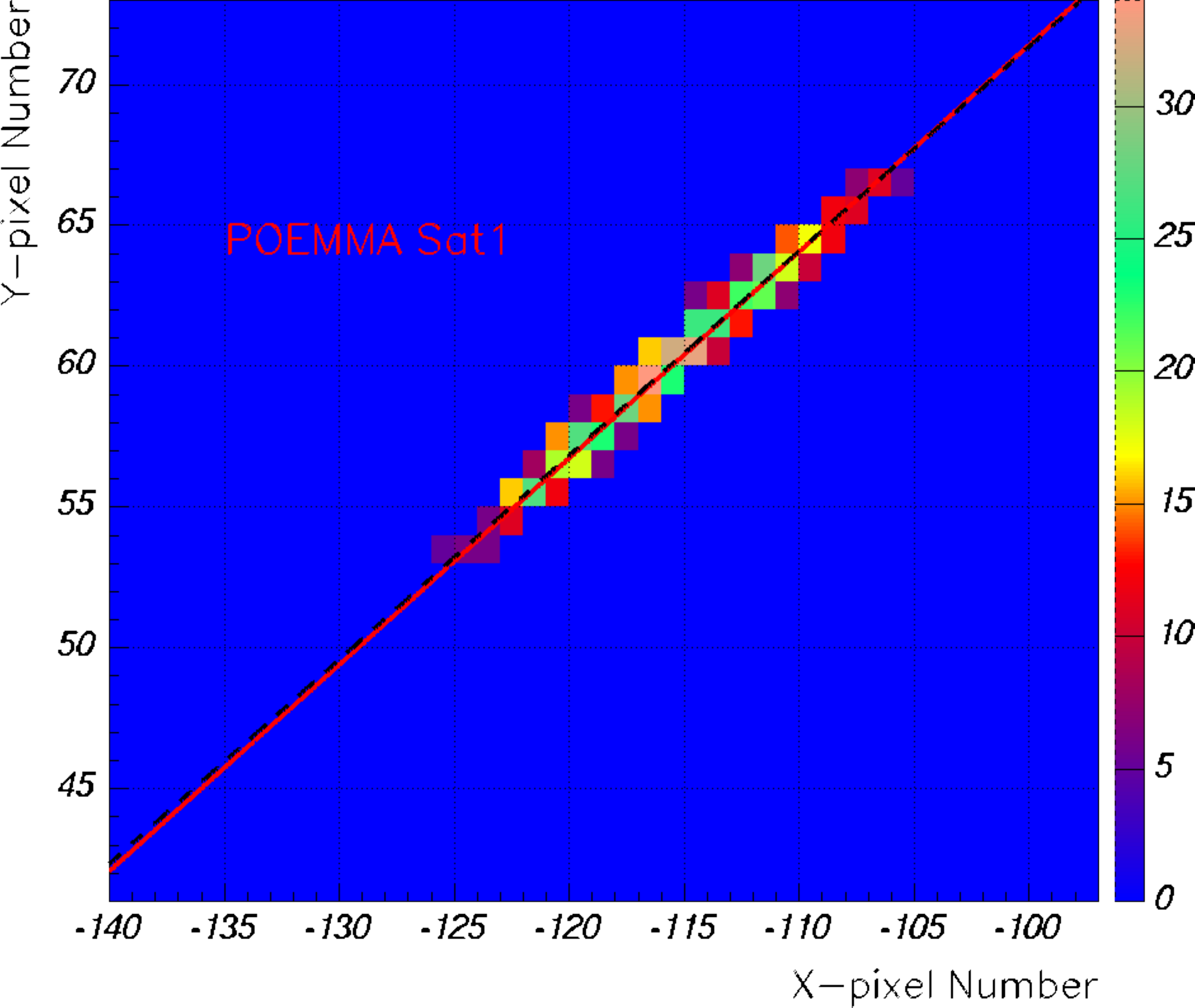}
\hspace{0.1 cm}
\includegraphics[width=0.98\columnwidth]{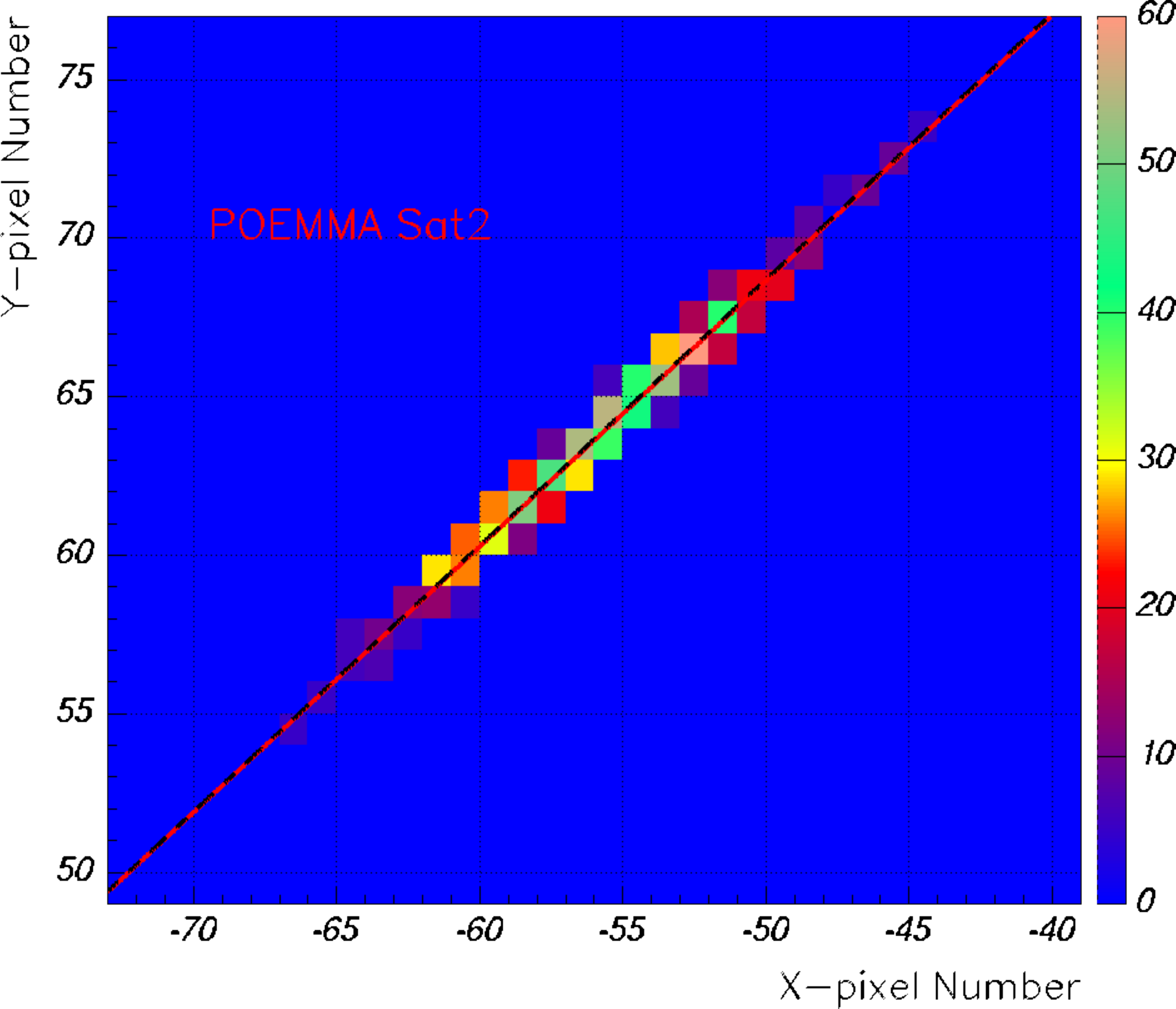}
    \caption{A stereo reconstructed 50 EeV UHECR in the two POEMMA focal planes. The solid line denotes the simulated trajectory while the dashed line shows the reconstructed trajectory. The color map provided the photo-electron statistics in each pixel \label{StereoEvent}}
\end{figure*}

\begin{figure}[tb]
  \centering
\includegraphics[width=0.99\linewidth]{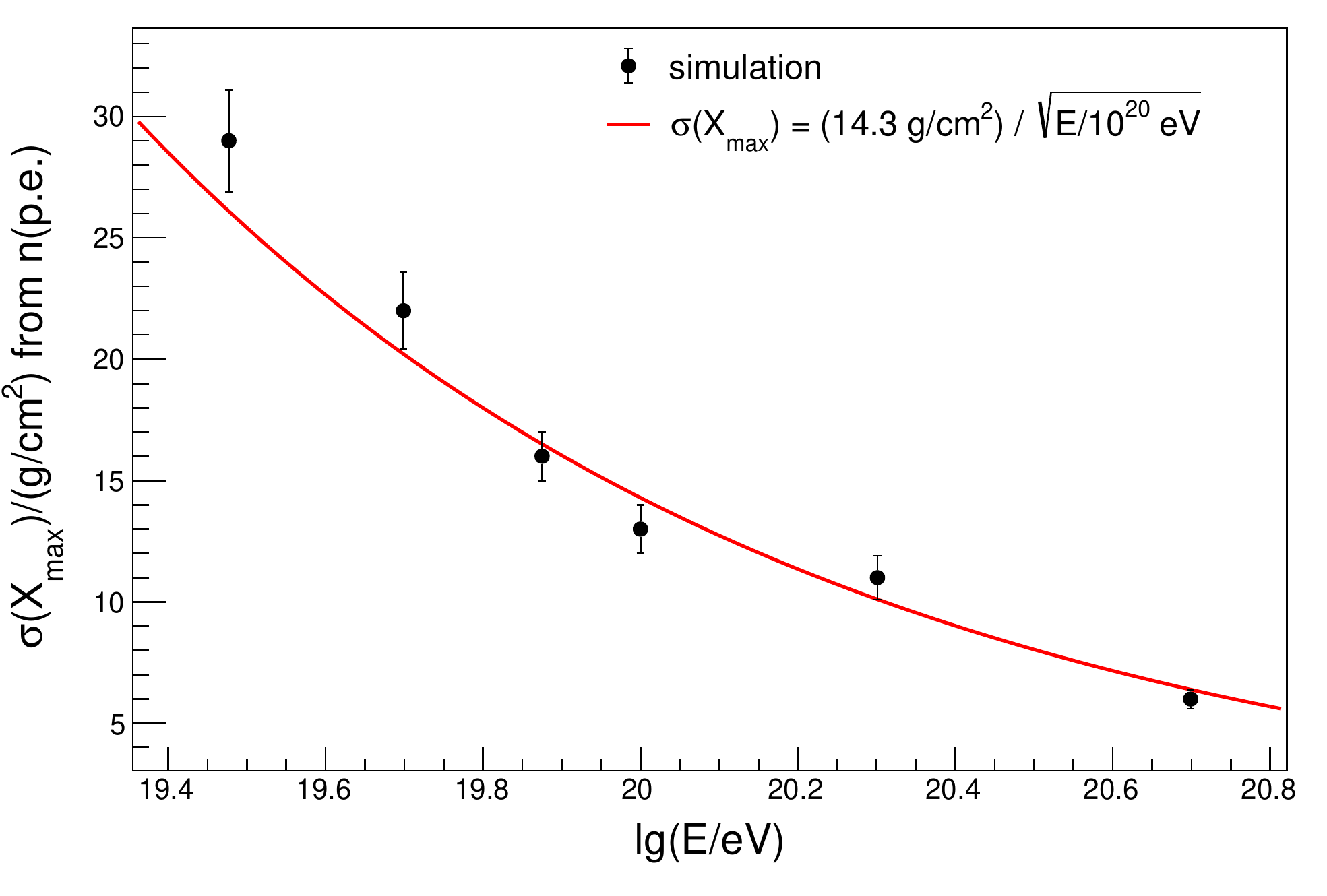} 
\caption{Single-photometer $X_\text{max}$-resolution from
  photo-electron statistics.}
\label{fig:xmaxResDetail}
\end{figure}

We have performed detailed simulation studies using POEMMA's optical
performance to determine POEMMA's UHECR exposure, angular resolution,
and nuclear composition ($X_{\rm max}$) resolution. For stereo UHECR
mode, we modeled a satellite configuration with a 300~km separation and
tilted ($\pm 12.2^\circ$) to view a common atmospheric volume between the two satellites.  The PFC instrument response model is derived from the ESAF simulations (described in Appendix~\ref{app:a}) which take into account the effects of air glow background, the PFC trigger, and electronic response.  A single POEMMA telescope pointing in nadir mode was modeled using the methodology of the stereo simulation. The UHECR event selection was then tuned by setting requirements on the number of pixels above a PE threshold to achieve a similar trigger UHECR aperture as reported by the ESAF simulations; see Appendix~\ref{app:a}. This condition was then used for the modeling of POEMMA's stereo response for the condition of separating the satellites by 300~km and tilting them to view a common volume of atmosphere between them. 

Each measured EAS trajectory in the POEMMA focal planes defines a unique geometrical plane. Simple geometry determines a line in 3-dimensional space where these two planes intersect, corresponding to the EAS trajectory. As long as the opening angle between these planes is larger than $\sim 5^\circ$, the reconstruction of the EAS trajectory is robust, due to POEMMA's excellent EAS pixel angular resolution, yielding superb UHECR angular resolution.  Figure~\ref{AngRes} shows POEMMA's stereo
reconstructed angular resolution, which is $\sim 1^\circ$ or better
above 30~EeV, and highlights the strength of the stereo reconstruction
technique when one has good pixel angular resolution. 
In Fig.~\ref{fig:dndcostheta} we show zenith angle distribution of triggered events above 50~EeV
and in Fig.~\ref{StereoEvent} we show an example 50~EeV UHECR as measured in the POEMMA focal planes.

The stereo
trigger condition in each satellite leads to a highly efficient
reconstruction fraction of $\sim 80\%$, with the losses due mainly to
the requirement of the $\sim 5^\circ$ opening angle between each EAS
geometrical plane. To estimate the energy resolution in stereo mode we performed simulations with ESAF in 
monocular mode (see Appendix~\ref{app:a}), assuming $\sigma = 1^\circ$ angular resolution in both zenith and azimuth angles based on the results of the stereo reconstruction (see Fig.~\ref{AngRes})
This leads to a resolution of 26\% and 
24\% at 50 and 100 EeV respectively. Since the two satellites provide an 
independent measurement of the same EAS, the resolution can be expected 
to be a factor of $\sqrt{2}$ better than the monocular one with fixed 
geometry, i.e. 18\% and 17\% at 50 and 100 EeV respectively. It is 
worthwhile noting that these numbers can be considered conservative, 
because the ESAF simulations used a lower quantum efficiency and a larger time binning (2.5 $\mu$s) than it will be 
used in POEMMA (1 $\mu$s). A partial estimate of the $X_{\rm max}$ resolution is evaluated by  considering the effects of the finite PE statistics when reconstructing the shower profile. The results, shown in Fig.~\ref{fig:xmaxResDetail}, give an $X_{\rm max}$ resolution of $20~{\rm g/cm^2}$ at 30~EeV. The $X_{\rm max}$ resolution is further degraded by
effects of angular resolution and acceptance. We will consider all these effects in Sec.~\ref{sec:4} to estimate the total $X_{\rm max}$ resolution.

\begin{figure}[tb]
\begin{center}
\includegraphics[width=0.45\textwidth]{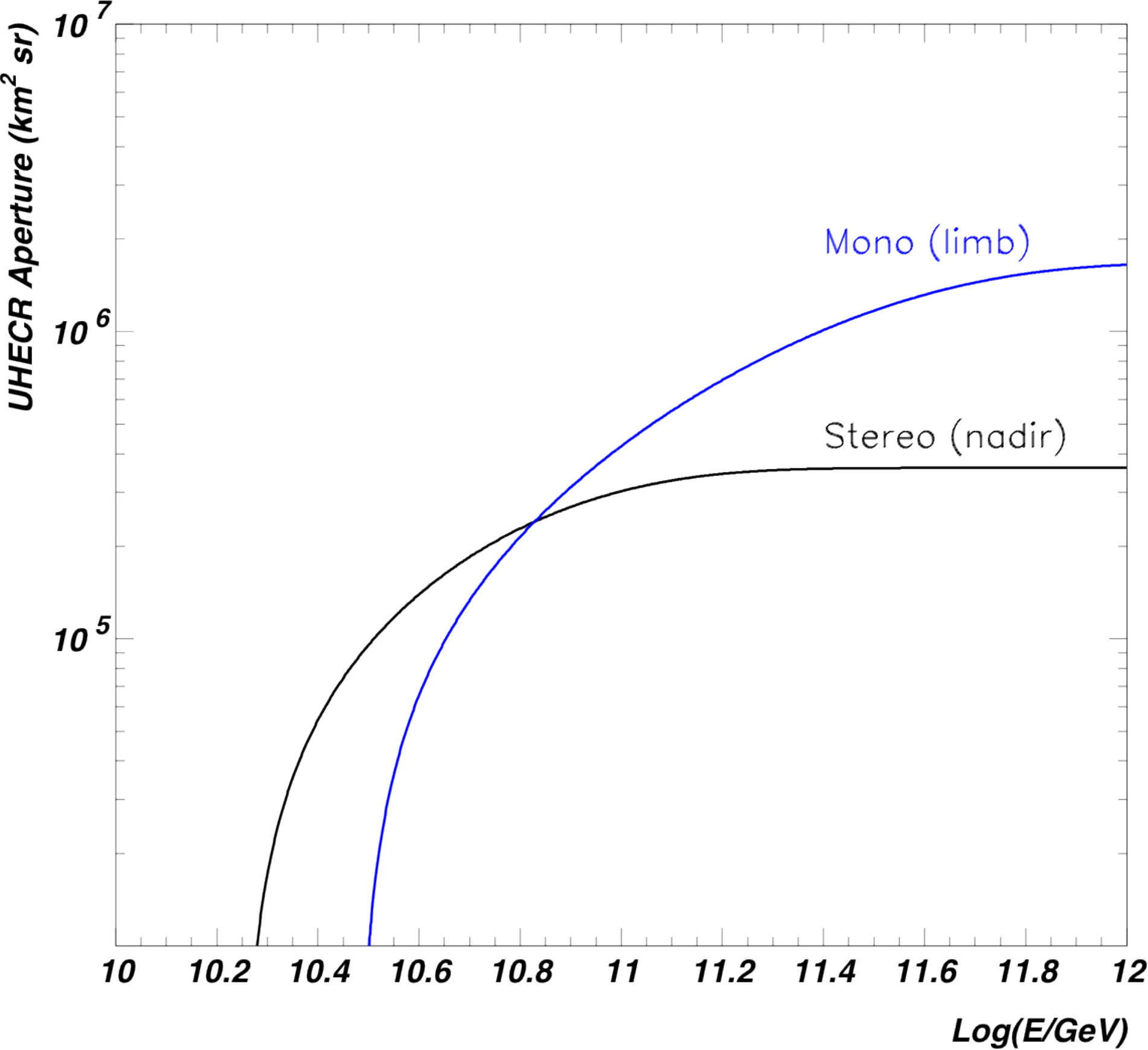}
\end{center}
\caption{The simulated UHECR aperture after event reconstruction for POEMMA for stereo mode and tilted mode.}
\label{trig}
\end{figure}

POEMMA is expected to operate also in tilt mode when observing the Cherenkov signal from tau
neutrinos and also to exploit different combinations of stereoscopic vision.
By tilting the instrument the EAS distance increases and therefore the energy 
threshold of the instrument increases as well. Moreover, it is expected that 
the background increases as the the column density of airglow emitting layer 
increases with the tilt angle. This is taken into account assuming that such 
increase is proportional to $\cos(\theta)^{-1}$. In Fig.~\ref{trig} we exhibit the UHECR proton aperture, after taking into account event reconstruction efficiency, for stereo viewing, as well as that when the satellites are tilted by 47$^\circ$ for the condition when POEMMA is viewing the limb of the Earth in neutrino mode. The stereo results are based on the stereo POEMMA simulation, whereas the 47$^\circ$ tilted mode results are based on the ESAF simulation using monocular reconstruction of POEMMA.

\subsection{Duty Cycle and Exposure}

The estimation of the UHECR exposure of a space-based experiment like
POEMMA requires accounting for:
{\it (i)}~the characteristics of the EAS
development in the atmosphere as observed from space, {\it (ii)}~the properties of
the telescope, including its orbit and FoV,
{\it (iii)}~the various
sources of {\it steady} background like night-glow and moonlight,
{\it (iv)}~the overall optical transmission properties of the atmosphere, in
particular the possible presence of clouds, and
{\it (v)}~the effect of anthropogenic light, or other light sources such as transient
luminous events (TLEs) and meteors. Topics {\it (i)} and {\it (ii)} are the principal factors
determining the threshold in energy and the maximum
aperture of the telescope. Topic {\it (iii)} limits the
observational duty cycle of the mission. Topics {\it (iv)} and {\it (v)}
affect the instantaneous aperture of the telescope.
The role of each of the above listed aspects has been studied in the past
to evaluate their contribution to the determination of the JEM-EUSO
exposure. A detailed description of such studies was reported 
in~\cite{Adams:2013vea}. 

The observational duty cycle of POEMMA is the fraction of time
in which the EAS measurement is not hampered by the brightness
of the atmosphere. The atmospheric brightness, which is mainly due to the night-glow and back-scattered
moonlight, is variable over time. We define the observational
duty cycle as the fraction of time $\eta_0$ in which the background intensity
$I_{\rm BG}$ is lower than a given value $I_{\rm BG}^{\rm thr}$.  The moonless
condition is assumed to be $I_{\rm BG}$ = 500~photons m$^{-2}$ sr$^{-1}$ 
ns$^{-1}$ in the range $300 < \lambda/{\rm nm} < 500 $. This produces a signal of 
$\sim$~1.5~photo-electron pixel$^{-1}$ $\mu$s$^{-1}$ for POEMMA. To remain conservative, herein we adopt  $I_{\rm BG}^{\rm thr}$ = 1,500~photons m$^{-2}$ sr$^{-1}$ ns$^{-1}$. In this condition, the signal of a 100~EeV shower is still more than 5 times brighter than the background level well around the shower maximum.
The back-scattered moonlight is calculated from the moon phase and
its apparent position as seen from the POEMMA orbit.
The zenith angle of the Sun is required to be greater than 109$^\circ$ for an 
orbiting altitude of 525~km.
The observational duty cycle $\eta_0~(I_{\rm BG}~<~I_{\rm BG}^{\rm thr}~)$ 
is of the order of $\sim$ 20\%. This value 
is actually conservative at $E \gtrsim 100~{\rm  EeV}$,
where it is possible to also operate in higher background levels.

Another source of background is the UV emission produced by direct particles
interacting in the detector, particularly with the corrector lens due to its
large size and transparency.  For the JEM-EUSO mission, which was designed to use two lenses, the increase of UV light due to this contribution was determined to be negligible ($\sim 1\%$). This will also be the case for POEMMA. A point worth noting at this juncture is that this estimate takes into account the UV emission in the corrector lens due to
trapped electrons in the center of the South Atlantic Anomaly, where
the flux of particles exceeds by orders of magnitude  the average value.

In addition to the diffuse sources of background, there are transient
or steady local sources, such as, lightning and TLEs, auroras or city lights.
To estimate the effect of lightning and TLEs, we scale the
rate of events detected by Tatiana satellite~\cite{Garipov}. We find this prevents observation by 
$\sim 4\%$. This scaling does not take into account the double counting
due to the fact that the presence of lightning is very often associated to the
presence of high clouds. This is explicitly done to reinforce the conservative
nature of our calculation. Because of the POEMMA equatorial orbit the presence of auroras is negligible.
This was evaluated for JEM-EUSO (ISS orbit) and even in the case
maximum solar activity, the effect is of the level of $\sim 1\%$.

To evaluate the effect of light sources on the Earth, which are
mainly anthropogenic, we use the Defense Meteorological Satellite Program
(DMSP) database. To remain conservative, in the study for JEM-EUSO it was assumed that
 no measurement of EASs is performed if, in a region viewed by a PDM,
there is at least one pixel which detects a light intensity which exceeds
the average level by a factor of 3 or more. The average level of intensity essentially corresponds to the typical 
condition on oceans. With this assumption, the inefficiency of the
instantaneous aperture is about $7\%$. For POEMMA, the FoV of a PDM is
4 times larger, but the trigger logic works at elementary cell (EC) level, which is 1/9 of a PDM.
Therefore, by assuming the same rule at EC level, the above results for 
JEM-EUSO remains anyway conservative also for POEMMA.

In order to quantify the reduction of the effective instantaneous
aperture of the telescope due to the presence of clouds, a study on the
distribution of clouds as a function of top altitude $H_{\rm C}$, optical depth $\tau_{\rm C}$ and
geographical location were performed using several meteorological data sets
and reported in~\cite{cloudy}. 
Showers were simulated using ESAF according to the
matrix of cloud occurrence determining the trigger efficiency in
the different conditions, and obtaining the corresponding
aperture to estimate the ratio $\kappa_{\rm C}$ between the aperture when the role of clouds is included, 
compared to purely clear atmosphere.
Selecting the cases of clouds with $\tau_{\rm C} <1$, or shower maximum above 
the cloud-top altitude (i.e. $H_{\max} > H_{\rm C}$) leads to $\kappa_{\rm C} 
\sim~72\%$ almost independently of energy~\cite{cloudy,Adams:2013vea}.

\begin{figure}[tb]
\begin{center}
\includegraphics[width=0.48\textwidth]{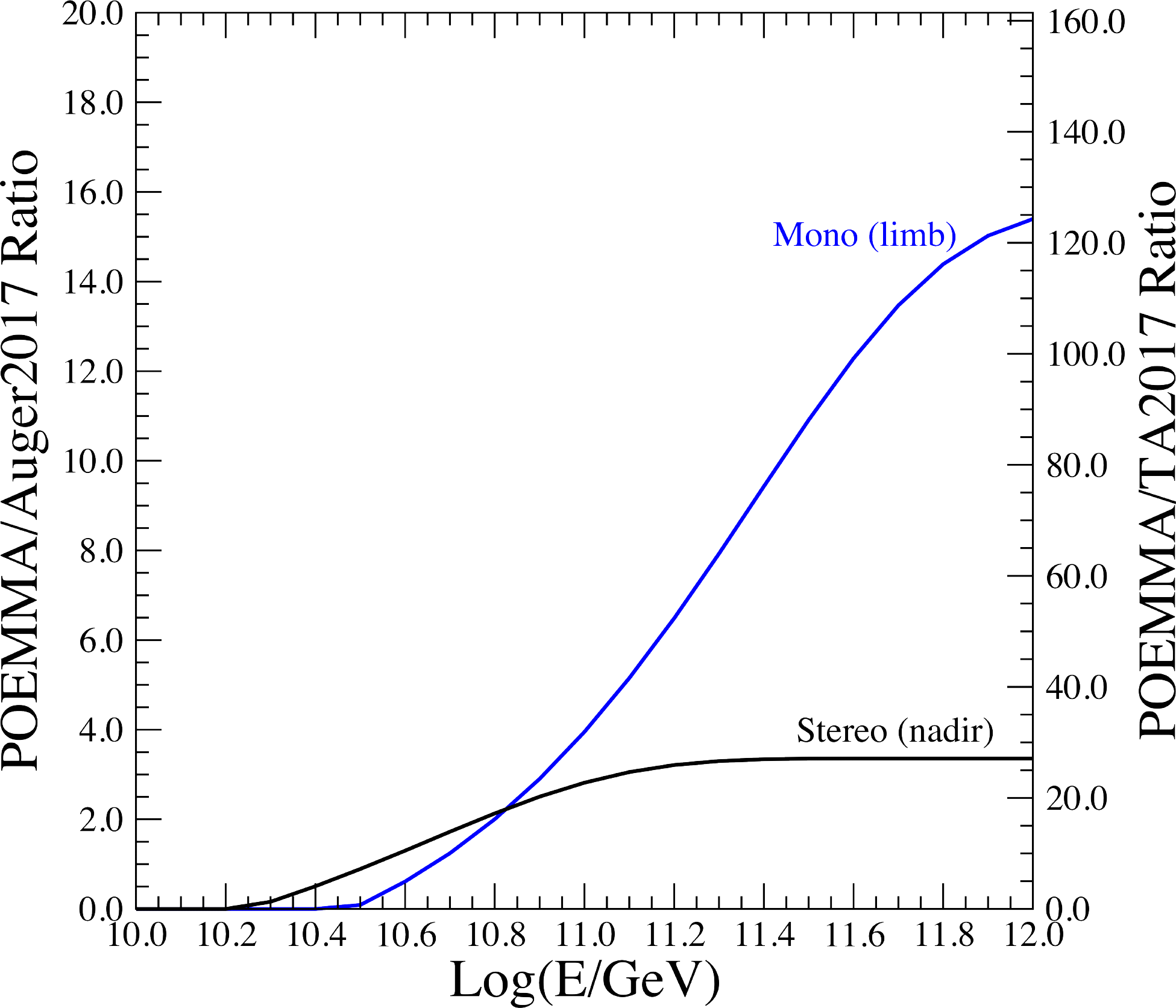}
\end{center}
\caption{Examples of the 5 year POEMMA stereo and tilted monocular UHECR exposure in terms Auger exposure and TA until ICRC-2017.}
\label{exposComp}
\end{figure}

All of the above factors give an overall conversion factor from geometrical 
aperture to exposure of about 13\% for POEMMA at 525 km. Based on the UHECR stereo and tilted monocular UHECR apertures, we calculate the POEMMA 5 year exposures, for both stereo (nadir) and mono (limb) configurations, compared to the Auger and TA exposures reported at the 2017 ICRC conference. Both the stereo and monocular reconstruction studies show an 80\% reconstruction efficiency and a 13\% duty cycle based on adapting JEM-EUSO studies assumed for POEMMA 
observations. In Fig.~\ref{exposComp} we show the 5 year POEMMA exposures in relation to Auger and TA. 

Apart from the effect of the clouds, the sky coverage of the POEMMA exposure
is determined by its orbit and the observable dark night time for a given
direction in the Celestial Sphere. The uniformity over the right ascension
slightly deviates from the uniformity due to the seasonal variation of the time
that POEMMA stays in the Earth's umbra per orbital period which is longest 
around equinoxes. An effect of the Earth's orbit eccentricity appears as the
excess in the observation time for the winter time in Southern Hemisphere when
the Earth revolves at the slowest velocity.

The differential distribution of the exposure is primarily expressed as a
function of the declination. As can be seen in 
Fig.~\ref{fig:dndcostheta}, at lower energies the trigger efficiency  
increases with zenith angles. On the other hand, the effective area of the 
instantaneous apertures is proportional to the cosine of the zenith angle. These 
effects compose the exposures in terms of the observable time and geometrical 
apertures. The zenith angle dependence is mostly irrelevant for the highest
energies. 

\begin{figure*}[tb]
 \includegraphics[width=0.9\textwidth]{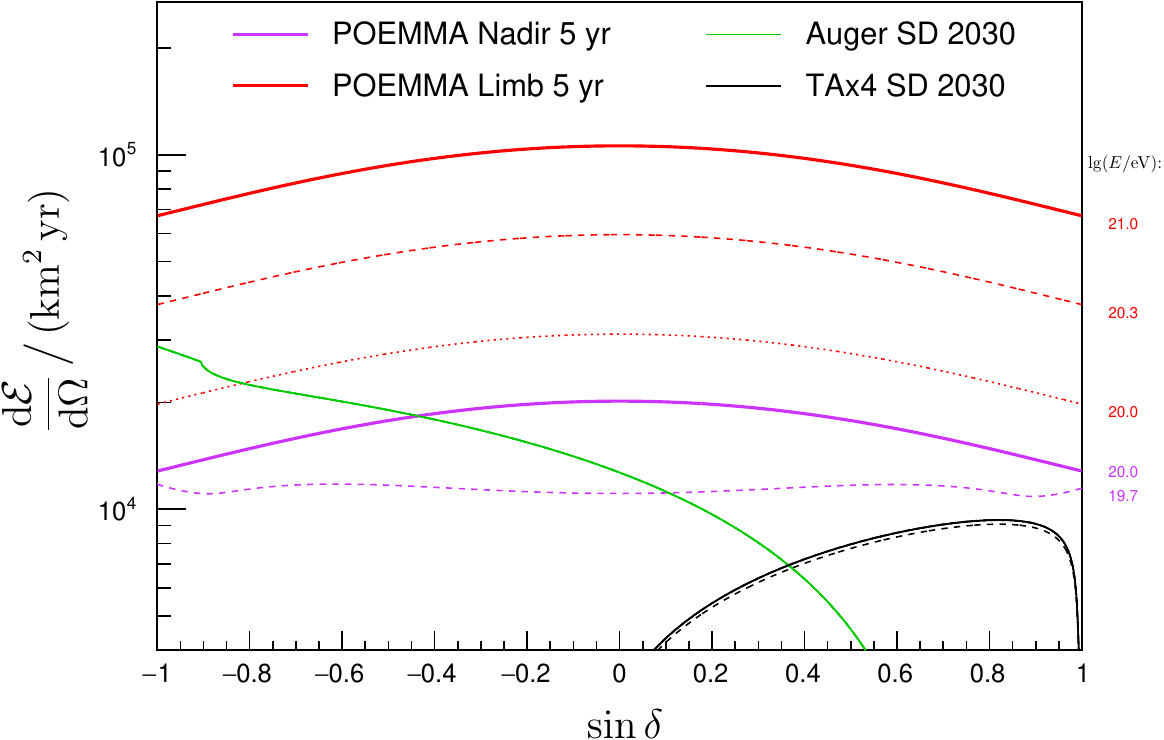}
  \caption{Differential exposure as a function of declination 
     in different modes, assuming a single-mode operation for the full 5-year  benchmark. Purple curves denotes the
    stereo (near-nadir) mode at $10^{19.7}$~eV~(dashed) and $10^{20}$~eV~(solid). Red curves
    denotes the limb-viewing mode at $10^{20}$~eV~(dashed), $10^{20.3}$~eV~(dash), and
    $10^{21}$~eV~(solid). The exposures of the surface detectors of Auger and
    TA (including the TA$\times$4 upgrade) assuming being in operation until
    2030 are shown as green and black curves, respectively.}
  \label{fig:declination2}
\end{figure*}

The differential exposure as a function of declination for five years of the
POEMMA operations in each of the two different modes is shown in Fig.~\ref{fig:declination2}.
Purple curves denote the stereo (near-nadir) mode at $50$~EeV~(dashed) and
$100$~EeV~(solid). Red curves denote the limb-viewing mode at
$100$~EeV~(dashed) 
and $200$~EeV~(dash), and $1$~ZeV~(solid). The exposures of the
surface detectors assuming being in operation until 2030 are shown as green and
black curves for Auger and TA including the TA$\times$4 upgrade, respectively.
In this figure, the absolute exposures in units of km$^2$~sr~yr have been normalized
considering the overall effect of the clouds as studied for
 JEM-EUSO in Refs.~\cite{cloudy,Adams:2013vea}. Compared
with geographically settled surface detectors, the major advantage of POEMMA
is the full sky coverage over the whole Celestial Sphere with the single
experiment that may reduce the systematic uncertainties, e.g., in energy scale,
for comparing the different part of the sky. 

In the same references, the climatology of cloud distribution and the
fundamental role to the overall exposure have been studied. According to
these studies, showers from the large zenith angles develop at higher altitudes
and thus, seen from POEMMA, they are more impervious to the presence of clouds.
The effects on the exposure map depend upon the event selection cuts applied on the analysis
of air showers with respect to the cloud characteristics such as cloud-top
height. For the different conditions compared in Fig.~\ref{fig:declination2},
different cuts should be optimized according to the science purpose. 

\begin{figure*}[tbp]
\includegraphics[width=0.49\textwidth]{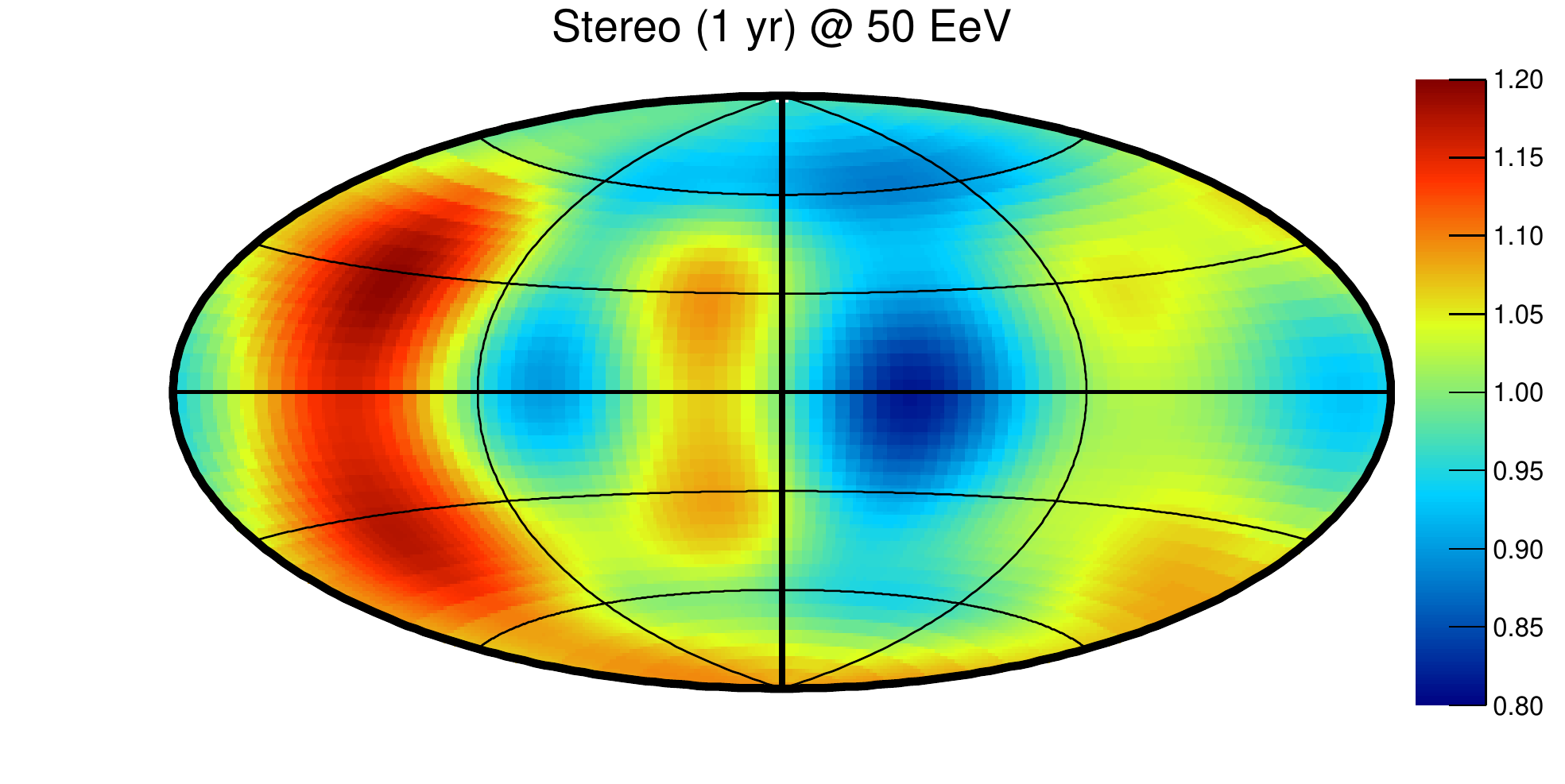}
\includegraphics[width=0.49\textwidth]{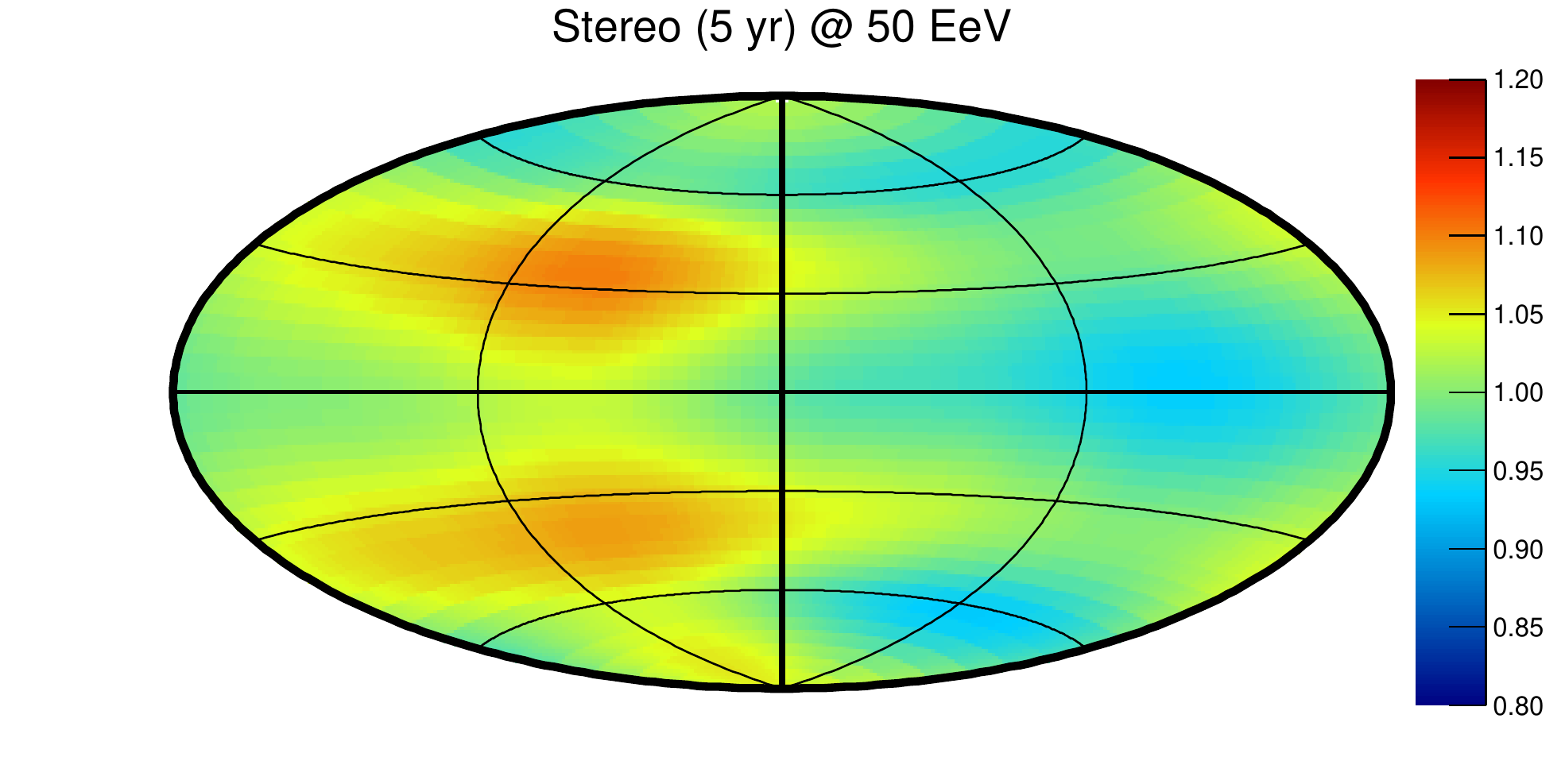}
\includegraphics[width=0.49\textwidth]{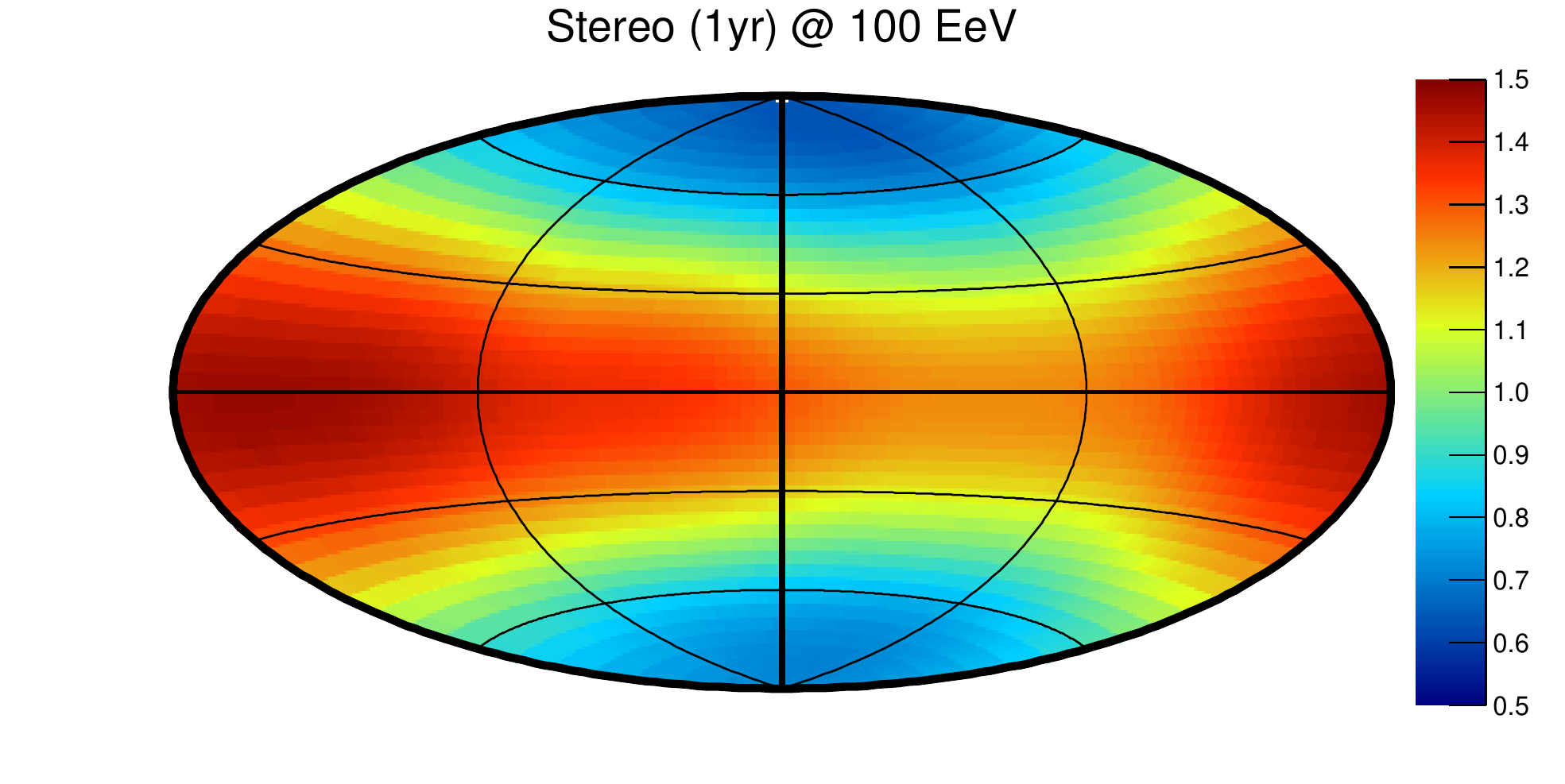}
\includegraphics[width=0.49\textwidth]{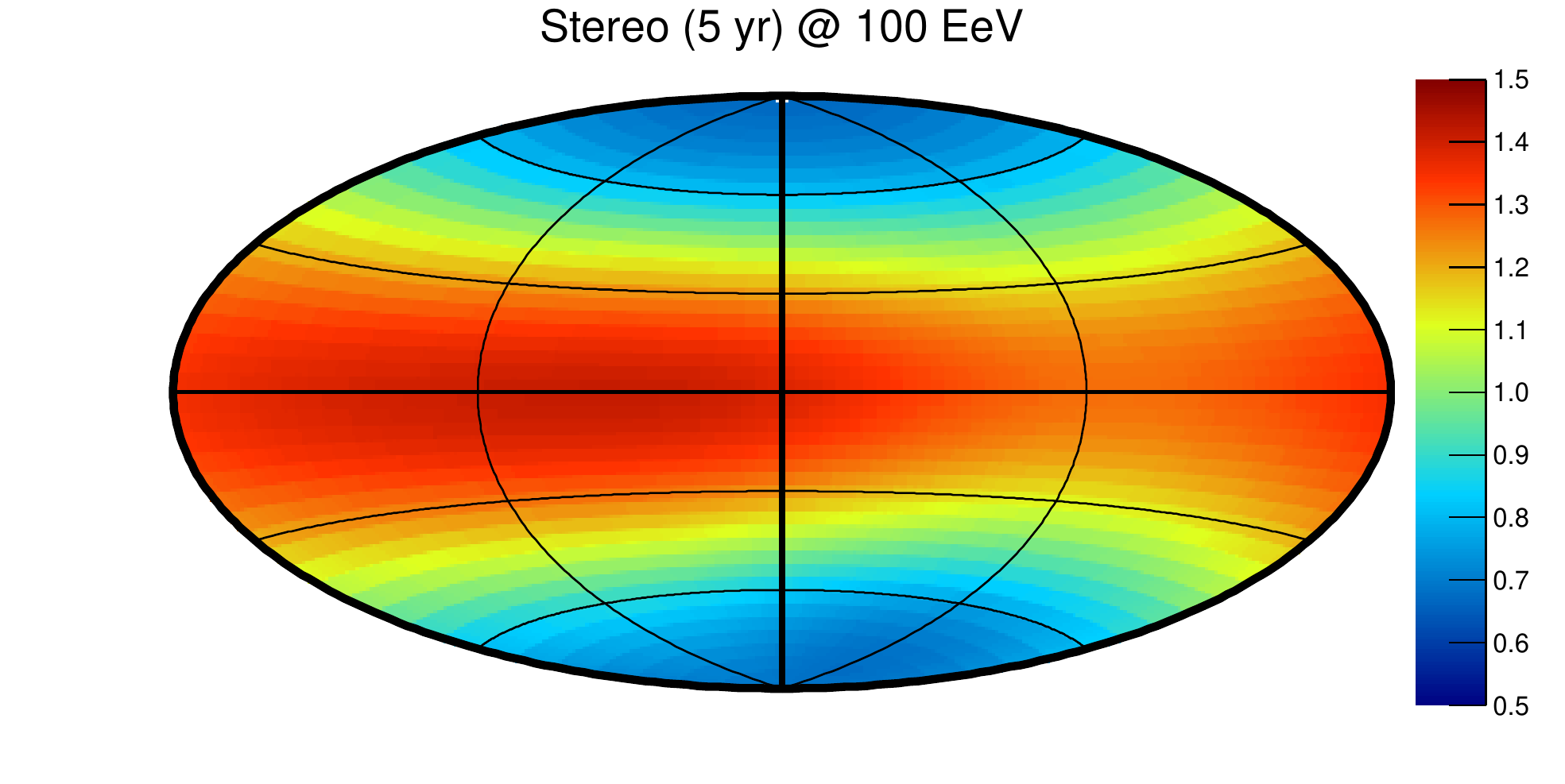}
    \caption{POEMMA's UHECR sky exposures in declination versus right ascension. The color scale denoting the exposure variations in
terms of the mean response taking into account the positions of the sun and the moon during the observation cycle. \label{fig:skymap}}
\end{figure*}

In Fig.~\ref{fig:skymap} we show celestial sky coverage maps, where a $\sim50\%$ variation is evident in
the uniformity of the UHECR sky exposure assuming an isotropic
flux. Thus, POEMMA is sensitive to UHECR sources in both the northern
and southern hemisphere.  POEMMA will measure the UHECR source
distribution on the full celestial sphere under a single experimental
framework with a well-defined UHECR acceptance, mitigating the issues
of cross-comparisons inherent to viewing different portions of the sky
with multiple experiments.  The response shown in each panel of  Fig.~\ref{fig:skymap} was
calculated assuming that the POEMMA telescopes point in a near nadir
viewing configuration used in stereo mode. The ability of the space-based POEMMA
telescopes to tilt towards the northern or southern hemisphere allows
for the sky exposure to enhance for a specific
hemisphere. Likewise, it is easy for POEMMA to change its pointing direction for
a sequence of orbital periods to further tailor the UHECR sky
coverage.

\section{UHECR Performance}

\label{sec:4}

POEMMA is designed to obtain definitive measurements of the UHECR spectrum, nuclear composition, and source location for $E \agt 20~{\rm EeV}$. UHECR events are well reconstructed by POEMMA when the two orbiting satellites stereoscopically record the waxing and waning of the EAS fluorescence signal. The video recordings with $1~\mu{\rm s}$ snapshots from each POEMMA satellite define a plane (the observer-EAS plane). The intersection of the two planes from both satellites determines a line in 3-dimensional space corresponding to the EAS trajectory. For opening angles between these two planes larger than $\sim 5^\circ$ , the reconstruction of the EAS trajectory is robust given POEMMA’s excellent pixel angular resolution, yielding superb UHECR angular resolution.  

\begin{figure}[tb]
\centering
\includegraphics[width=0.99\linewidth]{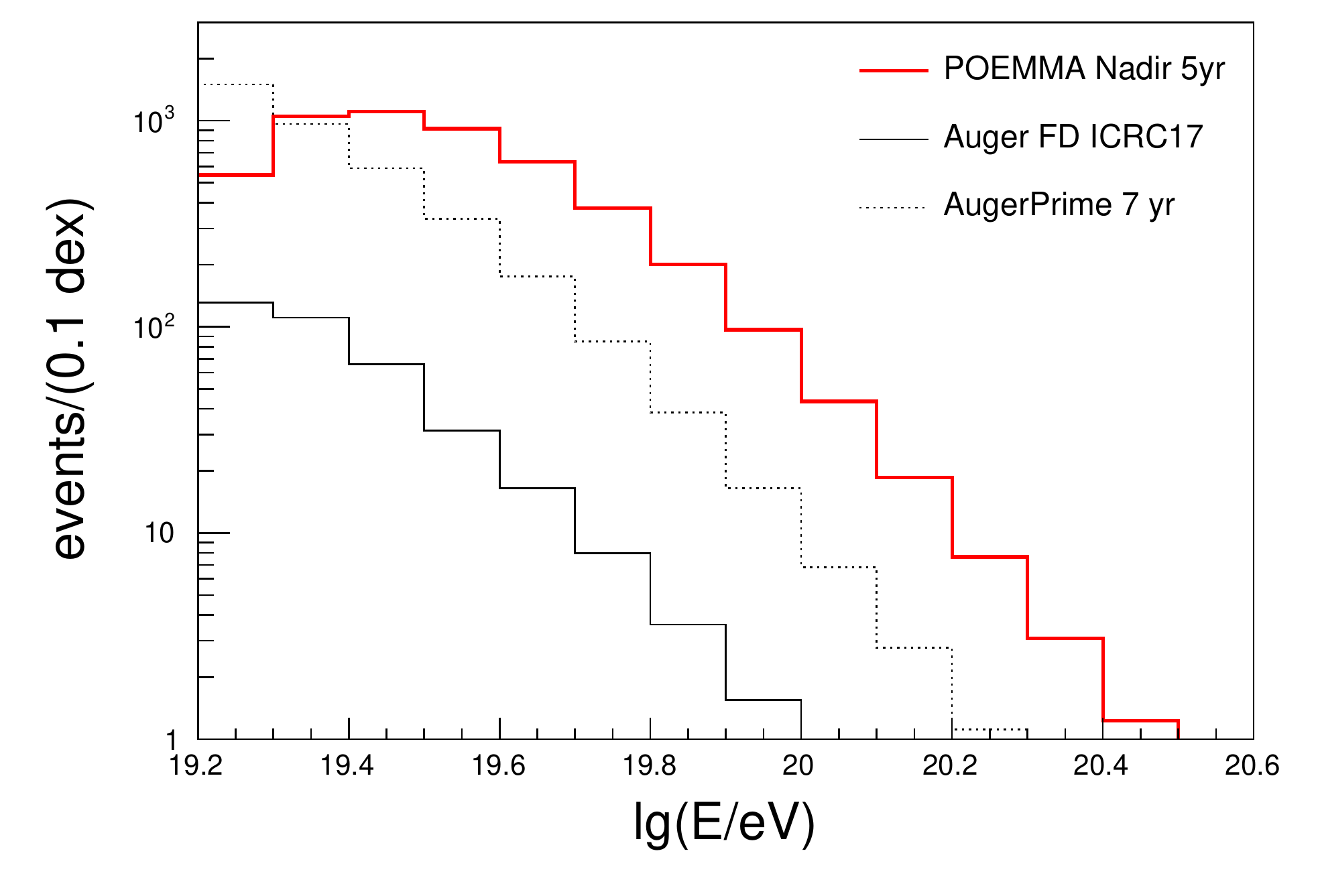}
\caption{Number of UHE events with composition information detected by
  POEMMA for five years of data taking in stereo (near-nadir) operational mode (red
  line).  For comparison, the current event statistics collected with
  the fluorescence detector of the Pierre Auger Observatory is shown
  as solid black lines and the expected number of events from
  AugerPrime are indicated by dashed black lines.}
\label{fig:events1}
\end{figure}
\begin{figure}[tb]
\centering
\includegraphics[width=0.99\linewidth]{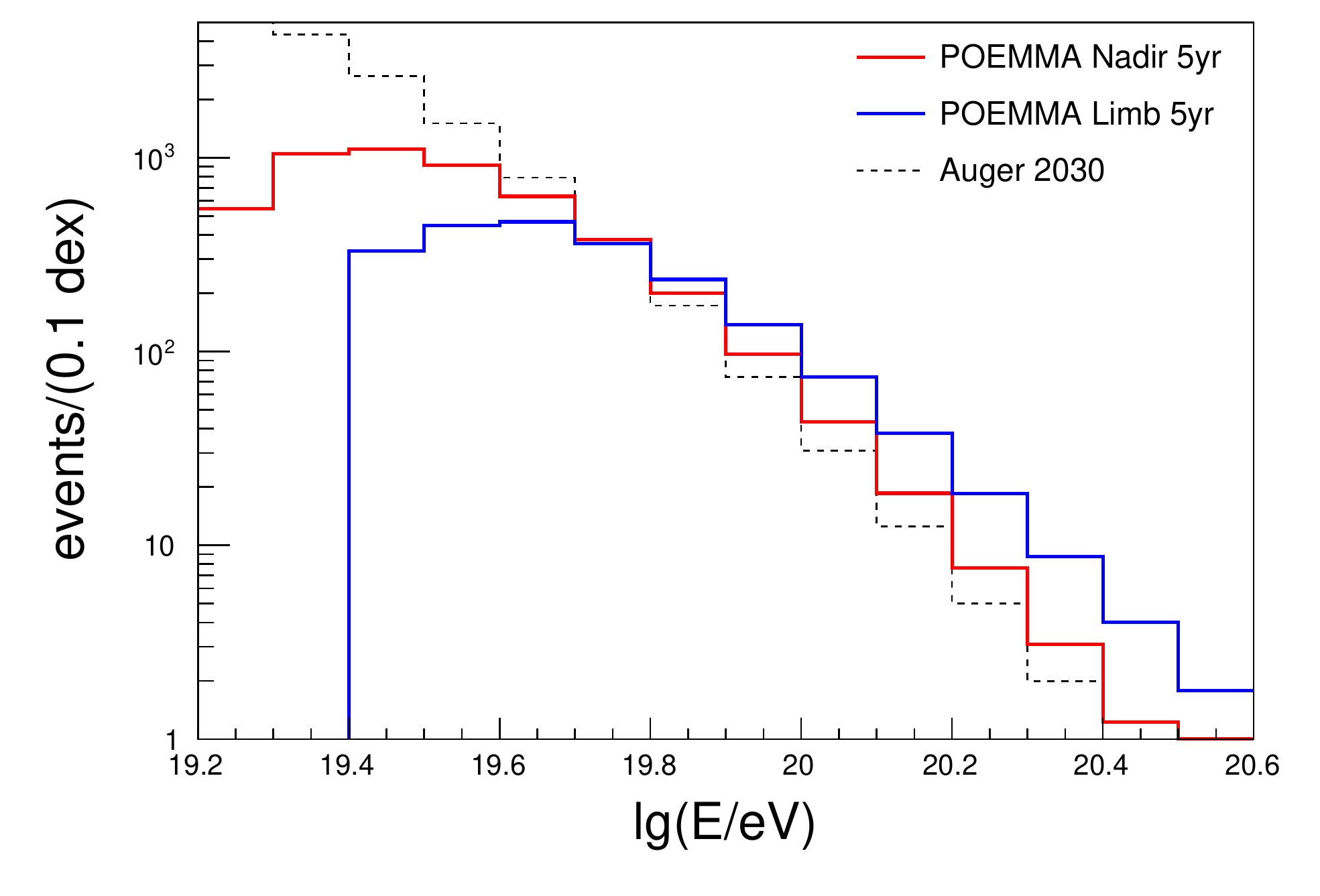}
\caption{Number of UHE events detected by POEMMA for five years of
  data taking in stereo (near-nadir) (red line) and limb-viewing (blue line) operational
  mode assuming the Auger energy spectrum. The expected number of events
  collected by Auger in case of a continued operation until 2030 is shown
  as dashed black line.}.
\label{fig:events2}
\end{figure}
\begin{figure}[tb]
\centering
\includegraphics[width=0.99\linewidth]{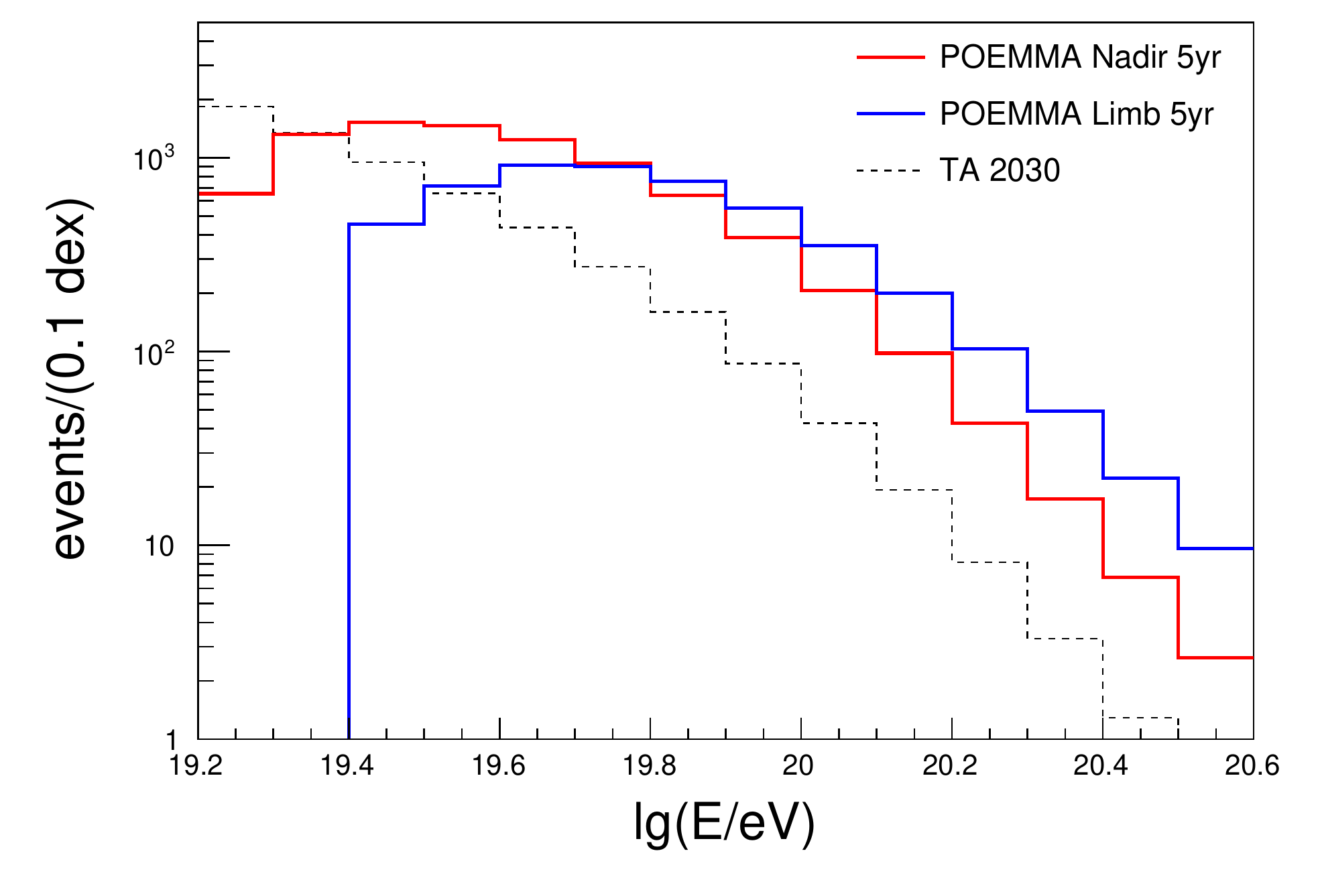}
\caption{Number of UHE events detected by POEMMA for five years of
  data taking in stereo (near-nadir) (red line) and limb-viewing (blue line) operational
  mode assuming the TA energy spectrum. The expected number of events
  collected by TA (including its upgrade TAx4) in case of a continued
  operation until 2030 is shown as dashed black line.}.
\label{fig:events3}
\end{figure}

The expected number of events detected during five years in stereo and
limb-viewing operational mode is shown in Figs.~\ref{fig:events1},
\ref{fig:events2}, and \ref{fig:events3}. To estimate the event rate in these figures we assumed
that the UHE flux follows the fit of the Auger combined spectrum given
in~\cite{Aab:2017njo} (Figs.~\ref{fig:events1} and \ref{fig:events2})
and the one from TA given in~\cite{Matthews:2017waf} (Fig.~\ref{fig:events3}).  The comparison of the events in stereo mode
with the current statistics of the Auger FD shows that POEMMA will be
a giant leap forward regarding the detection of cosmic rays with the
fluorescence technique. POEMMA will increase the number of UHE
cosmic-ray events with direct observation of $X_\text{max}$ and energy
from 62 (last integral bin of the Auger FD analysis above
$30$~EeV) to more than 2000.\footnote{Indirect estimates of
  $X_\text{max}$ with the surface detector of Auger were reported
  in~\cite{Aab:2017cgk} with 517 events above $10^{19.5}$~eV and a
  resolution of 45~g/cm$^2$. The AugerPrime detector could detect
  about 700 events above this energy within 7 years of running time
  and a resolution of about 30~g/cm$^2$~\cite{Aab:2016vlz}.}

The integral number of events will be a factor of 1.5 larger than the
one used by Auger to study correlations with starburst galaxies above
40~EeV and and 2.1 larger than the one for $\gamma$AGN above
60~EeV~\cite{Aab:2018chp}. Moreover,  contrary to the Auger data set, the
POEMMA exposure covers the full sky and each event detected by POEMMA
in stereo mode will have a measurement of the shower maximum and thus
allow to study the correlations for different cosmic-ray rigidities.

\begin{figure*}[tb]
\centering
\includegraphics[width=0.49\linewidth]{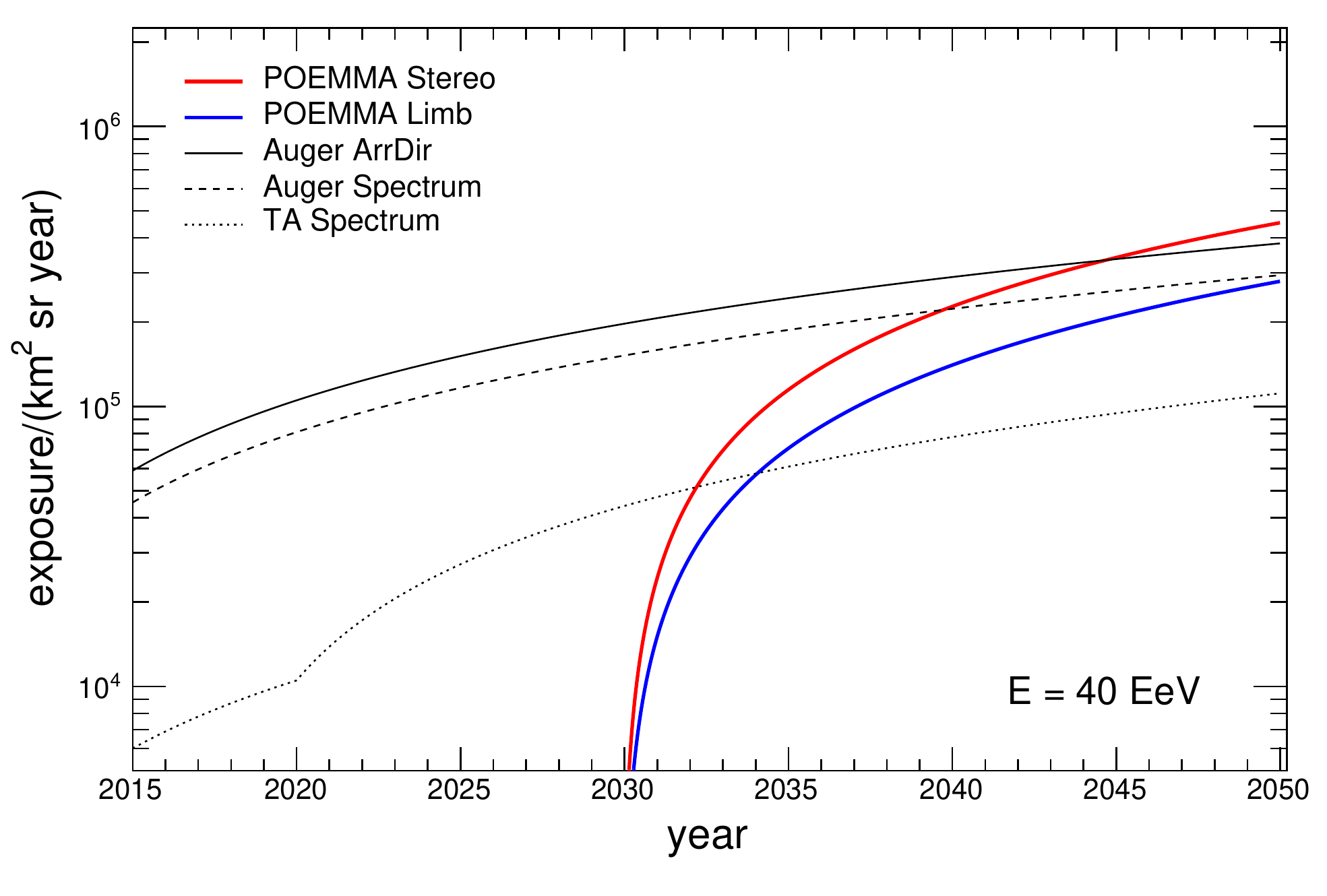}
\includegraphics[width=0.49\linewidth]{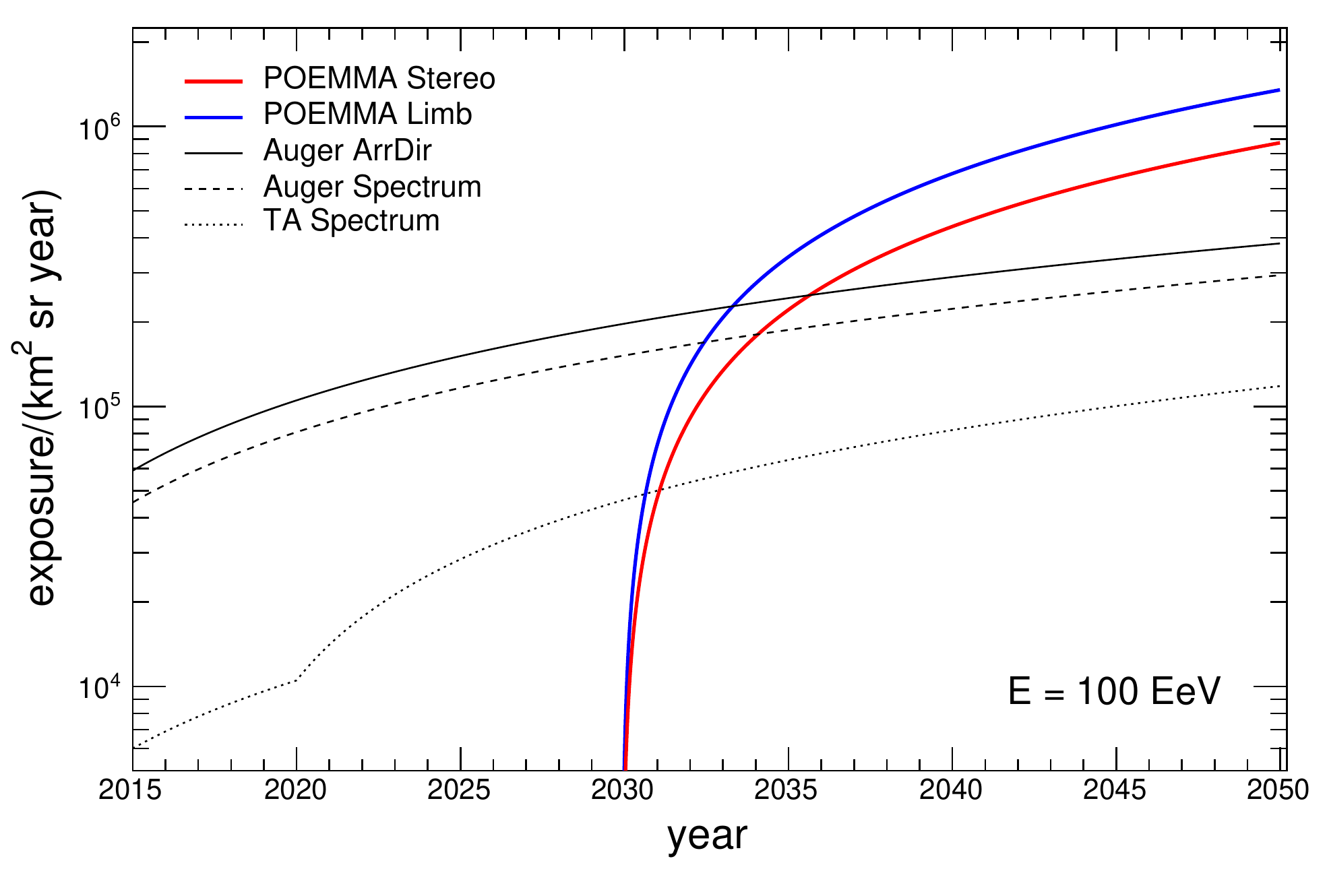}
\caption{Exposure as a function of time collected by Auger, TA
  (including TAx4) and POEMMA.
  For Auger the exposure for two different event selections are shown.
The left panel shows the exposures at 40 EeV and the right panel at 100 EeV.}
\label{fig:exposure}
\end{figure*}

The UHECR aperture in limb-viewing mode starts to outperform the stereo operation
above $60$~EeV. However, due to the steeply falling cosmic-ray flux
above the suppression, the expected number of  events beyond the crossover is
of the same order of magnitude (75 for stereo and 146 for limb-viewing assuming
the Auger spectrum). Given that stereo events have a better energy
resolution and provide information about the shower maximum, we
foresee that most of the UHECR data taking time will be spent in stereo (near-nadir)
mode. The collected exposures at 40 at 100 EeV after 5 years of operation
are shown in Fig.~\ref{fig:exposure}.

\begin{figure}[tb]
\centering
\includegraphics[width=\linewidth]{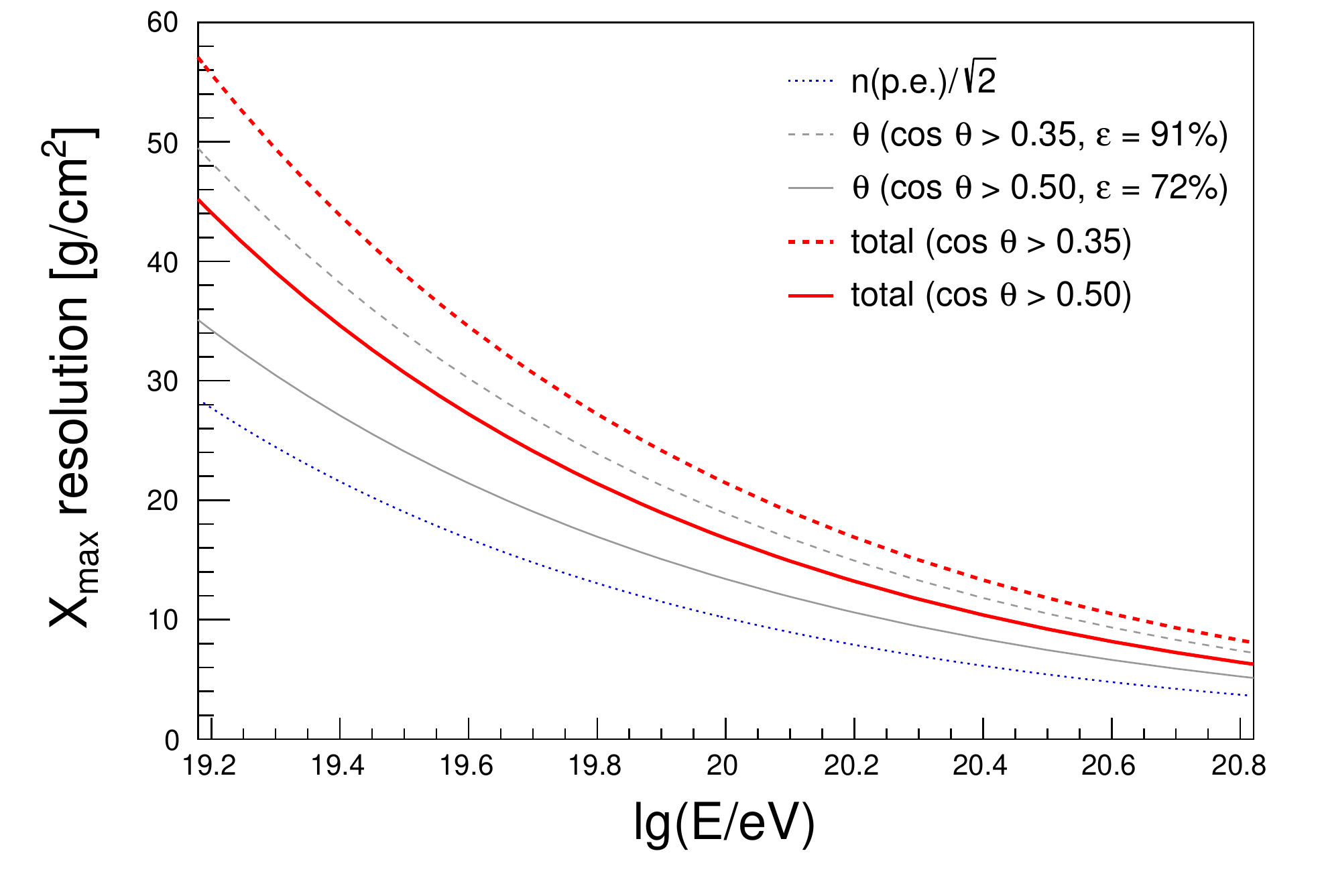}
\caption{Preliminary estimate of the $X_\text{max}$ resolution of
  POEMMA in stereo mode.  The contributions from the photo-electron
  statistics and angular resolution are shown in blue and gray
  respectively. The total resolution, obtained by adding both
  contributions in quadrature is shown in red for two cuts on the
  maximum zenith angle.}
\label{fig:xmaxRes}
\end{figure}

An estimate of the $X_\text{max}$ resolution of POEMMA in stereo mode is
shown in Fig.~\ref{fig:xmaxRes}.  The contribution from the
PE statistics was studied with a full simulation in Sec.~\ref{sec:3} and
amounts to about 20~g/cm$^2$ at $30~{\rm EeV}$ for one photometer and
decreases approximately with $\sqrt{E}$ (see Fig.~\ref{fig:xmaxResDetail}). The uncertainty in the
measured zenith angle of the shower affects the calculation of the
slant depth of the shower maximum. A preliminary analytic propagation
of this uncertainty to the resolution of $X_\text{max}$ is indicated
as gray lines in Fig.~\ref{fig:xmaxRes} averaged over the expected
arrival directions of triggered events. Since inclined events are more
affected than vertical ones, the overall resolution depends on the
maximum zenith angle of the data sample (a possible correlation
between the zenith angle and angular resolution has not been taken
into account yet). The total $X_\text{max}$ resolution of POEMMA
including both, angular resolution and PE statistics, is
about 31~g/cm$^2$ at $30$~EeV for events below 60$^\circ$ (72\% of the
data sample) and 39~g/cm$^2$ below 70$^\circ$ (91\% of the data sample).
At $100$~EeV the resolution is 17 and 21~g/cm$^2$ respectively.

\section{Science Reach}
\label{sec:5}

The typical observables for comparing data to model predictions are the energy spectrum, the mass spectrum, and the distribution of arrival directions of UHECR reaching the Earth. From these observables, the last one provides the most unambiguous conclusions about the locations of the sources.
In this section, we determine the sensitivity of POEMMA to measure the first two observables and discuss the discovery reach of anisotropy searches using the third observable.
We also investigate supplementary science capabilities of POEMMA. We first determine the sensitivity to probe particle interactions at extreme energies and after that we explore the potential for observing extreme energy photons produce in the decay of super-heavy  dark matter particles clustered in the halo of the Milky Way. We then present the UHE neutrino sensitivity based on stereo fluorescence measurements of neutrinos interacting deep in the atmosphere.

\subsection{Energy Spectrum}

The all-particle spectrum contains information  
about the source distribution, emission properties, nuclear composition, and propagation effects. Indeed, there is a fair amount of work devoted to deducing such fundamental information from details of spectral features. The standard approach involves establishing some hypothesis about source properties and, using either Monte Carlo simulations or analytic methods, inferring the mean spectrum one expects to observe here on Earth. Since at present we have a limited understanding of source distributions and properties, it is common practice to assume spatially homogeneous and isotropic UHECR emissions, and compute a mean spectrum based on this assumption. Obviously, in the real world this assumption cannot be correct, especially at the highest energies where GZK interactions severely limits the number of sources visible to us at Earth. However, one can quantify the possible deviation from the mean prediction based on the understanding we do have on the source density and the possible distance to the closest source populations. Such a next statistical moment beyond the mean prediction is referred to as the {\it ensemble fluctuation}~\cite{Ahlers:2012az}. It depends on, and consequently provides information on, the distribution of discrete local sources, source nuclear composition, and energy losses during propagation. The ensemble fluctuation in the energy spectrum is one manifestation of the cosmic variance, which should also come out directly via eventual identification of nearby source populations. 

\begin{figure*}[tb]
\centering
\includegraphics[height=7cm]{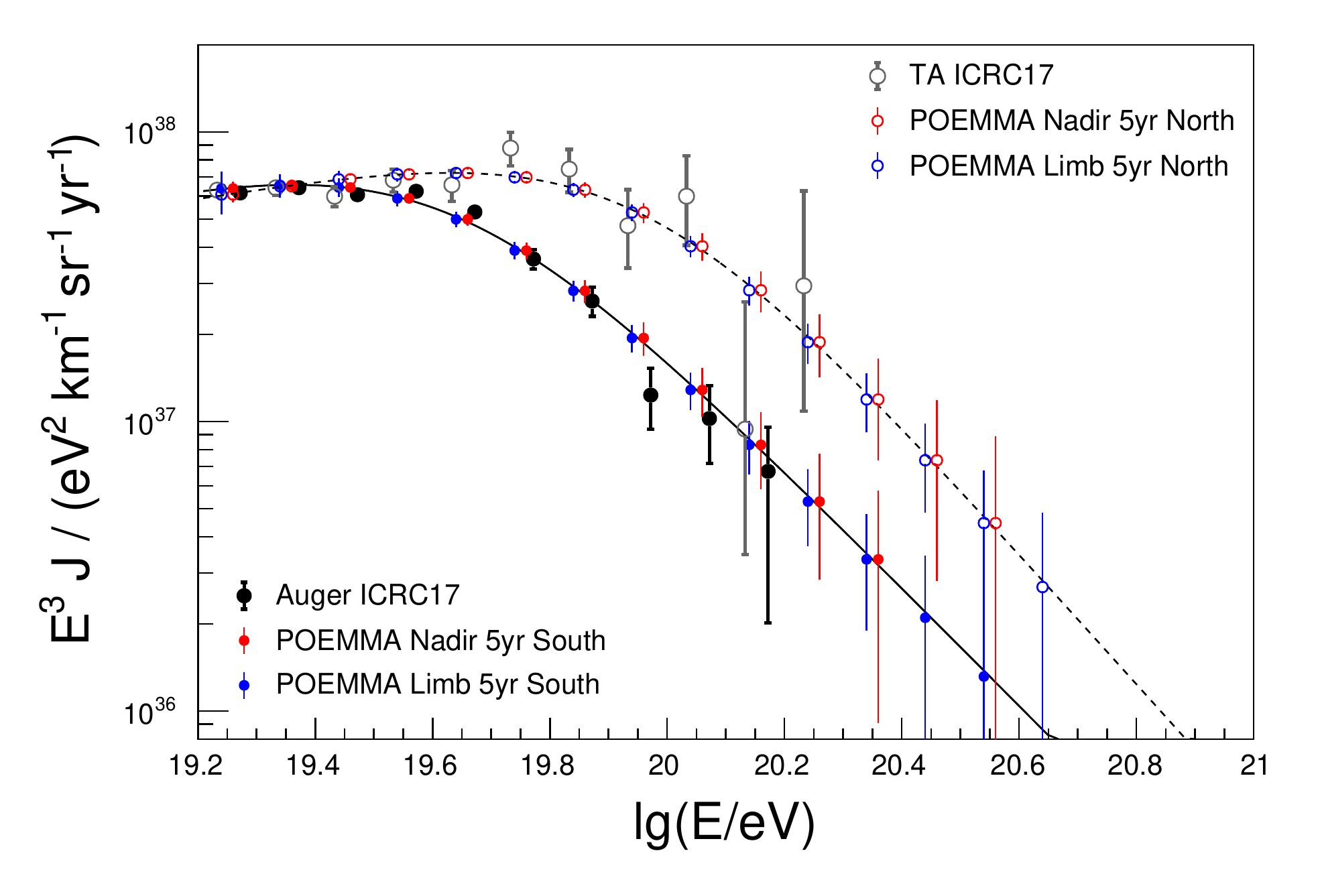}\includegraphics[clip,rviewport=-0 -0.08 1 1.04,height=7cm]{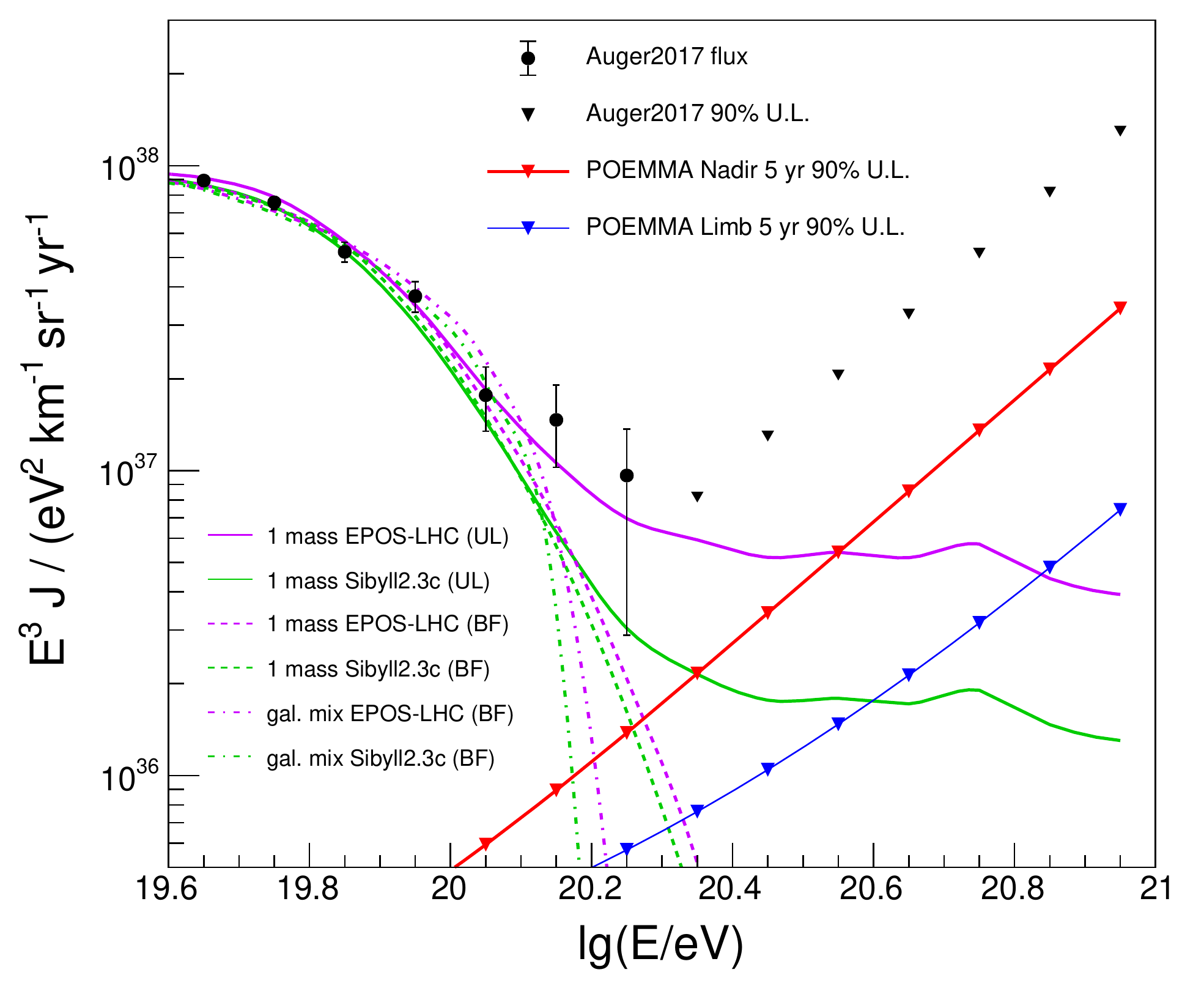}
\caption{{\itshape Left:} Energy spectrum of UHECRs as measured by TA
    and Auger in the Northern and Southern hemisphere
    respectively. The energy scale of the two experiments were
    cross-calibrated by $\pm5.2\%$ as derived by the UHECR Spectrum
    Working Group at low energies. Red and blue dots with error bars
    illustrate the expected accuracy reached with POEMMA in stereo and
    limb-viewing mode within 5 years of operation.
    {\itshape Right:} Flux suppression at UHE as measured by the Pierre Auger
  Observatory (data points)~\cite{Aab:2017njo}. 90\% confidence upper
  limits of the flux at UHE are shown as downward triangles (ideal
  limits without taking into account event migration due to the
  limited energy resolution of the observatories). Black: Pierre Auger
  Observatory 2017, red: POEMMA 5 year stereo mode, blue POEMMA 5 year
  limb-viewing mode. Various model predictions for the shape of the flux
  suppression from~\cite{Muzio:2019leu} are superimposed as black
  lines.}
\label{fig:spectrum}
\end{figure*}

POEMMA will have full-sky coverage for UHECRs due to the nature of the 525~km, $28.5^\circ$ inclination orbit and large $45^\circ$ full FoV for each telescope. Therefore, the satellites will observe the source distribution on the full celestial sphere under a single experimental framework with a well-defined acceptance. This implies that POEMMA will be sensitive to ensemble fluctuations in the energy spectrum for two ``realizations of the universe.'' For example, it will be able to detect spectral variations in the northern and southern skies, or else distinct ensemble fluctuations associated with the spatial anisotropy of the dipolar asymmetry observed by Auger. In the left panel of Fig.~\ref{fig:spectrum} we show the UHECR spectrum as observed by Auger and TA together with the expected accuracy reached with POEMMA in stereo and limb-viewing mode after 5~yr of data collection, assuming normalization to each of the experiments. As can be seen, the high-statistics  sample with high-resolution to be collected by POEMMA will provide a final verdict on whether the Auger and TA discrepancy in the measurement of the spectrum at the highest energies is due to physics, statistics, or systematic uncertainties in the energy calibration.

The abrupt softening of the  spectrum due to energy losses via photo-pion production and/or photo-disintegration in the CMB (a.k.a. GZK cutoff) is the only unequivocal prediction ever made concerning the spectral shape~\cite{Greisen:1966jv,Zatsepin:1966jv}. The discovery of a suppression at the end of the spectrum  was first reported by HiRes~\cite{Abbasi:2007sv} and Auger~\cite{Abraham:2008ru}, and later confirmed by TA~\cite{AbuZayyad:2012ru}; by now the significance in Auger data is well in excess of $20\sigma$ compared to a continuous power law extrapolation beyond the ankle feature~\cite{Abraham:2010mj}. Although the existence of the flux suppression has been firmly established, an alternative interpretation for the suppression region was put forward in~\cite{Allard:2008gj}, wherein it is postulated that the end-of-steam for cosmic accelerators is coincidentally near the putative GZK cutoff. Note that this alternative interpretation predicts an increasingly heavier composition from the ankle towards the suppression region, with a mix of protons and heavies undergoing acceleration to the same rigidity, so that their maximum energy scales as $E_Z^{\rm max} \sim Z E_p^{\rm max}$, where $E_p^{\rm max}$ is the maximum proton energy. Yet another model to explain the suppression postulates that the maximum energy  is constrained by GZK interactions at the source~\cite{Anchordoqui:2019mfb}. This model also yields a change towards a heavier composition at higher energies, but predicts a different scaling for the maximum energy because while the acceleration capability of the sources grows linearly in $Z$,  the energy loss per distance traveled decreases with increasing $A$. A general feature of the GZK cutoff is that of a slight recovery of the spectrum if the source emission spectra extend to energies far beyond 100~EeV. This is in sharp contrast to models that postulate that the suppression is primarily caused by 
the limiting acceleration energy at the sources, which makes the observe spectrum steeper than that at lower energy, developing a sharp cutoff.

In the right panel of Fig.~\ref{fig:spectrum} we show the sensitivity
of POEMMA to observe the GZK recovery. The sensitivity is given by the
90\% confidence level upper limit for the case of zero observation
with zero background~\cite{Feldman:1997qc}. Therefore, these
sensitivities are for the ideal case of perfect energy resolution and
the actual sensitivity will be somewhat worse due to the non-zero
probability of a net migration of events from lower energies towards
the low-flux UHE energy range. It should be noted that if a post-GZK recovery is observed in stereo mode, the POEMMA instruments could tilt to increase the sensitivity for the highest energy UHECRs.

The ideal POEMMA sensitivity is compared to generic model predictions of the spectral falloff~\cite{Muzio:2019leu} built on the UFA15~\cite{Unger:2015laa} model that explains the shape of the spectrum and its complex composition evolution via photo-disintegration of accelerated nuclei in the photon field surrounding the source, but also allowing for a subdominant purely protonic component which is constrained by UHECR composition measurements~\cite{Aab:2014kda}, limits on astrophysical neutrinos (IceCube~\cite{Aartsen:2016ngq} and ANITA~\cite{Gorham:2019guw}) and gamma-ray observations~(Fermi-LAT~\cite{TheFermi-LAT:2015ykq}).

\subsection{Nuclear Composition}

\begin{figure*}[tb]
\centering
\includegraphics[width=\linewidth]{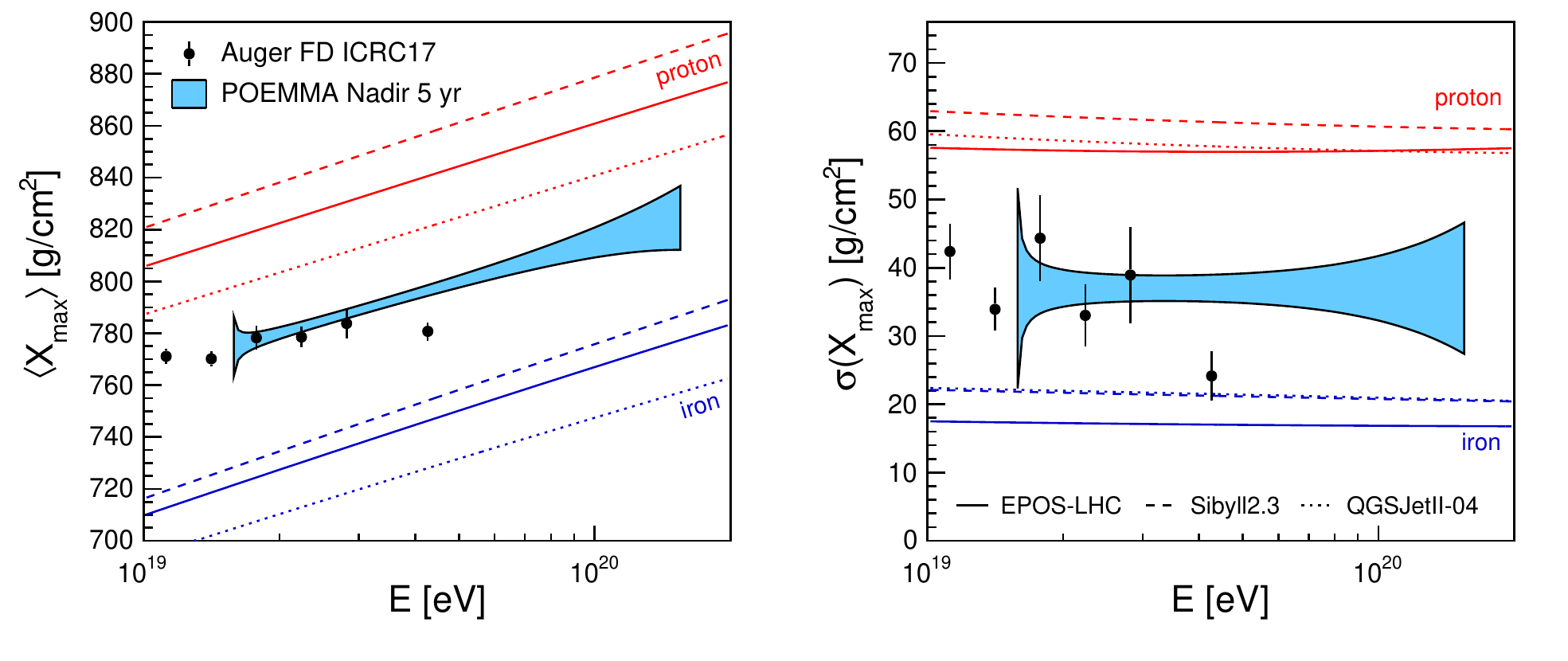}
\caption{Capability of POEMMA to measure $\langle X_\text{max}\rangle$
  and $\sigma(X_\text{max})$ for composition studies at UHE. The width
  of the blue band illustrates the expected statistical uncertainties
  given the number of events per 0.1 in logarithm of energy in five
  year stereo operation, the $X_\text{max}$ resolution and efficiency
  from Fig.~\ref{fig:xmaxRes} for $\theta < 70^\circ$ and intrinsic
  shower-to-shower fluctuations of 40~g/cm$^2$. The band spans the
  energy range for which more than 10 events are within a 0.1 dex
  bin. }
\label{fig:xmax}
\end{figure*}

The measurement of the composition of UHECRs is one of the key
ingredients to constrain their origin. The event-by-event measurement
of energy and $X_{\rm max}$ with fluorescence telescopes is well suited to
perform composition studies~\cite{Kampert:2012mx}.  As can be seen in
Fig.~\ref{fig:xmax}, the high statistics and good energy and $X_{\rm max}$ resolution of POEMMA
will allow for high-precision composition studies at hitherto
unexplored energies, while at the same time providing an overlap with
existing compositions measurements from Auger at low energies.

EASs initiated by photons have a larger $X_{\rm max}$ than showers initiated by nuclei with the same energy. This is because the radiation length is more than two orders of magnitude shorter than the mean free path for photo-nuclear interaction, and therefore this leads to a reduced transfer of energy to the hadronic component of the EAS~\cite{Risse:2007sd}. The development of EAS  initiated by a photon is hence delayed (compared to a baryon-induced shower) by the typically small multiplicity of electromagnetic interactions. On the other hand, for showers of a given
total energy $E$, heavy nuclei have smaller $X_{\rm max}$ than light nuclei  because they have a larger cross section and interact sooner, and because the energy is already shared among $A$ nucleons, so the shower develops more quickly. More concretely, $\langle X_{\rm max}
\rangle$ scales approximately as
$\ln(E/A)$~\cite{Linsley:1978bm,Linsley:1981gh}.  In addition, the standard
deviation $\sigma (X_{\rm max})$ is smaller for heavier nuclei because the sub-shower fluctuations average out.   By contrast, protons or He can have a deep or shallow first interaction, and the shower-to-shower fluctuations in subsequent development are larger.  Therefore, only light cosmic rays have large $X_{\rm max}$, permitting a fraction of events to be unambiguously identified as light nuclei. The high event statistics with good $X_{\rm max}$ resolution would allow POEMMA not only  isolate baryons from photon and neutrino primaries, but  also to distinguish between protons, light nuclei, medium mass nuclei, and heavies~\cite{Krizmanic:2013tea}.

\begin{figure}[tpb]
\includegraphics[width=0.9\linewidth]{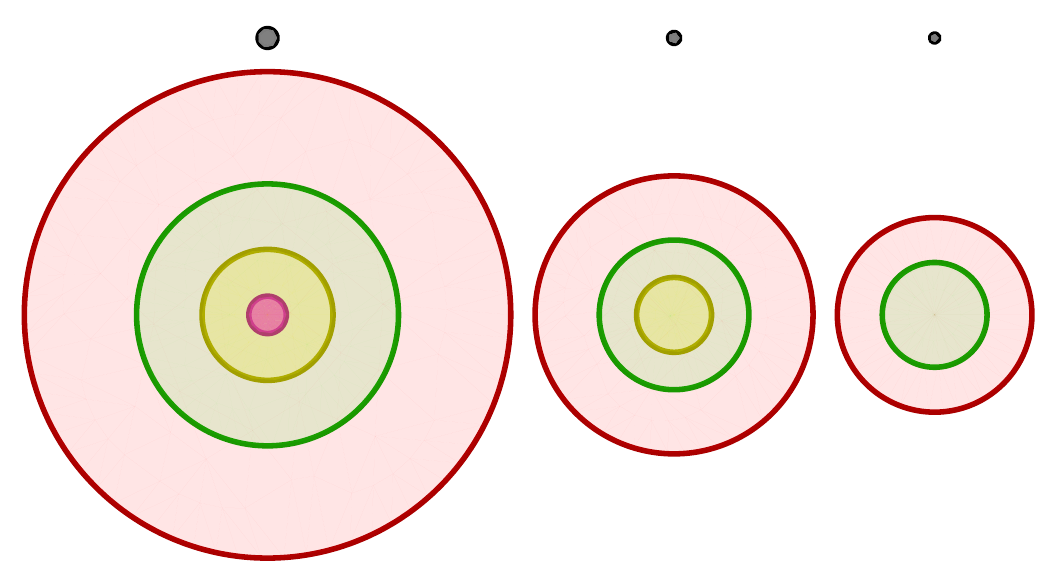}
\caption{Circles representing the composition-layered structure of hotspots at different energies, for proton sources (top) and nuclei sources (bottom). The radii of the circles respect the proportions of the angular sizes given by (\ref{eq:BG}), for protons (black), helium (magenta), nitrogen (yellow), silicon (green) and iron (red); and for $40\,\mathrm{EeV}$ (left), $70\,\mathrm{EeV}$ (center) and $100\,\mathrm{EeV}$ (right).
\label{fig:onion}}
\end{figure}

In addition, if a hot spot of a nearby source is identified, protons can be further discriminated from CNO and heavies by looking at the distribution of arrival directions. This is because  while sources of UHECR protons exhibit anisotropy patterns that become denser and compressed with rising energy,  nucleus-emitting-sources imprint layers on the sky that become more distant from the source position with rising energy~\cite{Anchordoqui:2017abg}. The peculiar shape of the hot spots from nucleus-accelerators is steered by the competition between energy loss during propagation and deflection on the Galactic magnetic field: for a nucleus of charge $Ze$ and baryon number $A$, the bending of the cosmic ray decreases as $Z/E$ with rising energy, while the energy loss per distance traveled decreases with increasing $A$. The potential for nucleus-proton discrimination is shown schematically in Fig.~\ref{fig:onion}, and can be understood as follows. If the source emits only protons, the size of the corresponding “spot” should decrease as $1/E$ with rising energy due to reduced deflection in magnetic fields. In contrast, if the source produce a mixed composition, a different quality emerges. Lighter compositions tend to shorter mean-free-paths at higher energies, so as their energy increases they begin to disappear from the sample leaving behind only the lower energy component. The latter suffers a relatively smaller magnetic deflection compared to heavier nuclei at all energies. One thus ends up with a hot spot  in which the energies of the species observed closer to the source have a lower rather than higher energy, as they would in the case that the source emitted only protons.

Despite the fact the Galactic Magnetic field is highly anisotropic, to anticipate the sensitivity of POEMMA herein we assume that the deflection of particles is isotropic around the line of sight, and given by (\ref{eq:BG}). We further assume that the fixed parameters of the statistical analysis should match the anisotropy clues provided by Auger data. Hence, we adopt a search angular window $\Delta =13^\circ$,  a source distance $\sim 4\,{\rm Mpc}$,  a threshold energy  $E_0=39~{\rm EeV}$, and source spectrum $\propto E^{-5.03}$ that 
consistent with both the energy spectrum above $40\,{\rm EeV}$
reported by the Auger Collaboration~\cite{Aab:2017njo} and the source
spectra of nearby starburst galaxies as estimated in~\cite{Anjos:2018mgr}. With this simplified picture in mind, we now assume that UHECRs are
\emph{normally} distributed around the source direction, which defines
the center of the hot spot. The deflection $\delta$, which characterizes the angle between
the arrival direction and the line of sight, is a random variable
distributed according to a one sided von Mises distribution,
bounded by a window size $\Delta$ with zero mean and a dispersion parameter $\kappa=1/\theta^2(E,Z)$. The probability density for an UHECR to have energy
in $[E,E+ dE]$ and deflection in $[\delta,\delta+
d\delta]$ is found to be
\begin{eqnarray}
f(E,\delta|A,Z) & = & \mathcal A
  E^{-5.03}  \exp\left[-\frac{E}{E_A}\right] \Theta(E-E_0)\nonumber \\
& \times & \exp\left[\frac{\cos\delta}{\theta^2(E,Z)}\right] \ \Theta(\Delta-\delta)
 ,\label{pdf} 
\end{eqnarray}
where $\mathcal A$ is a normalization constant and $E_A$ is the cutoff energy in the observed spectrum of the various species. Following~\cite{Anjos:2018mgr} we take $E_4 = 60~{\rm EeV}$, $E_{14} = 80~{\rm EeV}$, $E_{28} = 130~{\rm EeV}$, and $E_{56} = 210~{\rm EeV}$, which accounts for energy losses during propagation. Before proceeding, we pause to note that in the actual data analysis one should also consider the variations of the magnetic field for UHECRs arriving from different points of the sky.  A full consideration of the anisotropic magnetic deflections would require an adjustment of the distribution in (\ref{pdf})  to incorporate an azimuthal variable around the line of sight, and should also take into account the source direction in the sky. Notwithstanding, our assumption of isotropy around the line of sight provides a demonstration of the search procedure while keeping the complexity at a reasonable level at this stage.

To carry out the statistical analysis we the adopt the Kolmogorov-Smirnov test.  The power of a statistical test is the probability that the null hypothesis is rejected if it is actually false. The probability of rejecting the null hypothesis while it is true, depends on the significance level of the test $\alpha$. For a chosen null hypothesis ${\cal H}_0$ and significance level $\alpha$, there is a critical value for the test statistic, $t_c$, above which there is a fraction $\alpha$ of the data simulated following ${\cal H}_0$. For a given alternative hypothesis ${\cal H}_k$, the fraction $\beta_k$ of the data with test statistic $t<t_c$ is the probability of not rejecting the null hypothesis while it is false.  All in all, the power of the test for a given ${\cal H}_k$ is $\mathcal P_k=1-\beta_k$.

\begin{figure}[tb]
 \includegraphics[width=0.9\linewidth]{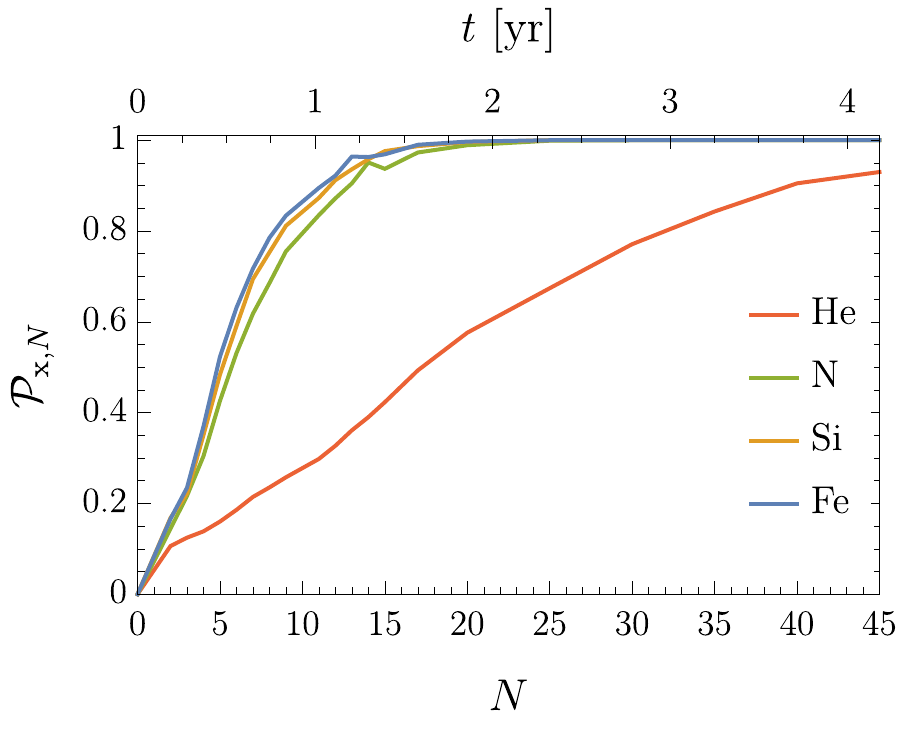}
 \caption{Power of the statistical test
   for different alternative hypotheses, that is different nuclei
   and number of events per hot spot. The horizontal axis on the top
   indicates the projected time-scale for POEMMA. \label{fig:powers}}
\end{figure}

We simulate datasets $\mathcal D_{x,N}$ following the distributions in (\ref{pdf}), where $x\in\{p, {^4{\rm
    He}}, {^{14}{\rm N}}, {^{56}{\rm Fe}}\}$ and $N=\dim\mathcal
D_{x,N}$ is the number of data points in the hot spot.  For each value of $N$, which
roughly corresponds to a given livetime of the POEMMA mission, we
consider as null hypothesis a pure proton composition, ${\cal
  H}_{\mathrm p,N}$, and the different nuclei as alternative
hypotheses ${\cal H}_{\mathrm x,N}$.  To estimate the performance of POEMMA we take an event rate above $E_0$
of $\Gamma \sim 280~{\rm yr}^{-1}$. A $13^\circ$ angular radius solid
angle covers a fraction $f_{\rm sky} \sim 0.013$ of the sky. Within a
hot spot, one expects both background and source contributions, with a
ratio $f_{\rm events}=n_{\rm ev}/n_{\rm bg}$. With this, the required
livetime of the experiment to measure a hot spot of $N$ events can be
roughly estimated to be \begin{equation} T\sim \frac{N}{\Gamma f_{\rm
      sky} f_{\rm events} }.\end{equation} For $f_{\rm events}\sim3,$
as observed in~\cite{Aab:2017njo} from the direction of the nearby radiogalaxy Centaurus A,
$T\sim0.09 N\,{\rm yr}$. The projected
sensitivity of POEMMA is shown in Fig.~\ref{fig:powers}. For hot spots
of 20 or more events, the discovery power (with $\alpha=0.05$) is
almost one for nuclei other than helium. Therefore, we conclude that
if the hot spot is composed of nuclei heavier than nitrogen, in two
years of operation POEMMA will be able to exclude a pure-proton origin
at the 95\%~CL.

\subsection{Anisotropy Searches}

\subsubsection{Large-scale Anisotropy Searches Using Spherical Harmonics}

The most direct way to determine the location of the sources is to
 search for anisotropy in the distribution of arrival directions. The distribution of arrival directions, like any field on the unit sphere, can be conveniently decomposed in spherical  harmonics~\cite{Sommers:2000us}. Specifically, we can decompose the angular distribution of events $\Phi$ in some direction $\hat n$ by separating the dominant monopole contribution from the anisotropic one, $\Delta(\hat n)$, according to
\begin{equation}
\Phi(\hat n) = \frac{N}{4 \pi f_1} W(\hat n) [ 1 + \Delta (\hat n)],
\end{equation}
where $W(\hat n)$ is the relative coverage of the experiment varying from 0 to 1, $f_1 = \int d\hat{n} \ W(\hat n)/(4\pi)$ is the fraction of the sky effectively covered by POEMMA, and $N$ the total number of observed UHECRs~\cite{Aab:2014ila,Aab:2016ban}. Along these lines,  the multipolar expansion of $\Delta( \hat n)$ into the spherical harmonics basis $Y_{\ell m}(\hat n)$ is given by
\begin{equation}
    \Delta (\hat n) = \sum_{\ell=1}^{\infty} \sum_{m=-\ell}^{\ell} a_{\ell m} \ Y_{\ell m}(\hat n) \,,
\end{equation}
where the $a_{\ell m}$ coefficients encode any anisotropy signal.
An unambiguous measurement of the full set of spherical harmonic coefficients requires full-sky coverage. Indeed, using the orthogonality of the spherical harmonics basis one can directly recover the multipolar moments $a_{\ell m}$. The partial-sky coverage of ground based experiments, encoded in the $W(\hat n)$ function, prevents direct determination of the $a_{\ell m}$ coefficients. Although the exhaustive information of the distribution of arrival directions is encoded in the full set of multipole coefficients, the characterisation of any important overall property of the anisotropy becomes evident in the angular power spectrum
\begin{equation}
C(\ell) = \frac{1}{2\ell +1} \sum_{m = -\ell}^\ell a_{\ell m}^2 \, ,
\end{equation}
which is a coordinate independent quantity. Any significant anisotropy of the angular distribution over scales $\simeq 1/\ell$ radians would be captured in a non-zero power in the mode $\ell$. For a 5~yr mission with $N=1400$ events, we have $\langle C(\ell) \rangle = (4 \pi N)^{-1} \simeq 6 \times 10^{-5}$~\cite{Anchordoqui:2003bx}.

\begin{figure}[tb]
 \includegraphics[width=0.9\linewidth]{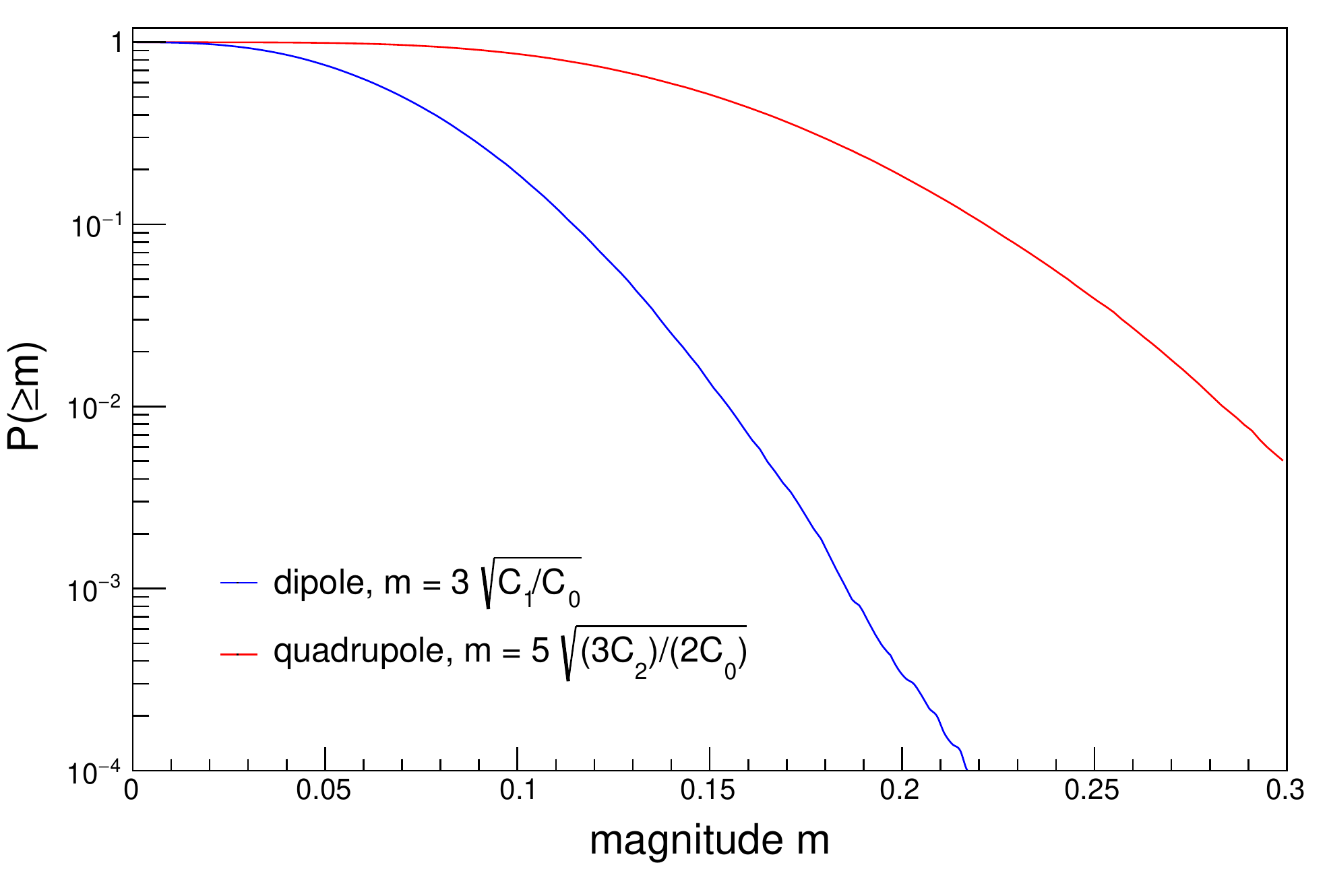}
 \caption{Cumulative distributions for the dipole and quadrupole moments for isotropically distributed skies with N=1400 events.}
\label{fig:cumu}
\end{figure}

In order to assess the sensitivity of POEMMA to dipolar and quadrupolar anisotropy patterns, we consider the distribution of dipole and quadrupole moment magnitudes for an isotropic $1400$ event sky, as shown in Fig.~\ref{fig:cumu}. From this distribution, it is straightforward to estimate that with POEMMA, dipole (quadrupole) moments above $\sim 11\%$ ($\sim 21\%$) and $\sim 18.5\%$ ($\sim 33\%$) can be probed at the 84\% and 99.9\% confidence level sensitivities at 40~EeV.

\subsubsection{Intermediate-scale Anisotropy Searches Through Cross-Correlations with Astrophysical Catalogs} \label{subsubsec:tsanalysis}

One commonly invoked test for anisotropy on intermediate scales is motivated by the expectation that UHECRs will point back to their sources above a given threshold energy. The exact value of the threshold energy is unknown due to uncertainties in the Galactic and extragalactic magnetic fields. However, the expectation is that the threshold energy occurs at roughly the same energies at which the flux of UHECRs is attenuated by cosmological photon backgrounds. UHECR attenuation results in a horizon distance for UHECRs within which astrophysical sources appear to be anisotropic; hence, the expectation is that above a given energy threshold, the arrival directions of UHECRs will be similarly anisotropic and will be, to a given degree, correlated with the positions of their sources on the sky, with angular separations corresponding to the degree of magnetic deflection (angular separations $\sim$ few--tens of degrees). As such, statistical tests cross-correlating arrival directions of UHECR events with astrophysical catalogs are effective in detecting anisotropy at intermediate scales and may also provide clues about the UHECR source population(s) and the amount of deflection due to intervening magnetic fields~\cite{Soriano:2019wfx}. Previous searches conducted by the Auger and TA collaborations utilizing this approach have provided hints of anisotropy~\cite{Aab:2018chp,Abbasi:2014lda}, with the strongest signal arising from cross-correlation with a catalog of starburst galaxies (significance $\sim 4.5\sigma$~\cite{Aab:2019ogu}). 

\begin{figure*}[tb]
\centering
\includegraphics[trim = 40mm 38mm 20mm 40mm, clip, width=1.0\columnwidth]{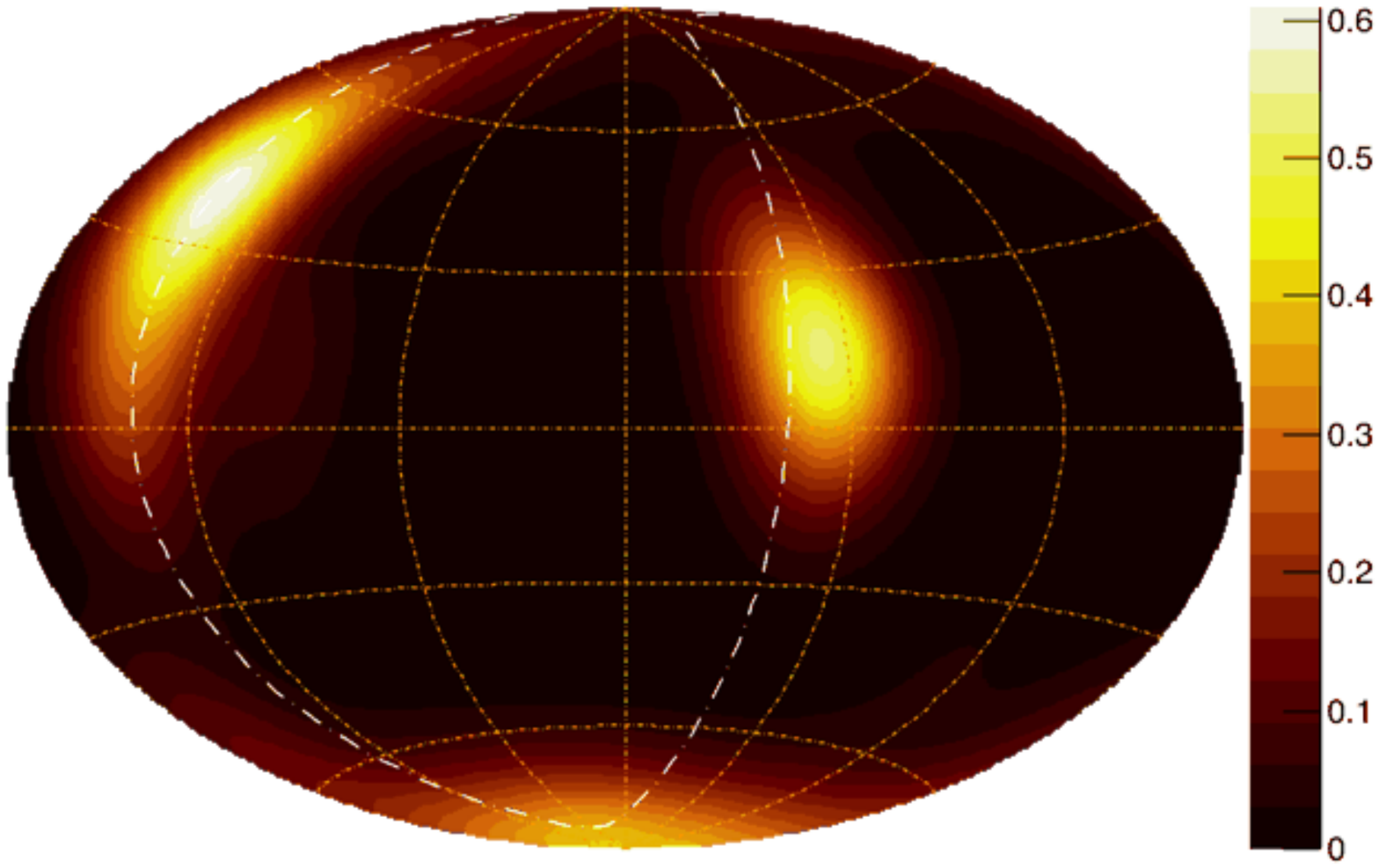}\includegraphics[trim = 40mm 38mm 20mm 40mm, clip, width=1.0\columnwidth]{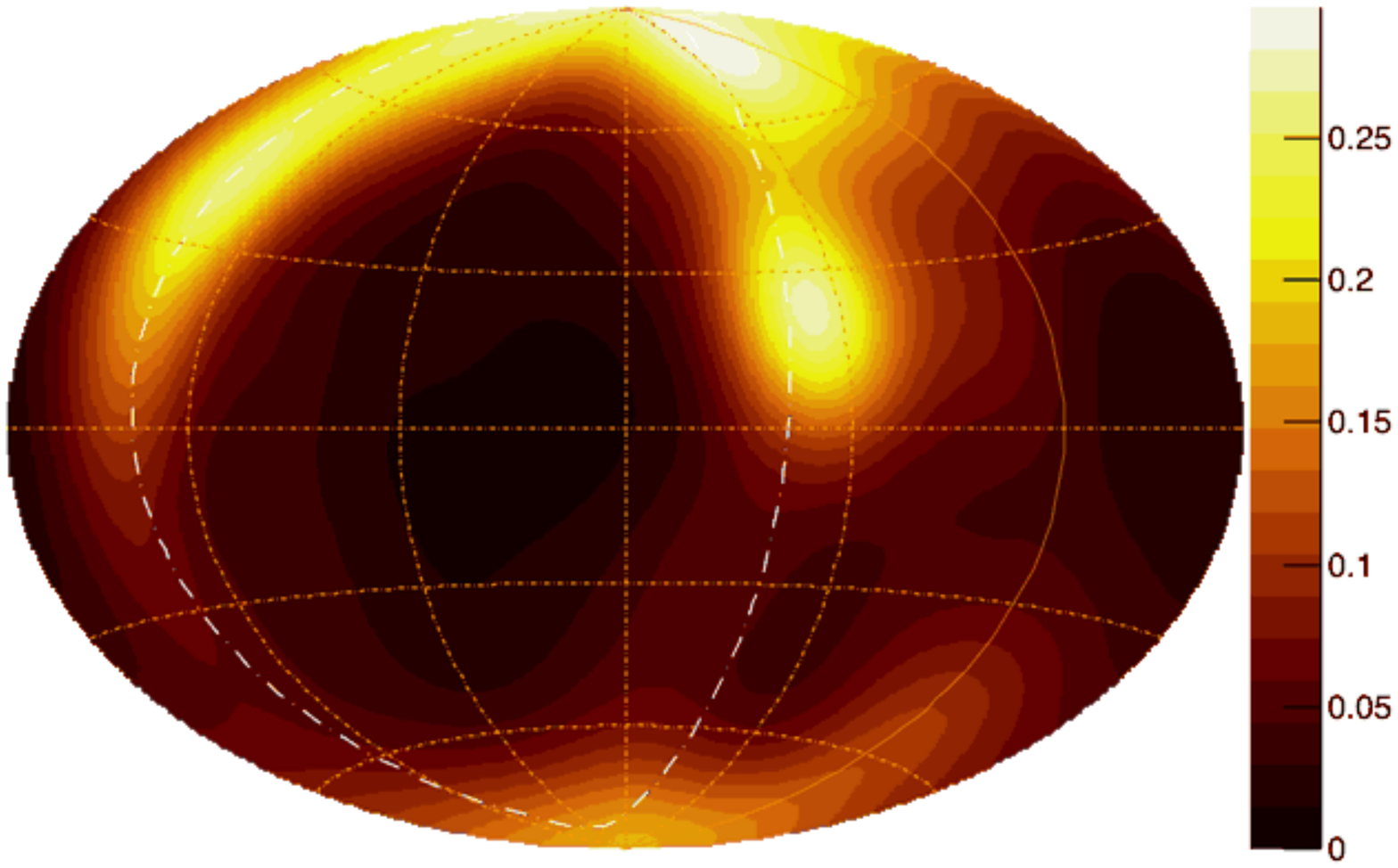}
\caption{\textit{Left:} Skymap of nearby starburst galaxies from Refs.~\cite{Aab:2018chp,Aab:2019ogu} weighted by radio flux at $1.4$~GHz, the attenuation factor accounting for energy losses incurred by UHECRs through propagation, and the exposure of POEMMA. The map has been smoothed using a von Miser-Fisher distribution with concentration parameter corresponding to a search radius of $15.0^{\circ}$ as found in Ref.~\cite{Aab:2018chp}. The color scale indicates $\mathcal{F}_{\rm src}$, the probability density of the source sky map, as a function of position on the sky. The white dot-dashed line indicates the Supergalactic Plane. \textit{Right:} Same as at left for nearby galaxies from the 2MRS catalog~\cite{2012ApJS..199...26H} and weighting by K-band flux corrected for Galactic extinction. \label{fig:sources_skymap}}
\end{figure*}

As can be seen in Fig.~\ref{fig:exposure}, POEMMA will attain an exposure of $\sim 1.5 \times 10^5~{\rm km^2~sr~yr}$ above 40~EeV within five years of operation. Furthermore, the POEMMA exposure will cover the entire sky, providing sensitivity to starburst galaxies that are not accessible to Auger or TA. As such, within its nominal mission lifetime, POEMMA will be capable of detecting anisotropy to high significances, achieving $5\sigma$ discovery reach for search parameters within the vicinity of the signal regions for anisotropy hints reported by the Auger~\cite{Aab:2018chp,Aab:2019ogu} and TA~\cite{Abbasi:2014lda} collaborations. In order to determine the reach of POEMMA in such cross-correlation searches, we implement a simple statistical study simulating POEMMA datasets assuming various astrophysical scenarios (i.e., starbursts and AGNs). Mock datasets are constructed by drawing a given fraction of events, $f_{\rm sig}$, from an astrophysical source sky map and drawing the rest ($1 - f_{\rm sig}$) from an isotropic skymap, where both skymaps are weighted by the variation in POEMMA exposure over the sky (see Fig.~\ref{fig:declination2}). We construct the astrophysical source sky maps from catalogs of candidate UHECR sources, weighting each individual source by its electromagnetic flux, accounting for UHECR attenuation due to energy losses during propagation, and smoothing with a von Mises-Fisher distribution with a given angular spread, $\Theta$ (see examples in Fig.~\ref{fig:sources_skymap}). For the purposes of this study, we use the same astrophysical catalogs as in Refs.~\cite{Aab:2018chp,Aab:2019ogu}, which include a catalog of starburst galaxies selected based on their continuum emission at $1.4$ GHz, a catalog of radio-loud and radio-quiet AGNs included in the 70 Month \textit{Swift}-BAT All-sky Hard X-ray Survey~\cite{2013ApJS..207...19B}, and a catalog of galaxies at distances greater than $1$ Mpc from the 2MASS Redshift Survey (2MRS) of nearby galaxies~\cite{2012ApJS..199...26H}. For calculating the UHECR attenuation factors, we adopted Composition Scenario A from Ref.~\cite{Aab:2018chp}, which best matches Auger composition and spectral measurements. For the threshold energy values, we adopted the values found in Ref.~\cite{Aab:2019ogu}, which corresponds to roughly $1400$ events with five years of POEMMA, assuming the Auger cosmic ray spectrum. We construct mock datasets several scenarios for each catalog, varying the signal fraction of events, $f_{\rm sig}$, and the angular spread, $\Theta$. For each mock dataset, we perform a statistical analysis testing the astrophysical hypothesis against the null hypothesis of isotropy. In so doing, we follow the likelihood ratio approach of Abbasi et al.~[\citenum{Abbasi:2018tqo}; see also, \citenum{2010APh....34..314A,Aab:2018chp,Anchordoqui:2018qom}], constructing profiles of the test statistic (TS) as a function of $f_{\rm sig}$ and $\Theta$ and finding the maximum TS value. Since TS values vary over realizations of the mock datasets, we simulate $1000$ datasets for each scenario and compute the average TS value at particular values of $f_{\rm sig}$ and $\Theta$ in order to construct the TS profiles. Motivated by reported search radii of $\sim 15^{\circ}$ found in Ref.~\cite{Aab:2019ogu}, we present results for selected scenarios in which $\Theta = 15^{\circ}$ in Table~\ref{tab:tsvaluesselect} and Fig.~\ref{fig:TS_SBG}. See Appendix~\ref{appendix:d} for more details on the maximum-likelihood methodology and a more complete table of results for all scenarios considered in this study.

\begin{table}
\caption{TS values for scenarios with $\Theta = 15^{\circ}$.}
\centering
\begin{tabular}{|C{1.0in}|C{0.5in}C{0.5in}C{0.5in}|}
\hline
\hline
Catalog & $f_{\rm sig}$ & TS & $\sigma$\\
\hline
\multirow{4}{1.0in}{\centering SBG} & $5$\% & 6.2 & 2.0 \\
    & $10$\% & 24.7 & 4.6 \\
    & $15$\% & 54.2 & 7.1 \\
    & $20$\% & 92.9 & 9.4 \\
\hline
\multirow{4}{1.0in}{\centering 2MRS} & $5$\% & 2.4 & 1.0 \\
    & $10$\% & 8.7 & 2.5 \\
    & $15$\% & 20.0 & 4.1 \\
    & $20$\% & 35.2 & 5.6 \\
\hline
\multirow{4}{1.0in}{\centering \textit{Swift}-BAT AGN} & $5$\% & 10.4 & 2.8 \\
    & $10$\% & 39.6 & 6.0 \\
    & $15$\% & 82.4 & 8.8 \\
    & $20$\% & 139.3 & 11.6 \\
\hline
\hline
\end{tabular}{}
\label{tab:tsvaluesselect}
\end{table}
\begin{figure*}[tb]
\includegraphics[trim = 32mm 30mm 20mm 27mm, clip, width=1.0\columnwidth]{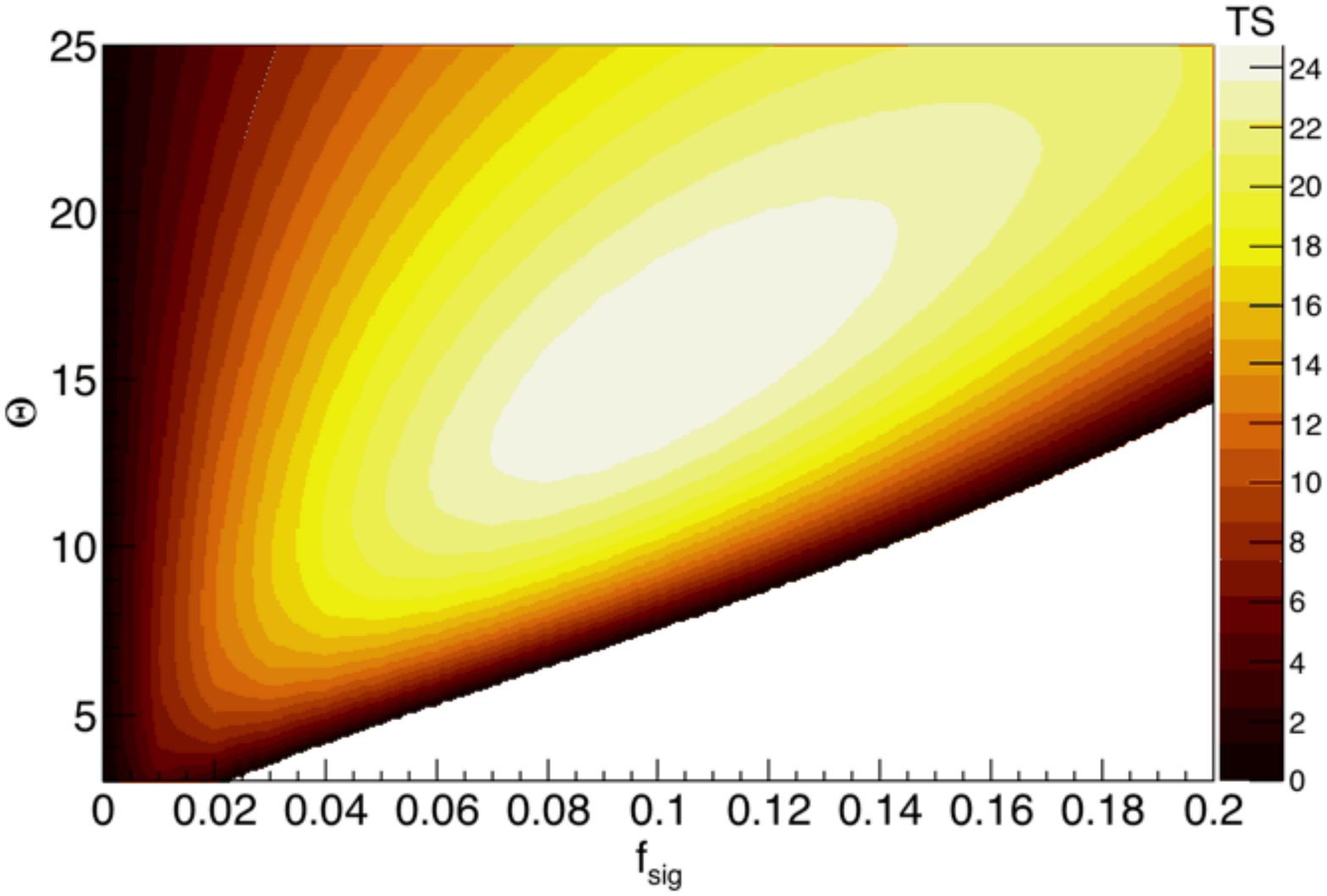}\includegraphics[trim = 32mm 30mm 20mm 27mm, clip, width=1.0\columnwidth]{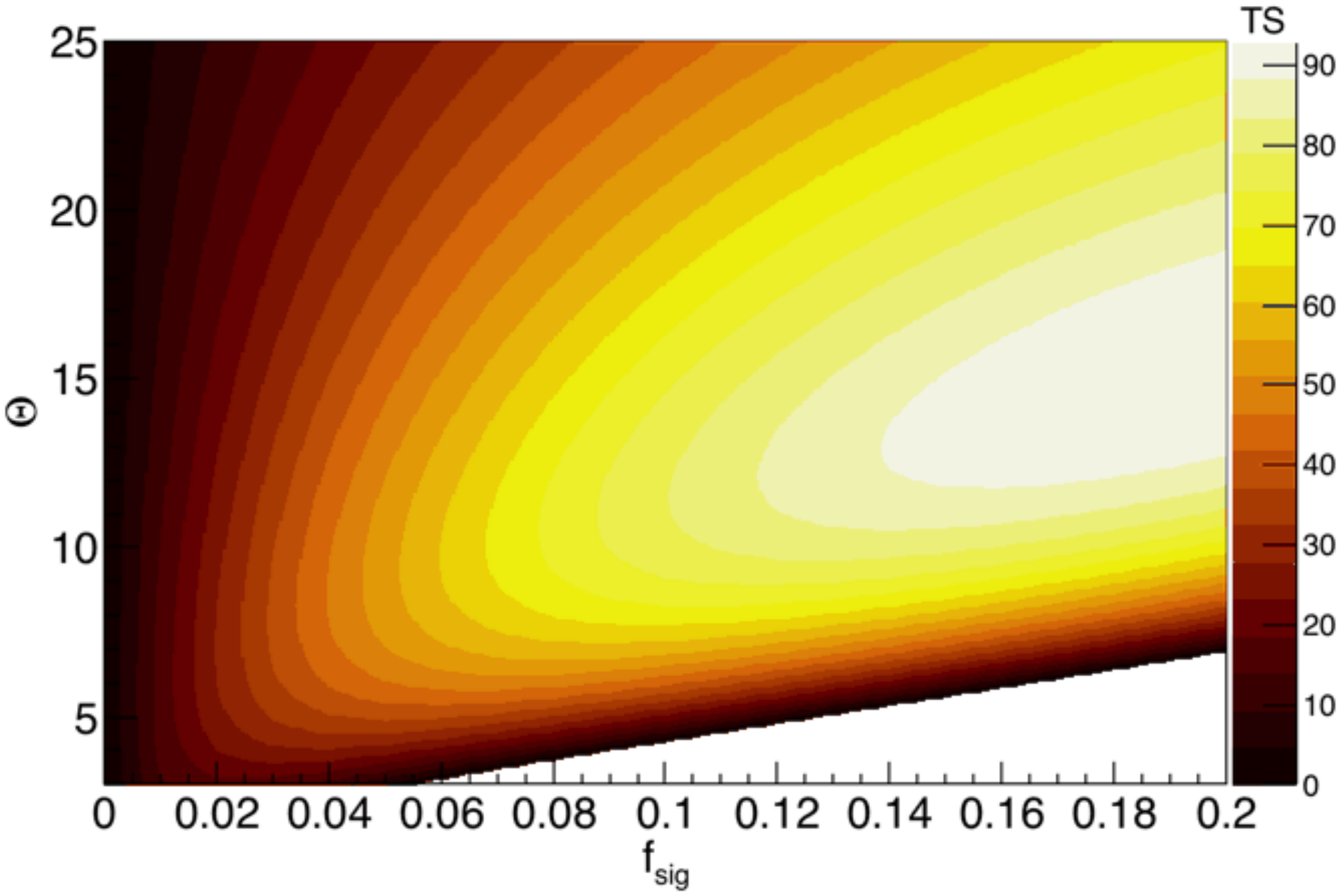}
\caption{TS profile for $1400$ events for a particular scenario using
  the starburst source sky map in Fig.~\ref{fig:sources_skymap}. In
  the scenario pictured here, the fraction of events drawn from the
  source sky map is $f_{\rm} = 10$\% (left) and 20\% (right), and the angular spread is $\Theta = 15^{\circ}$. \label{fig:TS_SBG}}
\end{figure*}
It is worth noting that though many of the scenarios included in this study are very similar to the maximum-likelihood search parameters obtained by the Auger collaboration~\cite{Aab:2019ogu}, the maximum TS values obtained from our simulations may be somewhat different than expected based on the maximum TS values obtained Auger. This is likely due to the fact that certain catalogs contain powerful sources in regions of the sky that are not accessible by Auger. The impact is that in simulations in which we assume the same signal fraction as found by Auger, the signal events are now distributed over more sources, spreading out the anisotropic events over a wider portion of the sky and making each individual source less significant. The result is that the TS values obtained from the simulations may be somewhat lower than expected, perhaps even lower than Auger found. This is most noticeable in the starburst scenario with simulation parameters $f_{\rm sig} = 10$\% and $\Theta = 15^{\circ}$. The Auger exposure map does not include M82, a nearby powerful starburst galaxy, that would be included in our simulations. The result is that the TS value predicted by the simulations ($24.7$; significance $\sim 4.6\sigma$) is somewhat lower than the TS value reported by Auger ($29.5$; post-trial significance $\sim 4.5\sigma$). However, if starbursts are the sources of UHECRs, we would expect that adding a powerful source like M82 would increase the fraction of events that would correlate with starburst galaxies. As such, we also present scenarios in which the signal fraction is higher, and in these scenarios, we see that POEMMA will detect the signal to very high significances.

\subsection{Fundamental Physics}

In this section we explore the potential of POEMMA mission to probe fundamental physics. We begin with a discussion of measurements of the $pp$ cross section beyond collider energies. After that, we study the sensitivity of POEMMA for two typical messengers of top-down models: photons and neutrinos. 

\subsubsection{Inelastic Proton-Air and Proton-Proton Cross Sections}
\label{sec:sigma}

\begin{figure*}[bt]
\centering
\includegraphics[width=0.49\linewidth]{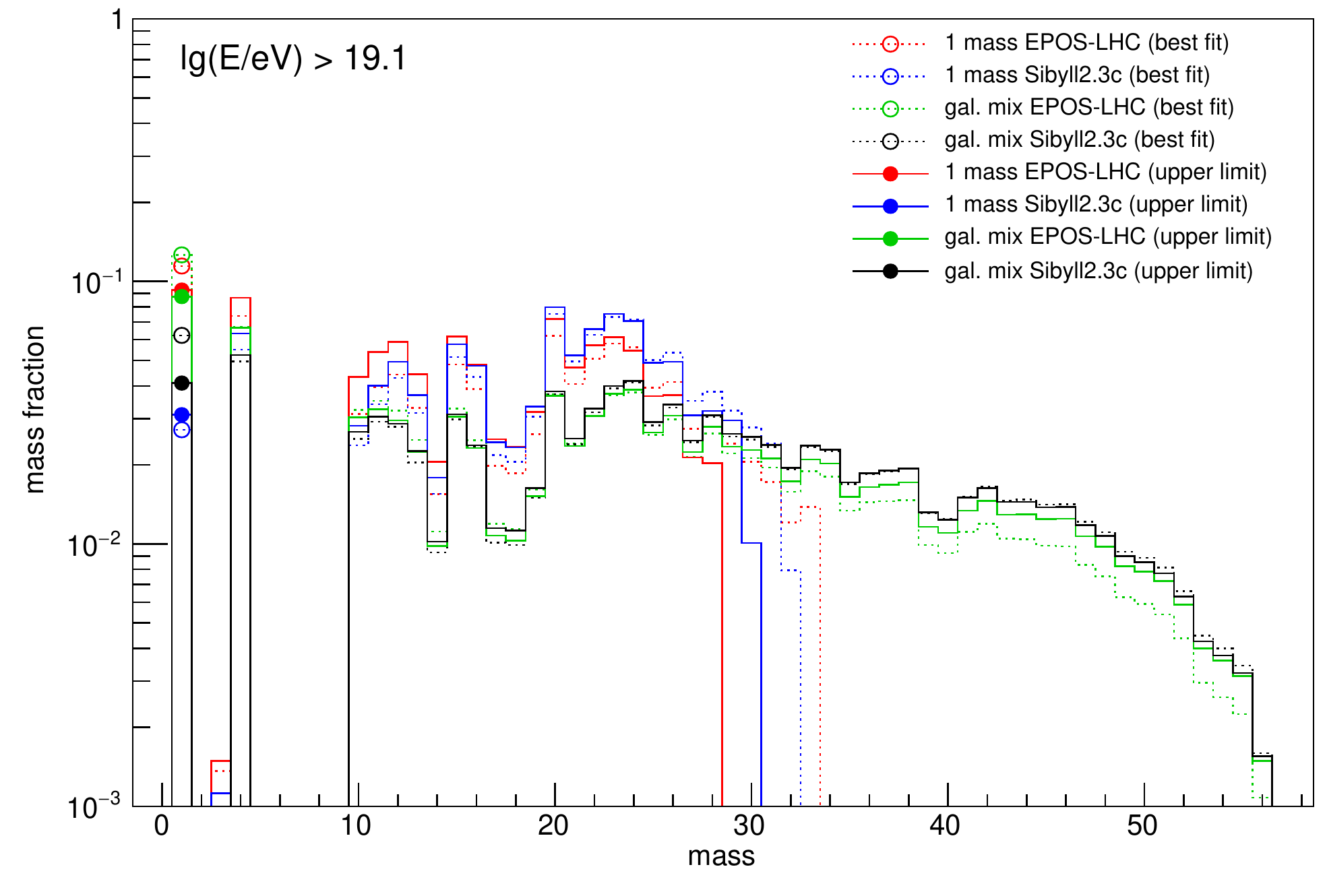}
\includegraphics[width=0.49\linewidth]{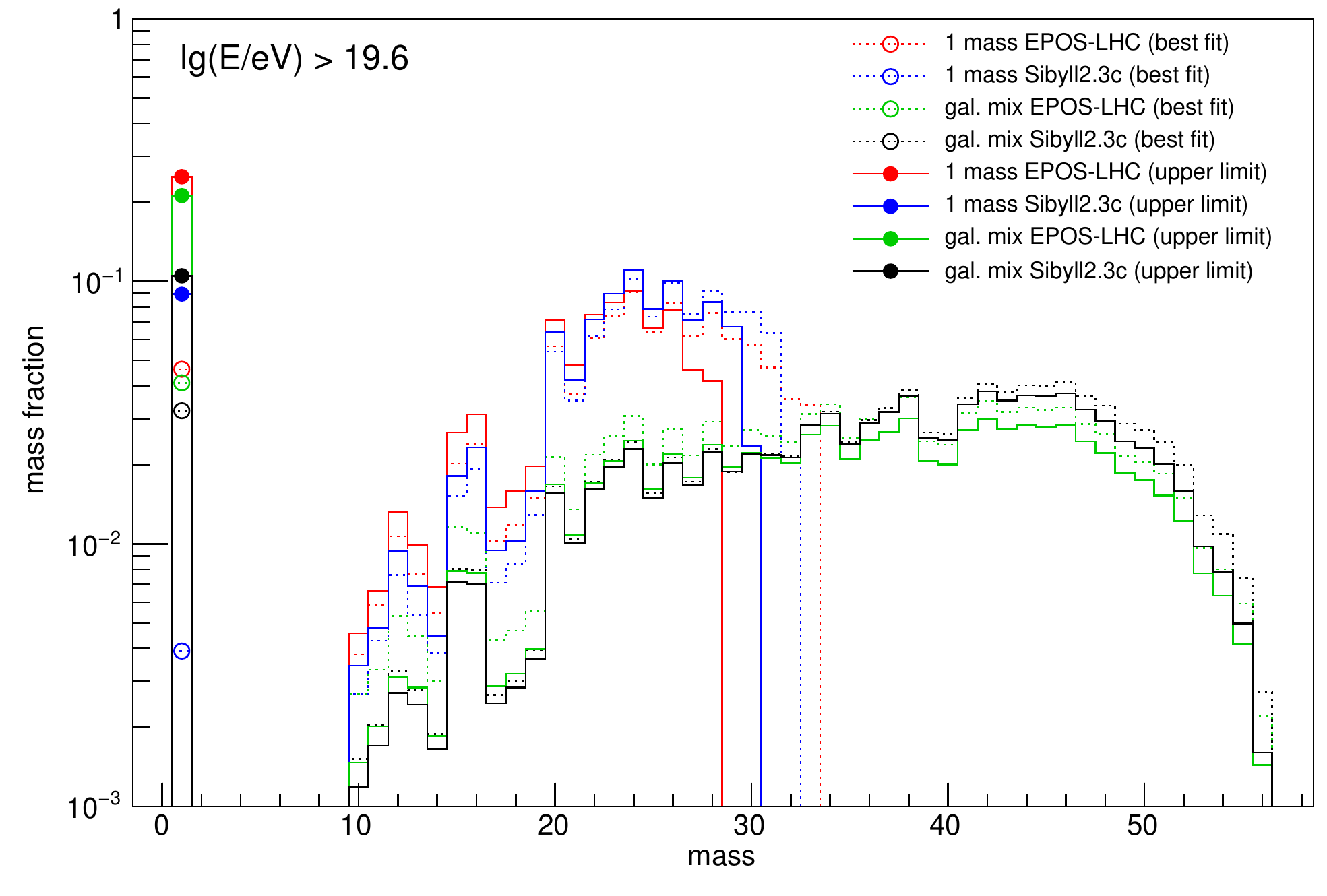}
\caption{Composition scenarios with UHE protons from~\cite{Muzio:2019leu} at
$10^{19.1}$~eV (left) and $10^{19.6}$~eV (right). There is no helium left above 40\,EeV and the mass distribution is broad with a peak at around $A\sim40$. }
\label{fig:composc}
\end{figure*}

\begin{figure*}[tb]
\centering
\includegraphics[width=.75\linewidth]{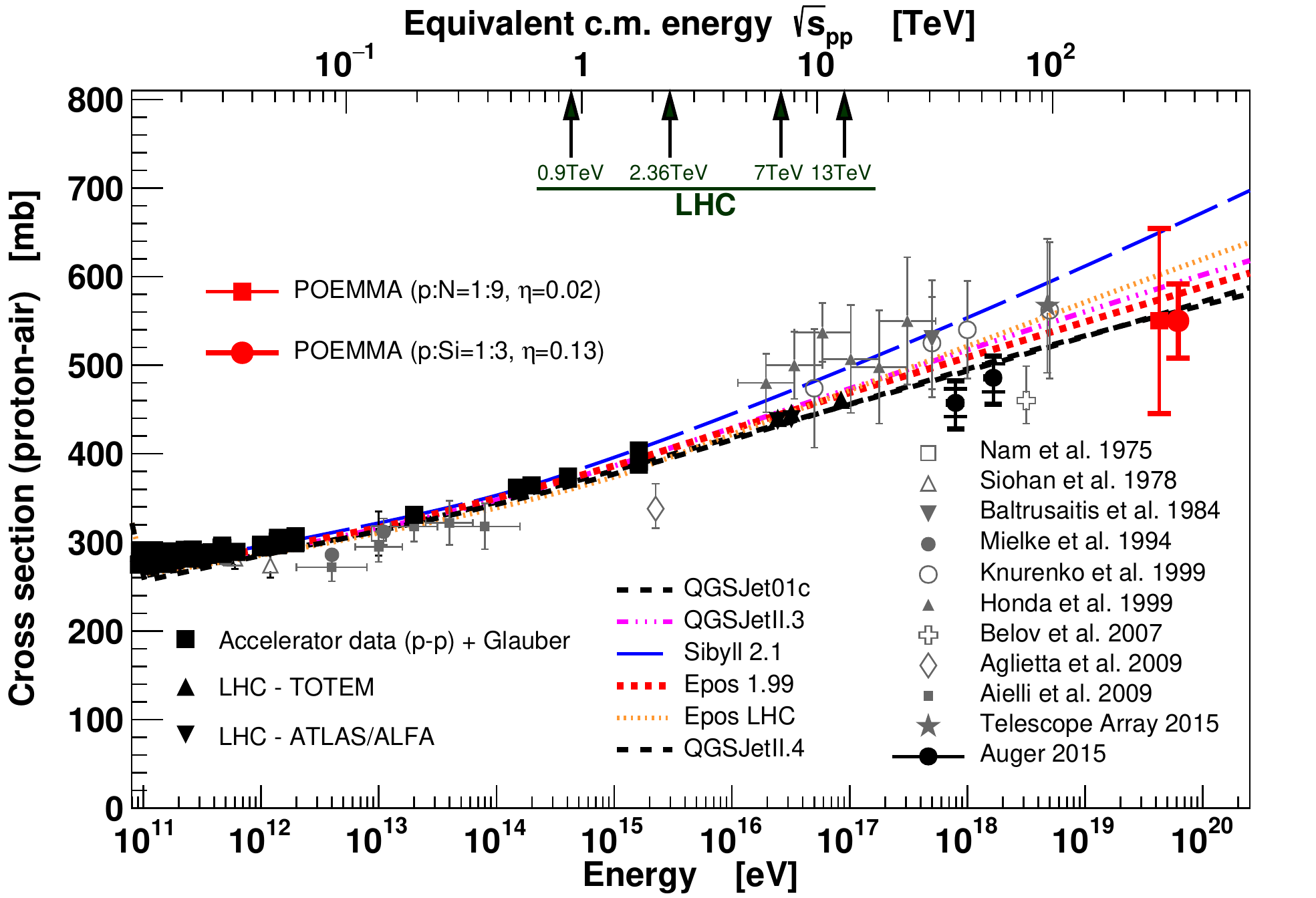}
\caption{Potential of a measurement of the UHE proton-air cross section
  with POEMMA. Shown are also current model predictions and a complete compilation of accelerator data converted to a proton-air cross section using the Glauber formalism. The expected uncertainties for two composition scenarios (left $p$:N=1:9, right $p$:Si=1:3) are shown as red markers with error bars. The two points are slightly displaced in energy for
  better visibility. }
\label{fig:xs}
\end{figure*}

The showers absorbed in the atmosphere observed by POEMMA correspond to a calorimetric fixed target experiment with $E_0>40\,$EeV. The collisions of the primary cosmic ray particles with atoms of the atmosphere happen at equivalent center-of-mass energies per nucleon-nucleon pair of above $\sqrt{s_{NN}}=\sqrt{2E_0m_{p}}=283\,$TeV. When those numbers are put into perspective with the capabilities of the LHC, where the beam energy is limited to 7\,TeV and the maximum center-of-mass energy to 14\,TeV, it is clear that there is an exciting opportunity to study fundamental particle physics at extreme energies.  
In this section we estimate the potential of POEMMA to measure the elemental inelastic cross sections at center-of-mass energies above $283\,$TeV. 
For this purpose we follow here the analysis procedure developed by the Pierre Auger Collaboration~\cite{Collaboration:2012wt}, where the exponential slope of the high-$X_{\rm max}$ tail is measured and related to the proton-air cross section $\sigma_{p-{\rm air}}$. 
The exponential slope $\Lambda_{\eta}$ is defined via 
\begin{equation}
\frac{dN}{dX_{\rm max}} \propto \exp(-X_{\rm max}/\Lambda_{\eta}) 
\end{equation}
for all $X_{\rm max}$ values above a minimum grammage $X_{\rm max}>X^{\rm start}_{\rm max}$. The slope $\Lambda_\eta$ is inversely proportional to the proton-air cross section via  
\begin{equation}
\sigma_{p-{\rm air}}=\frac{\langle m_{\rm air}\rangle}{k\Lambda_\eta},
\end{equation}
with $\lambda_{p-{\rm air}}=k\Lambda_\eta$ being the interaction length of protons in the air, and $\langle m_{\rm air}\rangle$ is the mean target mass of air.
Using the measurement of $\Lambda_\eta$ automatically enriches the proton contribution in a mixed mass scenario since protons are the most deeply penetrating cosmic ray nuclei. Thus, this approach is well suited for a first demonstration of the POEMMA capability. 

The choice of $X^{\rm start}_{\rm max}$ is critical and determines the statistical power of the analysis as well as the systematic effects of model- or mass-dependence. The value of $X^{\rm start}_{\rm max}$ can be defined via the fraction $\eta$ of the most deeply penetrating events. It was shown that $\eta=0.2$ yields optimal results in the case of the Pierre Auger Observatory for events with an average energy of $\sqrt{s_{NN}}=57\,$TeV and an assumed maximum helium contamination of 25\%~\cite{Collaboration:2012wt}. In the following, we investigate estimates of  $\eta$ at higher energies for POEMMA, 
informed by the $p-$air cross section analysis of the Pierre Auger Collaboration. 

In general, the distribution $dN/dX_{\rm max}$  depends to a large extend on the cosmic ray mass composition. We adopt as benchmark composition models the recent fits to Auger data from~\cite{Muzio:2019leu} shown in Fig.~\ref{fig:composc}. As can be seen, there is no helium in these models above 40~EeV. Indeed there is almost no helium production at sources, and the small attenuation length of $^4$He on the CMB~\cite{Soriano:2018lly} further suppresses the flux on Earth. Inspired by these models we investigated two simple scenarios as potential limiting cases of the mass composition at Earth  above 40 EeV: an optimistic one with a flux composition ratio of $p$:Si = 1:3 and a conservative one with $p$:N = 1:9. Obviously, a larger proton fraction directly leads to a better statistical precision and a better handle to limit systematic uncertainties. Heavier nuclei contribute less to the high $X_{\rm max}$ tail than lighter nuclei do, so nitrogen contributes more to the $X_{\rm max}$ tail than silicon, even with equal composition ratios. Thus, 90\% nitrogen corresponds to the most conservative assumption studied here. The more realistic mass distributions (for the galactic mixture) shown in Fig.~\ref{fig:composc} are significantly more favorable for a $\Lambda_\eta$ measurement than this extreme 90\% nitrogen scenario. 

It is anticipated that $N=1400$ events integrated over an energy above 40\,EeV will be observed by POEMMA,
when the measured shape of the cosmic ray spectrum above the ankle by the Pierre Auger Observatory is used~\cite{Aab:2017njo}. The flux as measured by the TA Collaboration could yield a much higher number of events, c.f.\  Figs.~\ref{fig:events2} and \ref{fig:events3}, but we do not include this option here.  

Since the measurement is entirely focused on the exponential slope of the tail the expected Gaussian detector resolution on the order of $35\,$g/cm$^2$ in $\xmax$ and 0.2 in $\Delta E/E$ does not affect this. The exponential slope $\Lambda_\eta$ is determined by using an unbinned $\log L$ fit~\cite{Ulrich:2015yoo}
approximated by the result in the large $\xmax$ limit, as described in Appendix \ref{appendix:b}. Thus, the relative statistical uncertainty of $\Lambda_\eta$ is simply $1/\sqrt{N_{\rm tail}}$, where $N_{\rm tail}=N\eta$ is the number of events in the tail of the $\xmax$ distribution. 

In the following, we use two different choices for $\eta$ following the guidance of the Pierre Auger Observatory: for the $p$:N = 1:9 case, $\eta$ reflecting the proton content of 0.1 is reduced by an additional fraction 0.2 to minimize the potential impact of the non-proton primaries that contaminate the high $\xmax$ tail; for the $p$:Si case, since Si is a heavier primary that affects $\Lambda_\eta$ much less, we use a fraction of 0.5 the proton content 0.25 for the tail measurement. We arrive with a very conservative effective $\eta=0.1\times0.2=0.02$ for the pessimistic scenario of $p$:N = 1:9, and $\eta=0.25\times0.5=0.13$ for the more optimistic one with $p$:Si = 1:3. 

Now, using the estimate of the overall number of events above $40\,$EeV of $N=1400$ and combining it with expectations from cosmic ray propagation simulations indicating possible mass composition scenarios, we can determine a projected measurement of the proton-air cross section as shown in Fig.~\ref{fig:xs}. In this plot the uncertainties of the left point for POEMMA corresponds to the $p$:N = 1:9 and the right point to $p$:Si = 1:3 proton fraction scenarios. The analysis described here is not yet optimized for the actual POEMMA observations and we study two very different potential scenarios. For illustration purposes, the central value of the projected POEMMA points in Fig.~\ref{fig:xs} are located at the lower range of the model prediction. This is what some of the recent data from the Pierre Auger Observatory and also LHC suggest. 

In the final step, these data will also be converted into the fundamental inelastic proton-proton cross section $\sigma_{pp}^{\rm inel}$ using an inverse Glauber formalism.

\subsubsection{Searches for super-heavy dark matter}

One of the leading objectives of the particle physics program is to identify the connection between dark matter (DM) and  the Standard Model (SM). Despite the fact there is ample evidence for DM existence, the specific properties and the identity of the particle DM remain elusive~\cite{Bertone:2004pz}. For many decades, the favored models characterized DM as  relic density of weakly interacting massive particles (WIMPs)~\cite{Feng:2010gw}. Theoretical ideas and experimental efforts have focused mostly on production and detection of thermal relics, with mass typically in the range of a few GeV to a hundred GeV. However, despite numerous direct and indirect detection searches~\cite{Klasen:2015uma,Undagoitia:2015gya}, as well as searches for DM produced at particle accelerators~\cite{Buchmueller:2017qhf,Penning:2017tmb}, there has thus far been no definitive observation of the WIMP particle. Moreover, as of today, there have been no definitive hints for new physics beyond the SM at any accessible energy scale~\cite{Rappoccio:2018qxp}, suggesting that nature does not too much care about our notion of naturalness. Anthropic reasoning, bastioned by probabilistic arguments of the string theory landscape, seems to indicate that if the universe is fine-tuned then the natural mass range for the particle DM would be the Planck scale~\cite{Garny:2015sjg,Garny:2017kha,Alcantara:2019sco}. Without DM, the epoch of galaxy formation would occur later in the universe, thus galaxies would not form in time for our existence. However, it is only the DM abundance and not any other details of dark sector which is critical for life to exist. Then it is quite reasonable to expect that the DM sector would not be as fine tuned as the visible SM sector. Production of non-thermal super-weakly interacting super-heavy dark matter (SHDM) particles is ubiquitous in string theory~\cite{Chung:1999ve,Kannike:2016jfs}. While SHDM must be stable over cosmological timescales, instanton decays induced by operators involving both the hidden sector and the SM sector may give rise to observable signals in the  spectrum of UHECRs~\cite{Berezinsky:1997hy,Kuzmin:1997jua}.

\begin{figure}[tb] 
    \centering
\includegraphics[width=\linewidth]{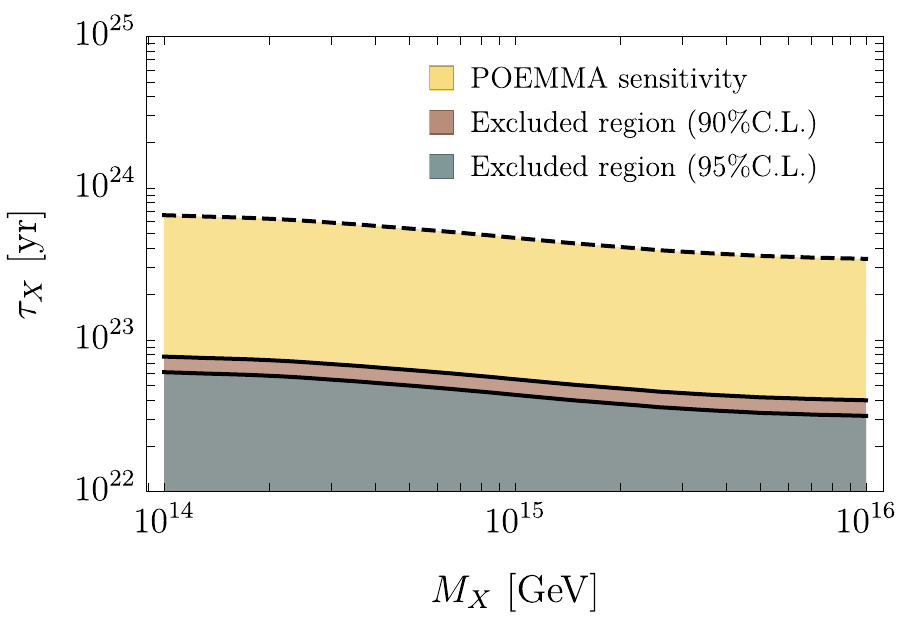}
  \caption{Lower limit on the lifetime of SHDM particles together
  with the sensitivity (defined as the observation of one photon
  event above $10^{11.3}~{\rm GeV}$ in 5~yr of data collection) of POEMMA operating in stereo mode~\cite{Alcantara:2019sco}. \label{fig:SHDM}}
  \end{figure}

When SHDM decays into SM particles, photons and neutrinos dominate the
final state.  The energy spectra depend on the exact decay mechanism
and is somewhat model dependent.
There are several computational schemes proposed in the
literature~\cite{Birkel:1998nx,Sarkar:2001se,Berezinsky:2000up,Barbot:2002gt},
which predict accurately the secondary spectra of SM particles
produced in the decay of SHDM $X$-particles and agree well each other. The
expected energy distribution on Earth follows the
initial decay spectrum, whereas the angular distribution incorporates
the (uncertain) distribution of dark matter in the Galactic halo via
the line-of-sight integral~\cite{Dubovsky:1998pu, Evans:2001rv,Aloisio:2007bh,Kalashev:2017ijd}. The photon energy flux is estimated to be
$\Phi_{\gamma} \propto (M_X \tau_X)^{-1},$ where $M_X$ is the mass of the
particle and $\tau_X$ its lifetime~\cite{Aloisio:2015lva,Kalashev:2016cre}. The
non-observation of extreme-energy photons can be used to constrain the
$\tau_X-M_X$ parameter space.  In Fig.~\ref{fig:SHDM} we show the lower limit on the
lifetime of SHDM particles from the non-observation of UHECR photons
at Auger together with the sensitivity of POEMMA operating in
stereo mode~\cite{Alcantara:2019sco}. CMB observations set a bound  $M_X \alt 10^{16}~{\rm GeV}$ at 95\%~CL~\cite{Garny:2015sjg}. Detection of a extreme-energy photon
would be momentous discovery. If this were the case POEMMA could be
switched into limb-mode to rapidly increase statistics. 
Note that for energies $E\agt 10^{20}~{\rm eV}$, the average $X_{\rm max}$
  of photon and proton showers differs by more than
  $100~{\rm g/cm^2}$~\cite{Risse:2007sd}. This implies that POEMMA
  operating in limb-viewing mode, with $X_{\rm max}$ resolution $\sim 100~{\rm g/cm^2}$ determined from simulated monocular reconstruction performance, will be able to deeply probe the SHDM
  parameter space.   
 
 \subsubsection{Searches for neutrinos from decaying topological defects}
 
 Beyond SM physics often also predicts phase transitions at which topological defects are formed~\cite{Kibble:1976sj}. The decay of  topological defects is also dominated by photons and neutrinos, but since these relics are expected to be distributed uniformly throughout the universe the photon signal would cascade down to lower energies and extreme-energy neutrinos
become the smoking gun.

POEMMA will be a powerful probe of extreme-energy neutrino emission. An example involves the flux neutrinos produced via decay of highly boosted (strongly coupled) moduli, radiated by relic cosmic strings~\cite{Berezinsky:2011cp}. These strongly coupled moduli are scalar fields, which appear to be quite generic in the string theory landscape~\cite{Conlon:2007gk,AbdusSalam:2007pm}.
 
\begin{figure*}[tpb]
 \centering
\includegraphics[width=\linewidth]{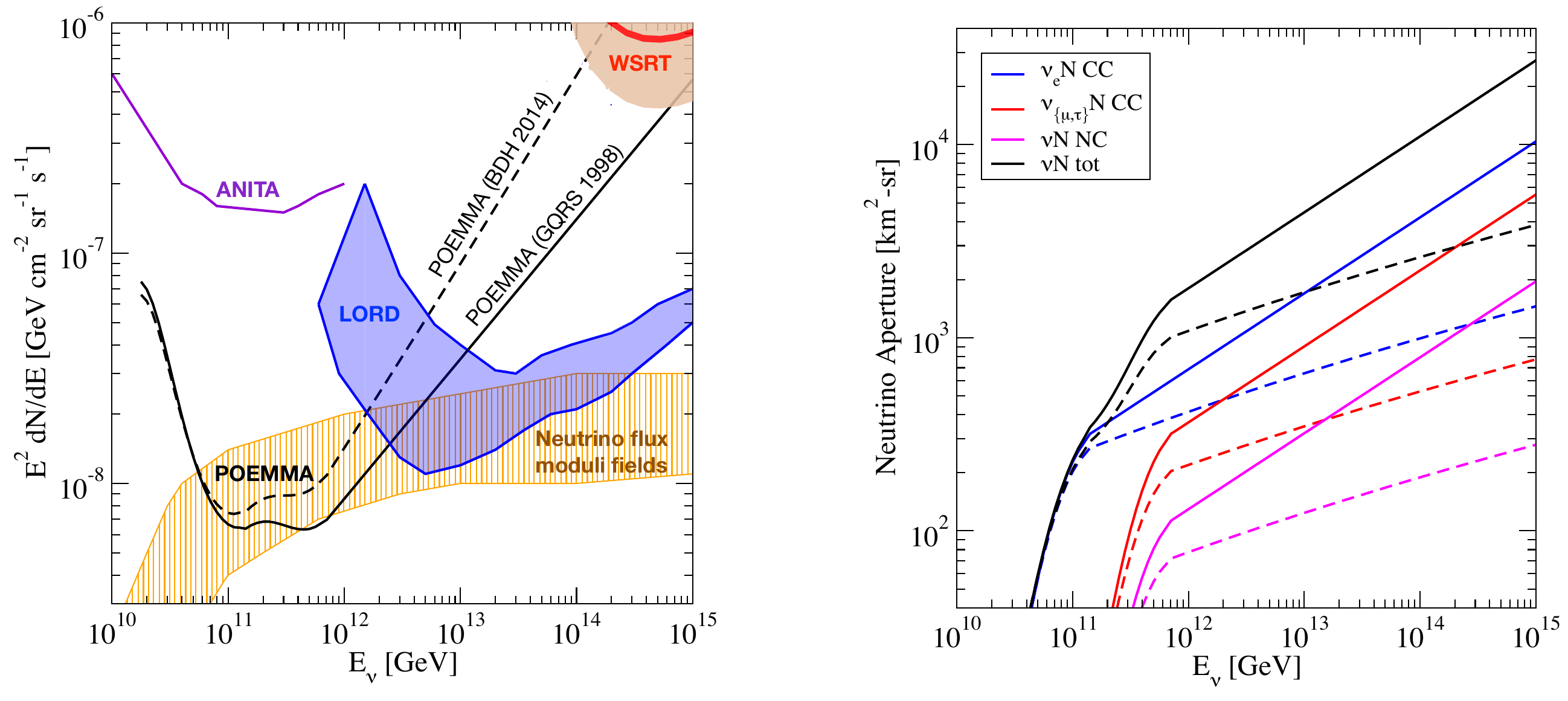}
\caption{POEMMA sensitivity to EAS showers produced by neutrinos interacting in the atmosphere. Left: Black curves are POEMMA sensitivity using~\cite{Gandhi:1998ri} (solid) and ~\cite{Block:2014kza} (dashed) cross sections. Predictions for strongly coupled string moduli (orange range)~\cite{Berezinsky:2011cp}, LORD sensitivity (blue band)~\cite{Ryabov:2016fac}, ANITA I-IV limits (purple)~\cite{Gorham:2019guw}, and WSRT~\cite{Scholten:2009ad}). Right: POEMMA neutrino aperture from fluorescence with~\cite{Gandhi:1998ri} (solid) and \cite{Block:2014kza} (dashed) cross sections of neutrino-nucleon scattering for CC interactions of $\nu_e$ (blue), $\nu_\mu$ and $\nu_\tau$ (red); NC interactions (pink); and total (black). \label{fig:eenus}}
\end{figure*}

In Fig.~\ref{fig:eenus} we show the sensitivity of both POEMMA and the future Lunar Orbital Detector (LORD) ~\cite{Ryabov:2016fac}, and current limits from ANITA I-IV~\cite{Gorham:2019guw}  and WSRT~\cite{Scholten:2009ad}. We also show the neutrino flux range resulting from models of strongly coupled moduli in a string theory background with $G \mu \sim 10^{-20}$, where $G$ is the Newton's constant and $\mu$ the string tension~\cite{Berezinsky:2011cp}. POEMMA sensitivity was estimated assuming the stereo configuration observations with 10\% duty cycle and 5 years of observation time. The neutrino aperture is the result of simulations of isotropic events interacting deep in the atmosphere (observed starting point $X_{\rm Start} \ge 2000~{\rm g/cm^2}$).
The sensitivity includes all flavors, charged and neutral current interactions, for two different cross-sections as estimated in~\cite{Gandhi:1998ri} and~\cite{Block:2014kza}.
POEMMA UHE neutrino simulated neutrino air fluorescence response is detailed in Appendix \ref{appendix:c}.

It is evident that POEMMA will probe a significant part of the parameter space, providing a method of searching for strongly coupled moduli, which complements searches based on gravitational effects of strings, like structure formation, cosmic microwave background, gravitational radiation, and gravitational lensing. The strongest current bound from lensing effects
is estimated to be $G\mu \lesssim 10^{-7}$~\cite{Christiansen:2010zi}, while millisecond pulsar
observations lead to
$G\mu \lesssim 4 \times 10^{-9}$~\cite{vanHaasteren:2011ni}. Next generation gravitational wave
detectors are expected to probe $G\mu \sim 10^{-12}$~\cite{Damour:2004kw,Olmez:2010bi}. Thus, POEMMA will attain sensitivity to a region of the parameter space more than 10 orders of magnitude below current limits and about 8 order of magnitude smaller than next generation gravitational wave detectors. It should be noted that these stereo fluorescence neutrino measurements effectively ``come for free'' when POEMMA is in stereo UHECR observation mode.

\section{Conclusions}
\label{sec:6}
The transformational UHECR physics performance of POEMMA above 40 EeV is based upon the  large, all-sky exposure enabled by a stereo UHECR space-based experiment. POEMMA is designed with wide FoV Schmidt optics coupled to relatively fine pixel resolution to provide excellent angular, energy, and $X_{\rm max}$ resolutions, using stereo UHECR reconstruction. This performance translates into high sensitivity for UHECR composition determination, UHECR anisotropy and source determination, and providing fundamental physics measurements on dark matter and proton-proton cross sections along with remarkable UHE neutrino sensitivity at the highest energies ever achieved.
We have shown that POEMMA will provide new Multi-Messenger Windows onto the most energetic environments and events in the universe enabling the study of new astrophysics and particle physics at these otherwise inaccessible energies.

\acknowledgements

We thank our colleagues of  POEMMA, Pierre Auger, and JEM-EUSO collaborations for some valuable discussion.
This work has been supported by NASA (grants 80NSSC18K0464, NNX17AJ82G, 17-APRA17-0066), the U.S. National Science Foundation (NSF grant PHY-1620661), and the 
 U.S. Department of Energy (grant DE- SC-0010113).

\appendix

\section{Modeling POEMMA using ESAF}
\label{app:a}

A first estimation of POEMMA performance in terms of trigger exposure and 
quality of event reconstruction has been assessed using the ESAF code assuming a clear atmosphere. 
Because the POEMMA PFC is baselined to use the photodetector modules (PDMS) and electronics developed for the JEM-EUSO mission, it is reasonable to adopt the JEM-EUSO trigger algorithms and
reconstruction procedures to evaluate the performance of POEMMA.

Here we provide an overview of ESAF. The Greisen-Ilina-Linsley (GIL) function \cite{Ilina:1992qv}
is used as parametric generator to reproduce the profile as a function of
slant depth. The GIL function is optimized to reproduce EAS from hadronic
particles simulated by CORSIKA \cite{1998cmcc.book.....H} with the QGSJET01 hadronic interaction
model \cite{Kalmykov:1993qe}. Proton showers have been simulated for the analyses presented
in this paper. This is motivated by the fact that they develop deeper in
the atmosphere, which results in a higher atmospheric absorption and
higher cloud impact. Therefore, the results that are discussed in the
following sections constitute a conservative estimation on the performance
of the instrument.
In the present work, the fluorescence yield is taken from \cite{Nagano:2004am}. In the
atmosphere, UV photon propagation is strongly affected by Rayleigh
scattering and absorption by ozone for wavelengths $< 320$ nm. These
processes along with the atmospheric profile are modeled with the LOWTRAN
package \cite{LOWTRAN7}.

\begin{figure}
\begin{center}
\includegraphics[width=0.45\textwidth]{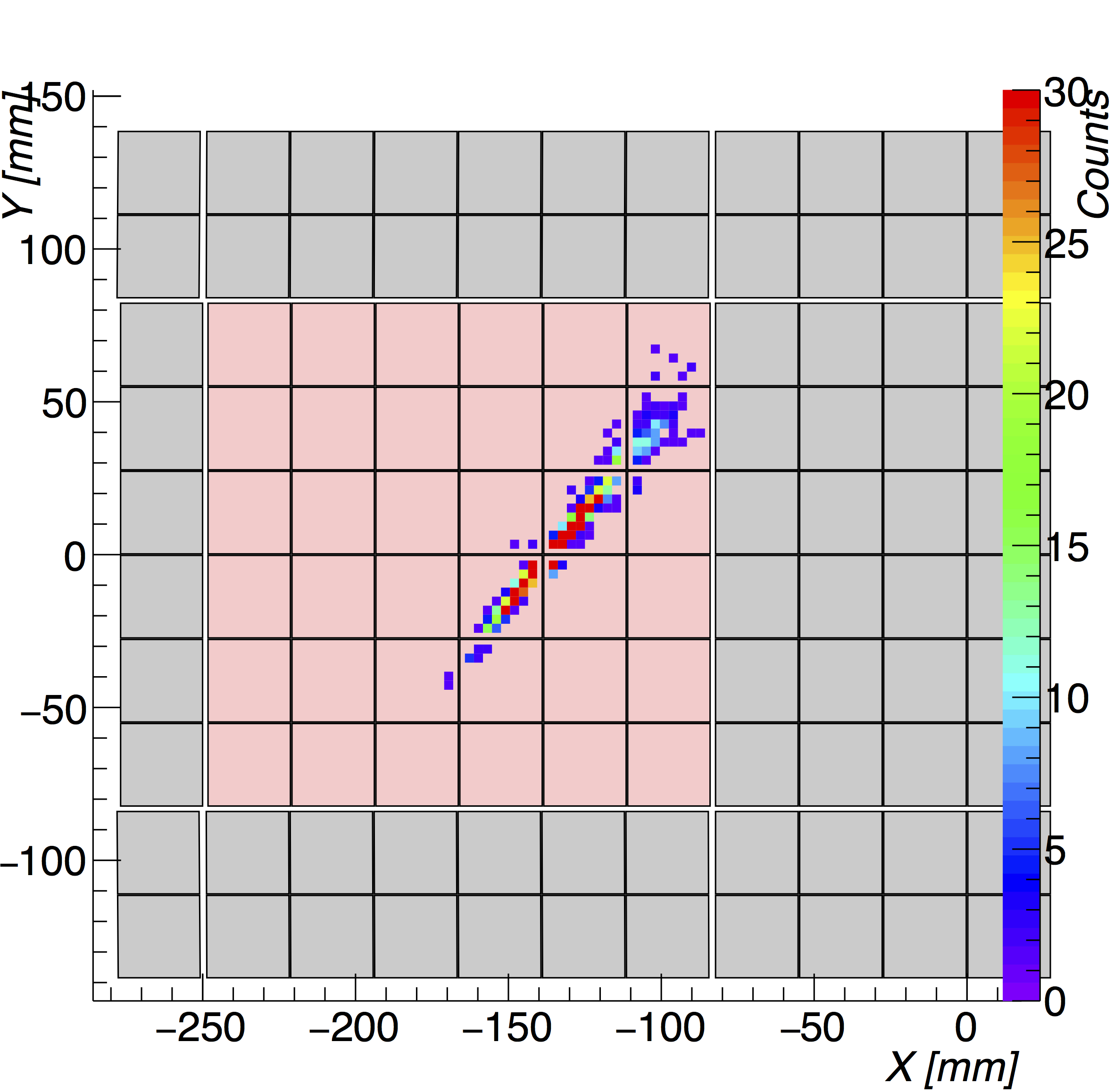}
\end{center}
\caption{EAS fluorescence image on one POEMMA PDM for 
a 100~EeV, 60$^\circ$ inclined shower. Background is not shown.}
\label{fig:event}
\end{figure}

The POEMMA detector was implemented in ESAF~\cite{Berat:2009va} using the 
parametric optical model shown in Fig.~\ref{ParamOptics}. A single telescope, which allows for the calculation of POEMMA's monocular mode performance, was considered since ESAF does not allow 
yet a stereoscopic vision.  The trigger efficiency is determined at single
telescope level, because POEMMA's telescopes operate independently regardless of mode of operation.
The simulations were performed assuming a standard UV nightglow background level of
500~photons/m$^2$/ns/sr  in the 300 - 500 nm
band~\cite{Adams:2013vea}, which is appropriate considering the use of a BG3 UV filter in the PFC. Taking into account the POEMMA detector response this corresponds to an
average equivalent count rate 1.54 counts/$\mu$s/pixel. 
 In Fig.~\ref{fig:event} we show the track image in a PFC focal plane of a typical proton EAS of 100~EeV,  60$^\circ$ inclined from nadir.

\subsection{Trigger efficiency}

\begin{figure*}
\begin{center}
\includegraphics[width=0.9\textwidth]{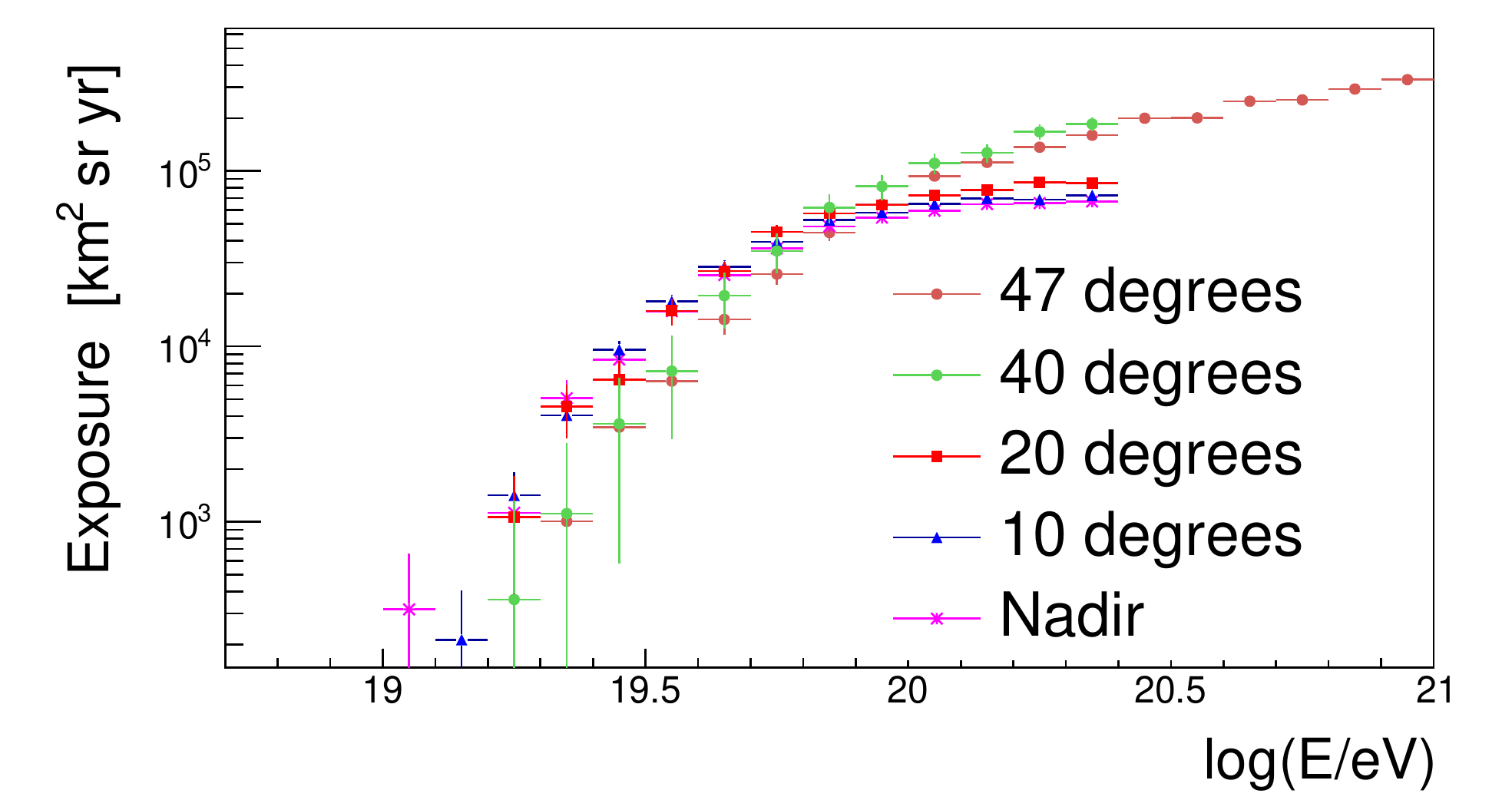}
\end{center}
\caption{Exposure curve of POEMMA monocular observation. Different curves 
correspond to different tilting angles. }
\label{fig:expo}
\end{figure*}

To estimate the trigger efficiency curve of POEMMA, an overall set of 20,000 
proton EAS were simulated with ESAF using the following parameters: primary energy ($E$) in the range $5 \leq
		E/{\rm EeV} \leq 500$, zenith angle $0 \leq \theta 
\leq \pi/2$, and azimuth angle $0 \leq \phi \leq 2\pi$. 
To avoid border
effects EAS were simulated on an area which is almost twice bigger; namely, 
$S_\text{FoV} \sim 145,000$~km$^2$ and $S_\text{sim} \sim 271,000$~km$^2$.
The exposure is then calculated according to 
\begin{equation}
{\cal E}(E) = A_\text{geo}(E) \ t \ \epsilon \,,
\end{equation}
where 
\begin{equation}
A_\text{geo}(E) = \int_{S_\text{sim}} \!\!\!dS \int_0 ^{2\pi} \!\!d\phi \int_0 ^{\pi/2} \!\! d\theta
\cos\theta \, \sin\theta \, \epsilon(E)
\end{equation}
is the geometrical aperture,
 $t$ is the observation time, 
 \begin{equation}
\epsilon (E) = \frac{N_\text{trig}}{N_\text{sim}} 
\end{equation}
  is the average efficiency of
the detector at a given energy $E$, and $N_\text{trig}$ and $N_\text{sim}$ are respectively the number of triggering 
events and the number of simulated events at a certain energy $E$.
The events were passed through the electronics, and the response of the first 
JEM-EUSO trigger level was applied~\cite{Abdellaoui:2017huh}. This trigger level is based
on an excess of signal in a box of $3 \times 3$ pixels for 5 consecutive GTUs.
The thresholds for this signal excess reduce the rate of fake trigger at the level of
$\sim 7$~Hz/PDM. The second trigger level, which is not applied here, looks for a signal
excess on a track, lasting 15 GTUs. The thresholds are set to have a fake trigger rate of
$\sim 0.1$~Hz/FS. According to simulations for the JEM-EUSO mission, the application of this second trigger
level shifts the trigger efficiency curve by $\sim 10\%$ at higher energies.
In Fig.~\ref{fig:expo} we show the exposure determined for nadir mode and  
different tilt angles away from nadir. 

\subsection{Reconstruction performance in monocular nadir observation}

A first evaluation of the reconstruction performance in monocular mode and 
nadir pointing was 
performed using the same algorithms defined for JEM-EUSO. No optimization has
been introduced yet, like the use of a shorter GTU (1 $\mu$s). Therefore, the
results presented in this work have to be considered as conservative. Two 
different pattern algorithms have been defined in ESAF to reconstruct the
EAS parameters. The  ``LTTPatternRecognition'' was developed to collect as much
signal as possible from the EAS track to get a better estimation of the EAS
energy, even at the expenses of the uncertainty in the direction reconstruction.
The details of this method are reported in~\cite{fenu-expastro}.
On the other hand, the {\it PWISE} method developed for the track direction is
meant to make the track as narrow as possible to reduce the uncertainty on the
direction reconstruction. In the following, results are shown applying the
two chains independently. Before proceeding, it is important to underline that a failure of
the PWISE algorithm does not imply that the event can not be reconstructed, as
the chain for the reconstruction of the energy would anyway provide a first
estimation of the arrival direction. Naturally, the ideal case would be to use 
both algorithms in parallel in order to get as much information as possible for
the same event.

The triggered EAS were passed through JEM-EUSO reconstruction algorithms
discussed in~\cite{fenu-expastro} to evaluate the quality in the EAS
reconstruction for the POEMMA detector.

In JEM-EUSO two methods have been developed to determine the altitude of the shower maximum: the
Cherenkov method and the method based on the direction reconstruction (slant depth method).
The Cherenkov method uses the Cherenkov reflection
mark which identifies the location and time at
which the EAS reaches the ground. This method can be applied only for vertical EAS, for which
the Cherenkov mark is bright enough and it is located on a limited number of pixels. For
more inclined EAS, such peak is spread in time and space and can not be identified. The slant depth method
assumes a parameterization for the depth of the shower maximum
and relies on the direction delivered by the direction reconstruction.

Before applying the reconstruction algorithms,
a pattern recognition algorithm is applied to the data to identify the center of the spot
in each GTU. Indeed, for each event, the background is simulated on all the PDMs that were
crossed by
the EAS event. Therefore, the signal due to the EAS cascade has to be extracted from 
background. This is done to mimic the real conditions expected on flight.

\begin{figure*}
\centering
\includegraphics[width=0.49\textwidth]{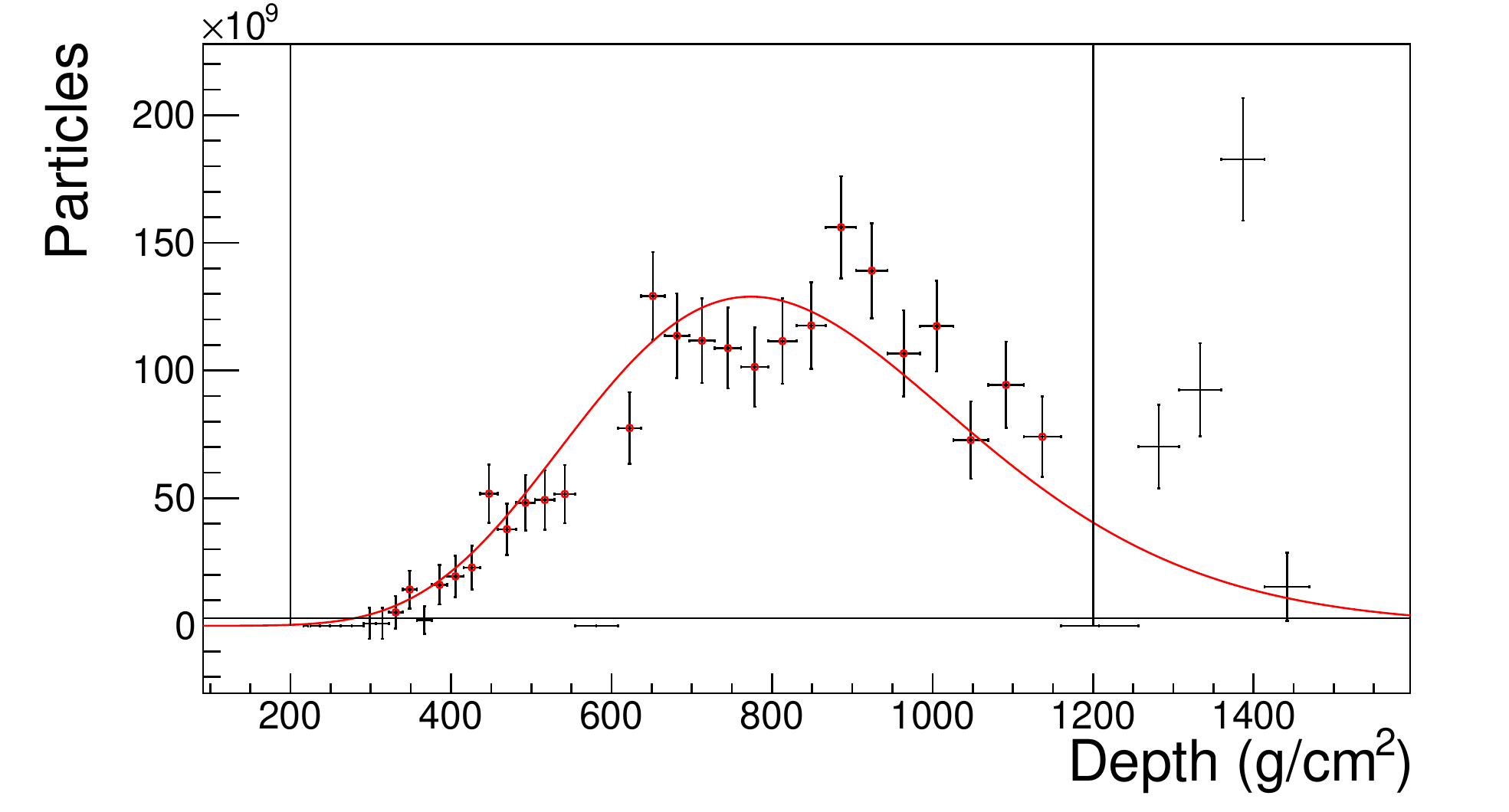}
\includegraphics[width=0.49\textwidth]{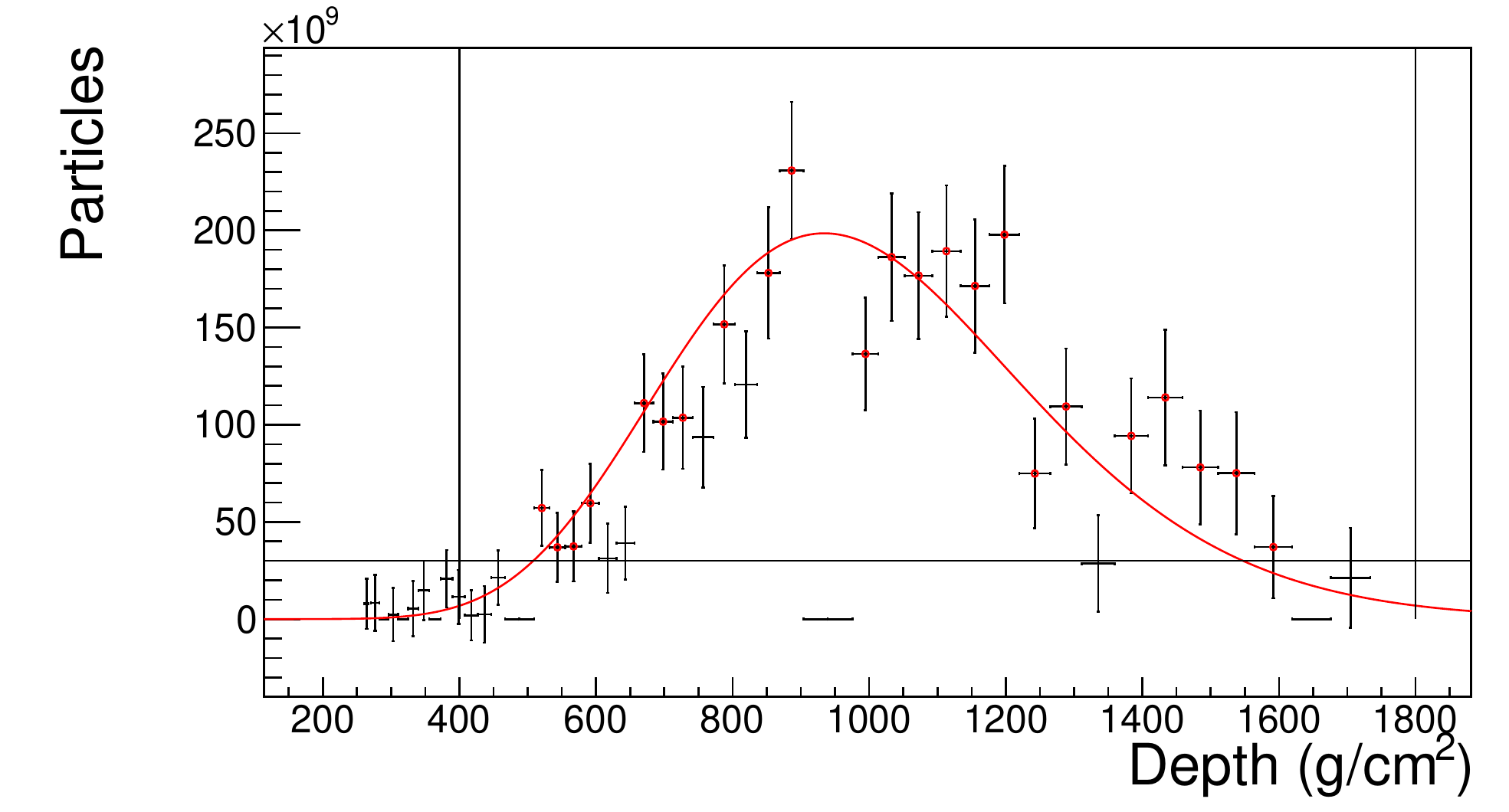}
\caption{Example of reconstructed events. The event on the left (E = 1.7$\times$10$^{20}$ eV,
$\theta$ = 49.6$^o$ and $\phi$ = 9.2$^o$) has been reconstructed with
the Cherenkov method (it is possible to recognize the Cherenkov peak -blue dotted line-). The
right one (E = 2.5$\times$10$^{20}$ eV, $\theta$ = 56.3$^o$ and $\phi$ = 344.0$^o$)
has been reconstructed by the method based on the direction reconstruction
(slant depth method). The red fitted lines are the GIL functions (an analytical approximation
of the EAS longitudinal development - details can be found in~\cite{fenu-expastro}).
The vertical black lines identify the fitting interval and the horizontal one the background
level.}
\label{Fig:CherNoCher}
\end{figure*}

The reconstruction of events were performed using both methods separately.
In Fig.~\ref{Fig:CherNoCher} we show two examples of reconstructed events. On the left panel, it is possible to recognize the Cherenkov peak. This event can be reconstructed using the Cherenkov
method and the slant depth method. Instead, on the right panel, the Cherenkov peak is not
present and the event can be reconstructed only using the slant method.

To perform a reconstruction, we carry out a chi-square goodness of fit test with a 
number of degrees of freedom (DOF) $>4$ and the likelihood
satisfying $0.1 < \chi^2 < 3$. For this analysis, 11,200 events were simulated in the energy range $19.2 \leq \log(E/{\rm eV}) \leq
20.8$. Out of the total, 3,879 events ($\sim$35\%) were triggered and passed through the pattern recognition
algorithm and EAS reconstruction chain. From this selected sample, 3,253 events (equivalent to $\sim$84\%) were
successfully reconstructed by the slant method and 1,472 events (equivalent to $\sim$38\%) by
the Cherenkov method. It is not surprising
that in the Cherenkov method only about half of the events were reconstructed. As mentioned
before this is due to the fact that this method is usable up to zenith angles $\sim 50^o$
(the value depends on energy and EAS location on the FS).
At higher zenith angles the Cherenkov signal is too dim to be isolated from 
background fluctuations.

\begin{figure*}
\centering
\includegraphics[width=0.49\textwidth]{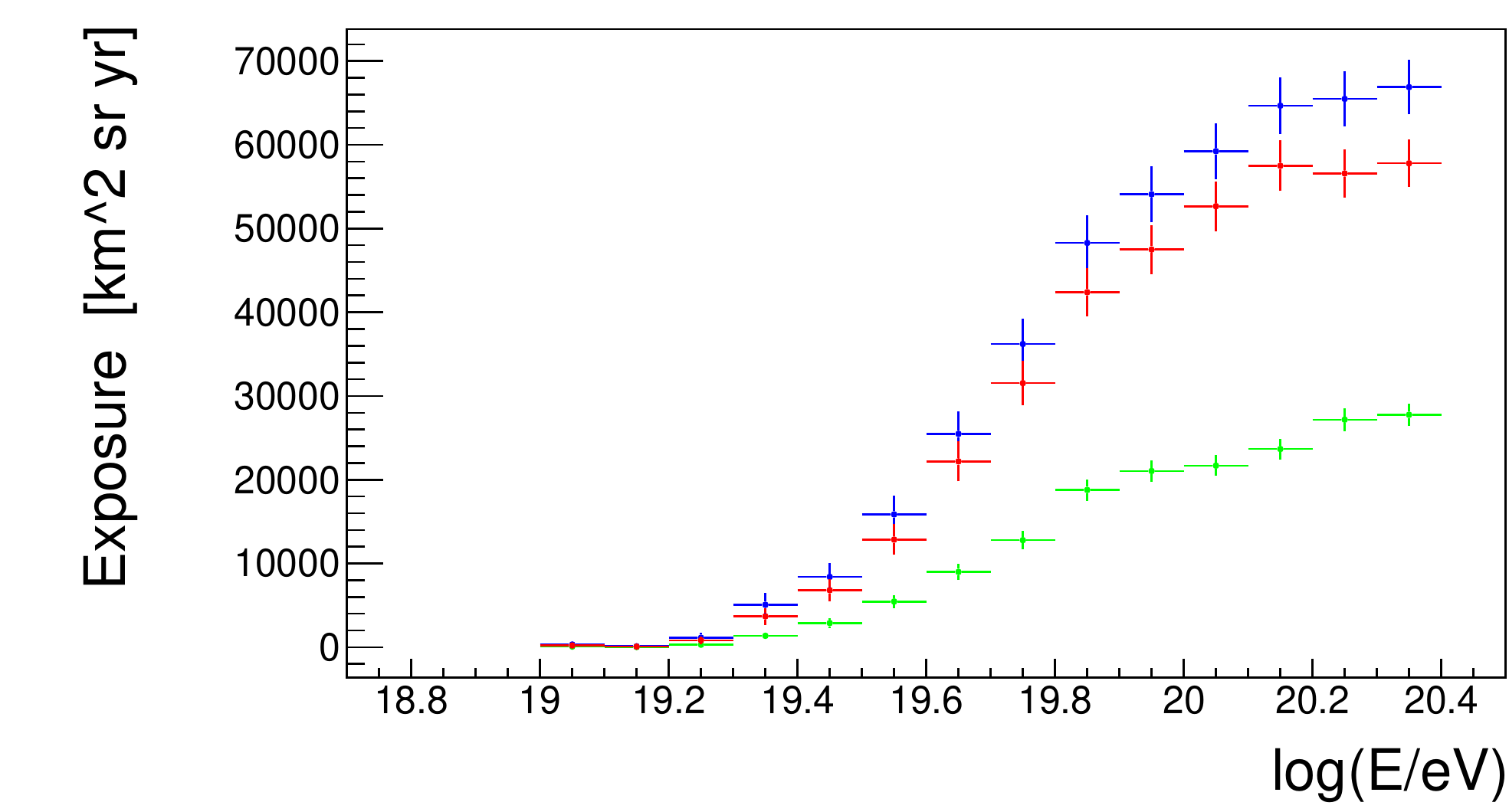}
\includegraphics[width=0.49\textwidth]{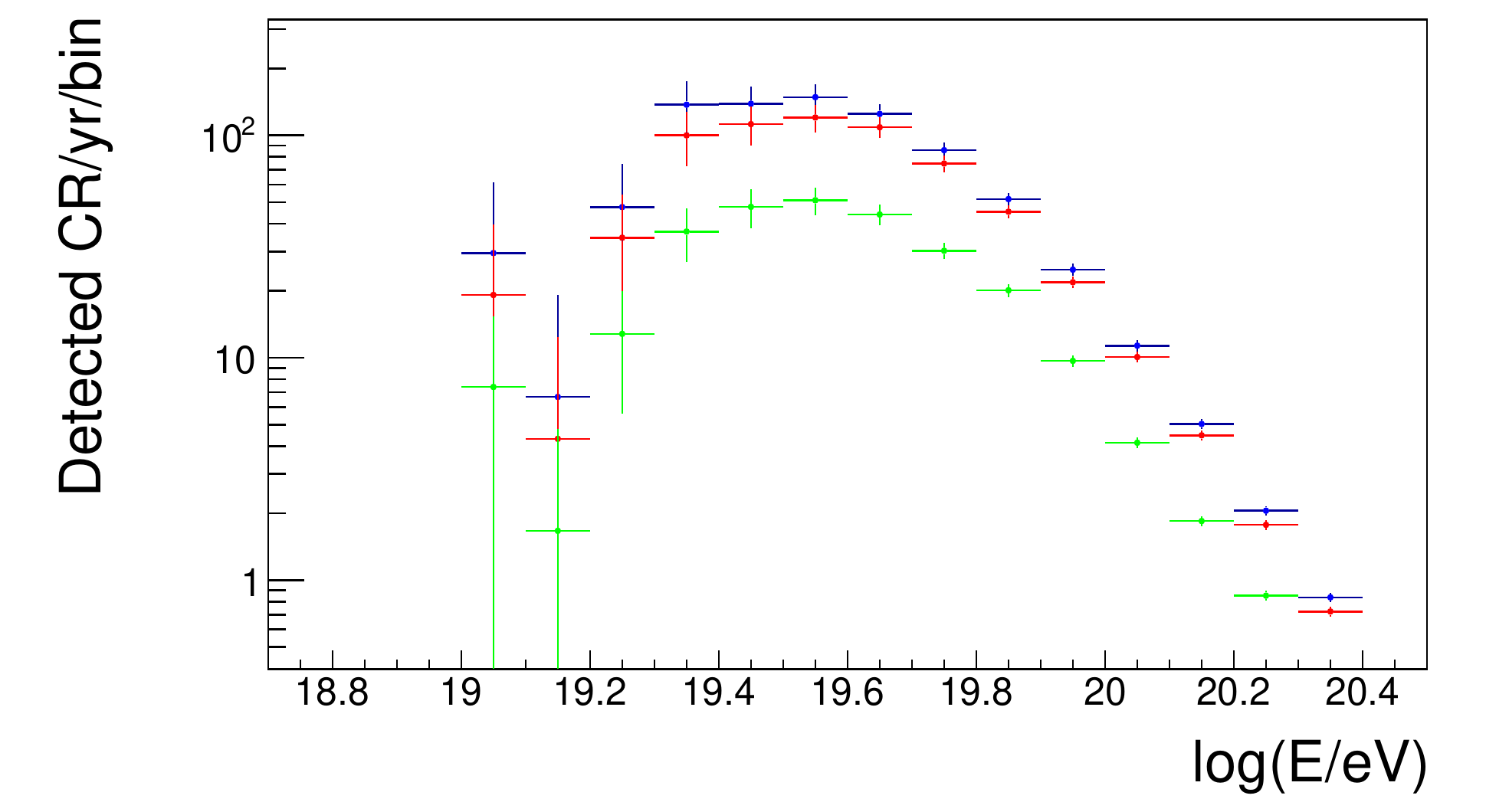}
\caption{Left: Comparison between the exposures obtained from POEMMA 
simulations (triggered events -blue-, events reconstructed by Cherenkov method 
-green-, events reconstructed by the slant depth method -red-). 
Right: Comparison between the triggered spectra obtained
from POEMMA simulations (triggered events -blue-, events reconstructed by 
Cherenkov method -green-, events reconstructed by the slant depth method 
-red-).}
\label{Fig:ExposureRecoTOT}
\end{figure*}

In the left panel of Fig.~\ref{Fig:ExposureRecoTOT} we show a comparison between the triggered and 
reconstructed
spectra using both reconstruction methods. In the right panel of this figure we show the expected number of detected
and reconstructed events per year assuming the Pierre Auger energy spectrum.

A comparison between the number of triggered events and the number of events automatically
reconstructed with the two reconstruction methods for
$log(E/eV) > 19.6$ is provided in~\tablename~\ref{Tab:NumberRECOandINTAutoManual}.
\begin{table}
\caption{Comparison between the number of triggered and reconstructed events 
above an energy $log(E/{\rm eV})=19.6$. }
\label{Tab:NumberRECOandINTAutoManual}
  \centering
\begin{tabular}{lcc}
\hline
\hline
    & \textbf{Cherenkov method} & \textbf{Slant depth method}       \\
\hline
Triggered      &     305           &   305                           \\
Reconstructed  &     110 $(\rightarrow 36\%)$ &   267 $(\rightarrow 88\%)$      \\
\hline
\hline
\end{tabular}
\end{table}
When a selection cut is placed at $log(E/eV)=19.6$, the overall fraction of
reconstructed events above this energy is slightly higher than using all the 
data sample.
In particular, the slant depth method approaches 90\% efficiency in EAS 
reconstruction.

\begin{figure*}
\centering
\includegraphics[width=0.46\textwidth]{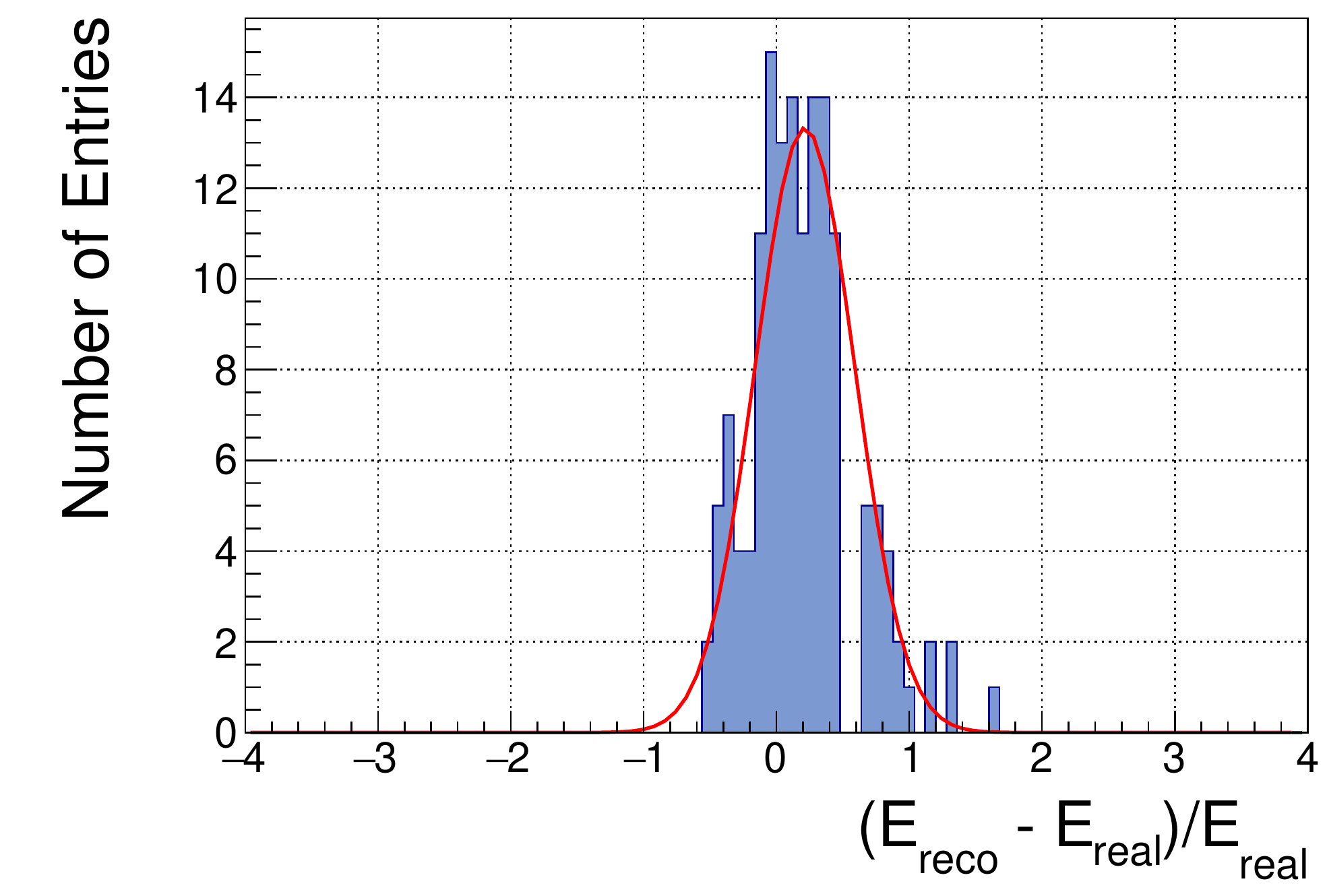}
\includegraphics[width=0.46\textwidth]{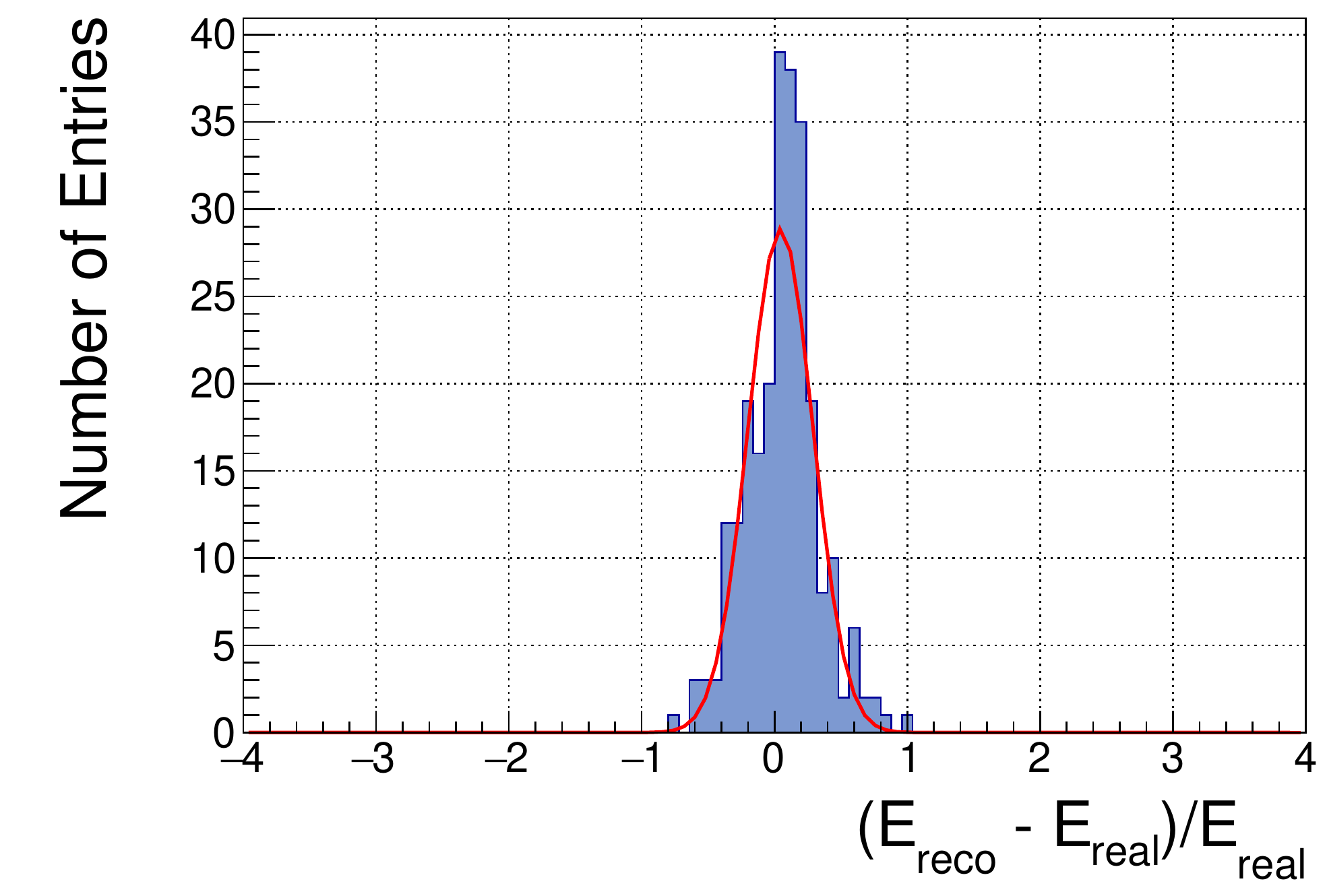}
\includegraphics[width=0.46\textwidth]{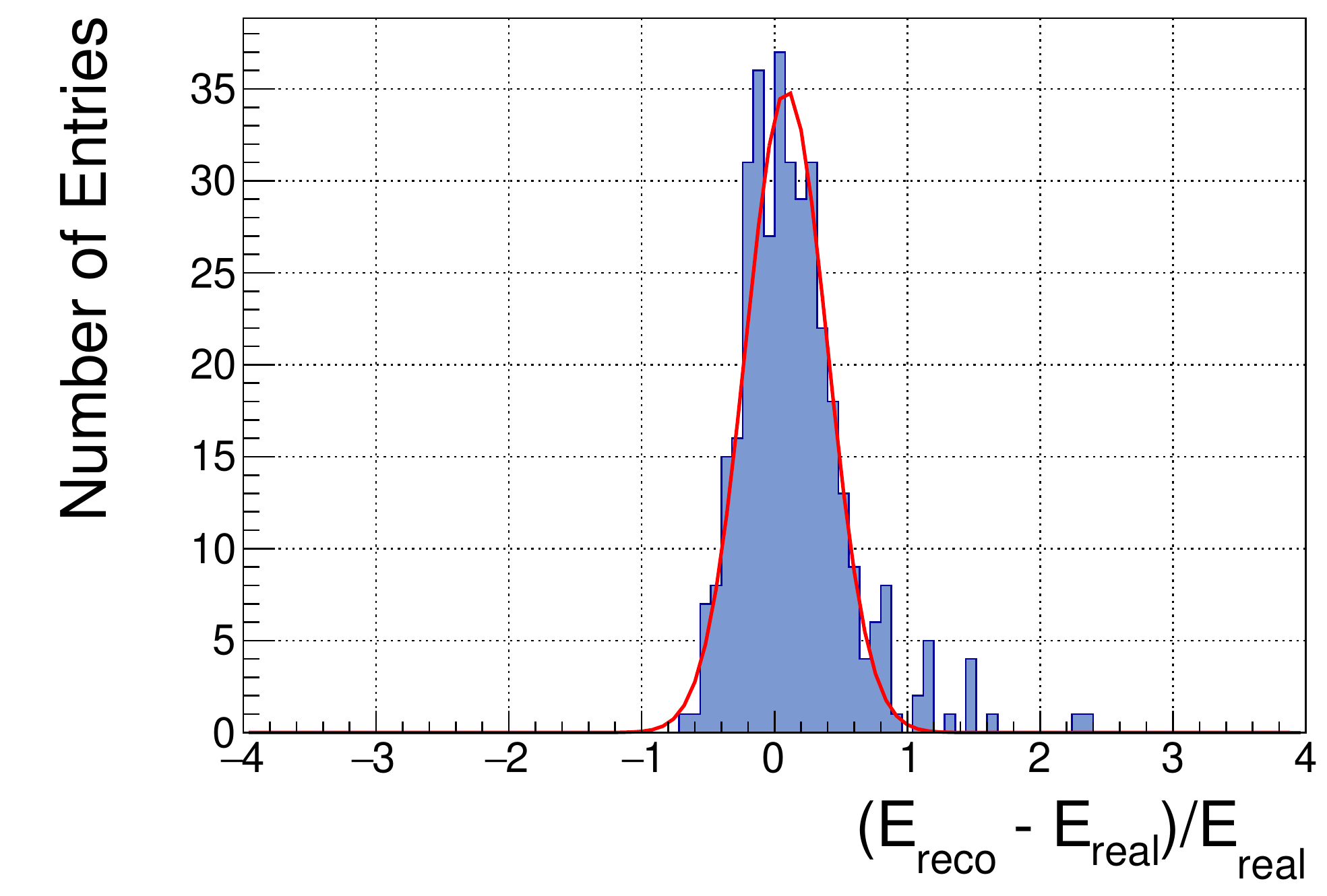}
\includegraphics[width=0.46\textwidth]{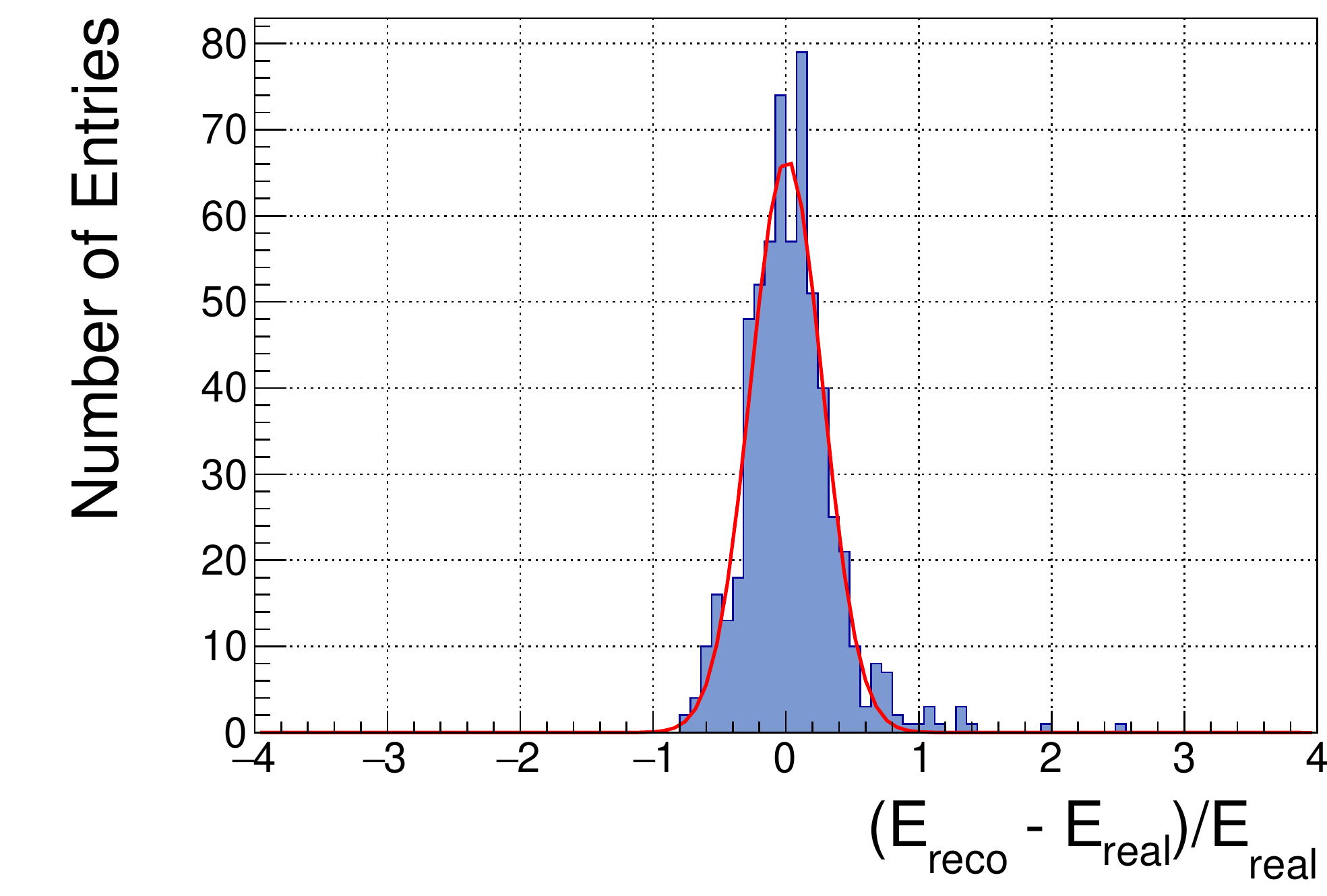}
\caption{Energy resolution $(E_\text{reco} - E_\text{real}) / E_\text{real}$ for events belonging to
two different energy intervals (low-energy on the left panels and high-energy on the right panels). The top plots refer to the performance of the Cherenkov method
while the bottom ones to the slant depth method.}
\label{fig:ene-reso}
\end{figure*}

In Fig.~\ref{fig:ene-reso} we show the relative energy resolution
$(E_\text{reco} - E_\text{real}) / E_\text{real}$ applying the two methods for two
different energy intervals. All zenith angles are included.
In Table~\ref{tabla:ene-reso} we describe the performance on $X_\text{max}$ and energy
reconstruction including all the reconstructed EASs using the two methods.
To test the validity of the automatic procedure, in parallel the same EASs have been
reconstructed manually. As one can easily verify in Table~\ref{tabla:ene-reso} the results of the automatic procedure are consistent with those obtained adopting the manual reconstruction. This indicates that the
automatic algorithm is quite effective on its own. However, the manual procedure allowed to
remove 4 events for which the result of the automatic procedure was providing a very bad
result. This might happen when the EAS goes through gaps between PDMs and the automatic
algorithm fails to reject those points for the fitting procedure.

\begin{table*}[tb]
    \centering
    \caption{$X_{\rm max}$ and energy resolutions (all EAS included) for the two methods adopting either an automatic or a manual reconstruction procedure.}
    \begin{tabular}{cc}
        \begin{tabular}{cc}
            \toprule
            \midrule
            \hline
            \hline
            \textcolor{red}{~~~~~~~~~~~~~~Lower $E$~~~~~~~~~~~~~~}  &        \\
            \hline
            \midrule
            Bias                               & 21\%   \\
            Resolution                         & 37\%   \\
            \hline
            \midrule
            \textcolor{red}{Higher $E$} &        \\
            \hline
            \midrule
            Bias                               & 4\%    \\
            Resolution                         & 20\%   \\
            \hline
            \midrule  
            \midrule
            \textcolor{blue}{Lower $E$}  &       \\
            \hline
            \midrule
            Bias                               & 9\%    \\
            Resolution                         & 30\%   \\
            \hline
            \midrule    
            \textcolor{blue}{Higher $E$ }&       \\
            \hline
            \midrule
            Bias                               & 0.5\%  \\
            Resolution                         & 27\%   \\
            \hline
            \hline
            \midrule  
            \bottomrule
        \end{tabular}
        &
        \begin{tabular}{cccc}
            \toprule
            \midrule
            \hline
            \hline
            \textcolor{red}{~~~~~~~~~~~~~~$X_{\rm max}$~(g/cm$^2$)  Cherenkov method}~~~~~~~~~~~~~~& \textcolor{red}{~~~~~~~~~~~~~~Bias~~~~~~~~~~~~~~} & \textcolor{red}{~~~~~~~~~~~~~~$\sigma$~~~~~~~~~~~~~~} \\
            \hline
            \midrule
            Automatic                          & $\phantom{-}12$            & 128               \\
            Manual                             & $-13$           & 107               \\
            \hline
            \midrule
            \textcolor{red}{$E$ (\%) Cherenkov method}           & \textcolor{red}{Bias} & \textcolor{red}{$\sigma$} \\
            \hline
            \midrule
            Automatic                          & $-10$           & 25                \\
            Manual                             & $-11$           & 25                \\
            \hline
            \midrule
            \midrule
            \textcolor{blue}{$X_{\rm max}$~(g/cm$^2$)  Cherenkov method}
             & \textcolor{blue}{Bias} & \textcolor{blue}{$\sigma$} \\
            \hline
            \midrule
            Automatic                          & 37            & 100               \\
            Manual                             & 34            & 110               \\
            \hline
            \midrule
            \textcolor{blue}{$E$ (\%) Slant depth method}           & \textcolor{blue}{Bias} & \textcolor{blue}{$\sigma$} \\
            \hline
            \midrule
            Automatic                          & 8             & 21                \\
            Manual                             & 11            & 21                \\
            \hline
            \hline
            \midrule
            \bottomrule
        \end{tabular}
    \end{tabular}
    \label{tabla:ene-reso}
\end{table*}

\subsection{Angular reconstruction}

\begin{figure}
\centering
\includegraphics[width=0.48\textwidth]{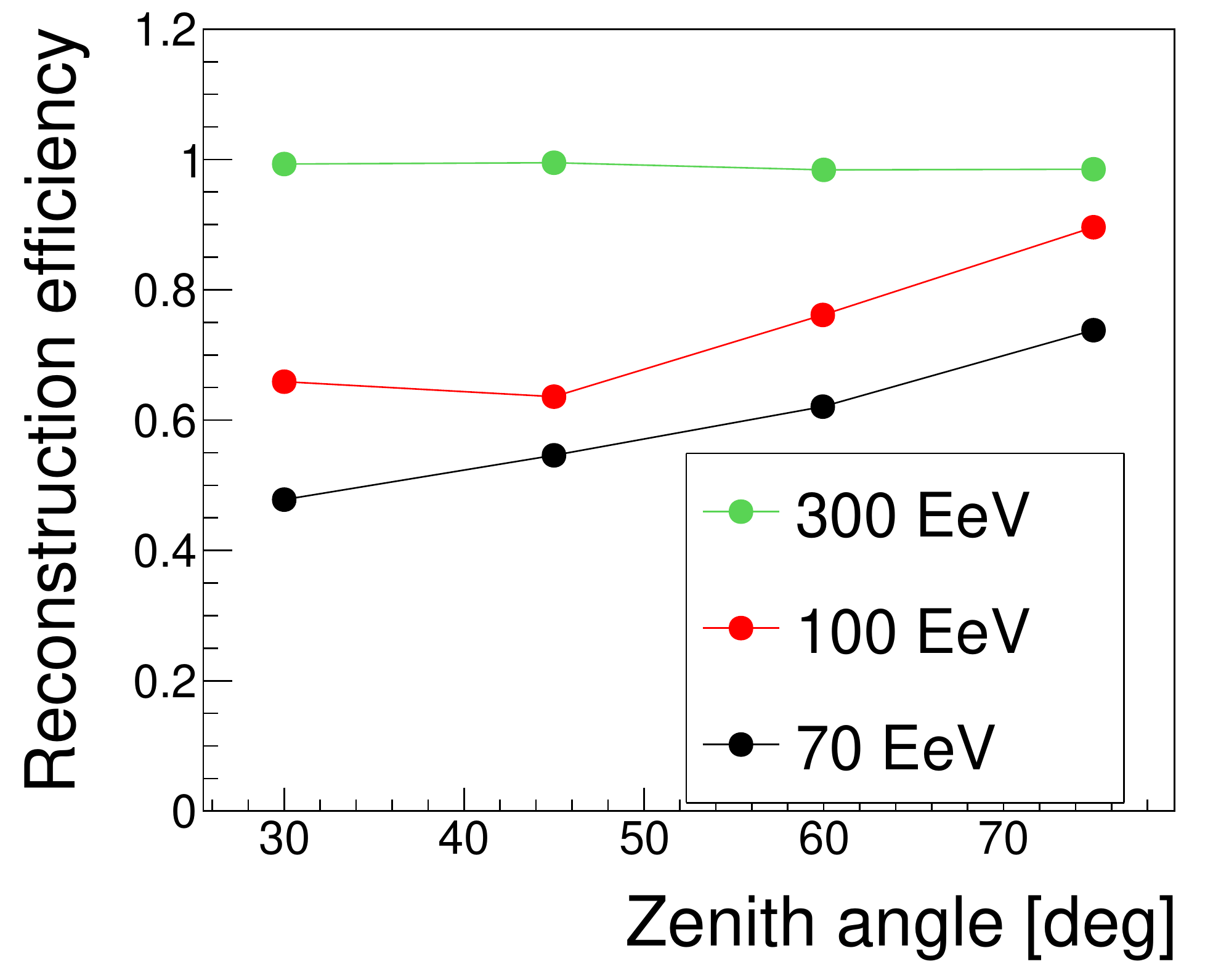}
\caption{Efficiency in the angular reconstruction for POEMMA as a function of zenith angle
for different EAS energies.}
\label{fig:eff_angular}
\end{figure}

\begin{figure*}
\centering
\includegraphics[width=0.45\textwidth]{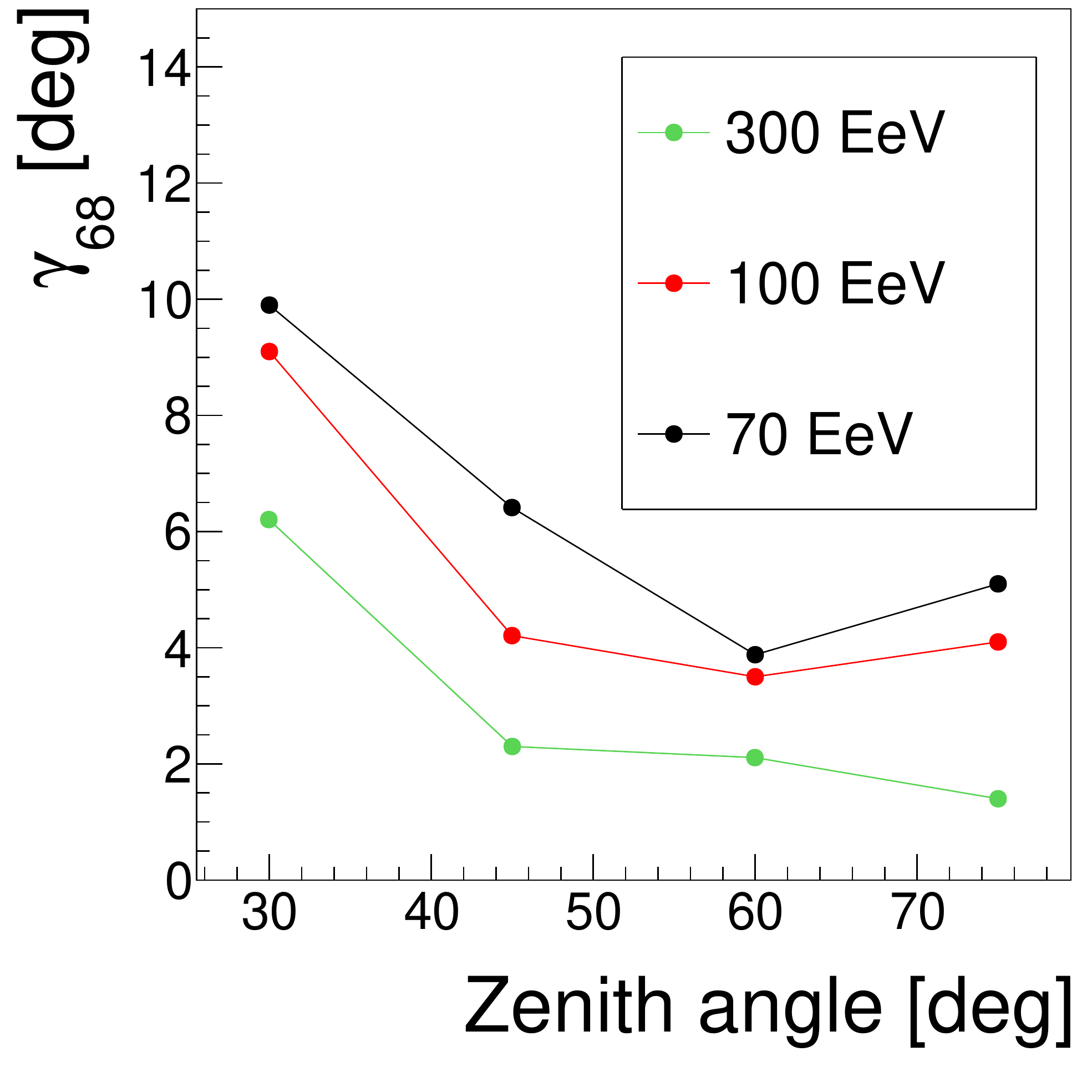}
\includegraphics[width=0.45\textwidth]{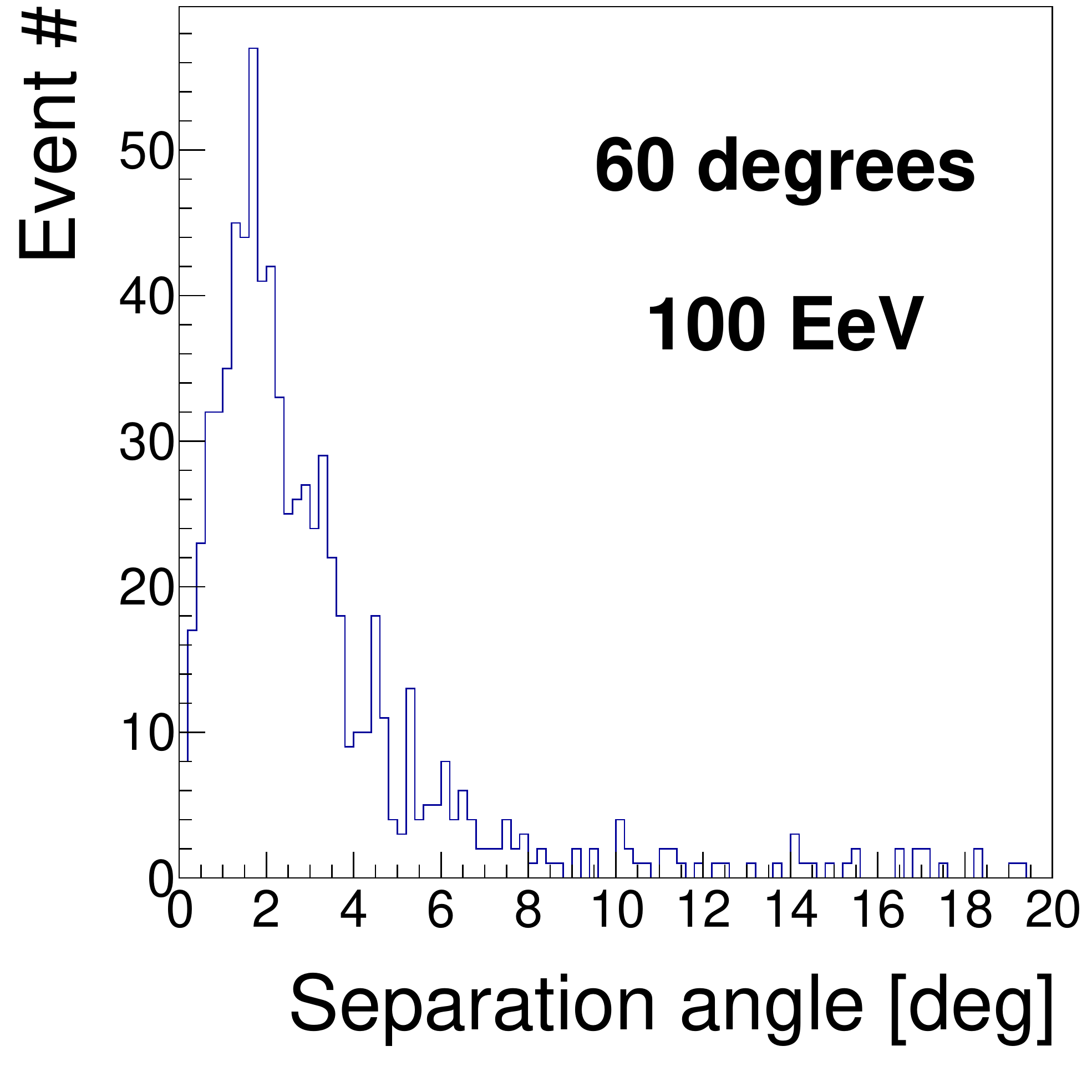}
\caption{Left: Angular resolution for POEMMA defined through the $\gamma_{68\%}$ parameter.
The circled point indicate that
a limited bias exist for these points, see table~\ref{tab:angular_reso} for details.
Right: Distribution of the $\gamma_{68\%}$ parameter for E = 10$^{20}$ eV and $\theta =
60^\circ$.}
\label{fig:angular_reso}
\end{figure*}

The angular reconstruction for POEMMA was evaluated at fixed zenith angles ($\theta$ =
30$^\circ$, 45$^\circ$, 60$^\circ$ and 75$^\circ$) for three different energies
(E = 7$\times$10$^{19}$ eV, 10$^{20}$ eV and 3$\times$10$^{20}$ eV). The same methodology
defined for JEM-EUSO reconstruction performance was applied to POEMMA, with a fine tuning of
the parameters of the {\it PWISE} algorithm.
Fig.~\ref{fig:eff_angular} shows the reconstruction efficiency as a function of zenith angle
for the three different energies. The reconstruction efficiency is defined as the ratio
$\epsilon_\text{reco} = \frac{N_\text{reco}}{N_\text{trig}}$.
About 1000 EAS are simulated in each condition (see table~\ref{tab:angular_reso} for details).
In order to define that the pattern recognition
is successful at least 10 pixels should have been selected in the track by the
{\it PWISE} algorithm.
The same pixel can be selected more than once in the track, i.e. in consecutive GTUs.
This selection criteria tends to make the angular reconstruction less efficient at lower
energies and for
lower zenith angles. This is an important parameter that needs to be explored with more details
in the future. Moreover, by reducing the GTU to 1 $\mu$s, the number of selected pixels might
increase considerably.
To estimate the expected angular resolution of POEMMA, as in JEM-EUSO, we compared the angle
($\gamma$) between the injected shower axis and the reconstructed one. We define 
$\gamma_{68\%}$ as the value at which the cumulative distribution of $\gamma$ reaches 0.68.
It is worth mentioning that both systematic errors and statistical fluctuations are included
within the definition of $\gamma_{68\%}$. We will use this parameter as a measurement of the
overall performance of the reconstruction capabilities. Fig.~\ref{fig:angular_reso} shows the
results in terms of $\gamma_{68\%}$ for different zenith angles and energies. Details can be found in table~\ref{tab:angular_reso}.
\begin{table*}
\caption{Angular performance for $\theta$, $\phi$ and $\gamma_{68\%}$. All numbers are
expressed in degrees. The number of triggered and reconstructed events and their ratio
are reported as well.}
\centering
\begin{tabular}{cccccccccc}
\hline
\hline
~~~~~E(eV)~~~~~&~~~~~$\theta_\text{sim}$~~~~~&~~~~~$N_\text{trig}$~~~~~&~~~~~$N_\text{reco}$~~~~~&~~~~~$\epsilon_{\rm reco}$~~~~~&~~~~~$<\Delta_{\theta}>$~~~~~&
~~~~~$\sigma(\Delta_{\theta})$~~~~~&~~~~~$<\Delta_{\phi}>$~~~~~&~~~~~$\sigma(\Delta_{\phi})$~~~~~&~~~~~$\gamma_\text{68\%}$~~~~~\\
\hline
7$\times$10$^{19}$ & 30 & 697 & 333 & 0.47 & 2.5 & 2.9 & 0.3 & 7.4 & 9.9\\
                   & 45 & 780 & 426 & 0.54 &  1.0 & 2.0 &  0.1 & 3.3 & 6.4\\
                   & 60 & 910 & 563 & 0.61 & -0.5 & 2.2 &  0.1 & 2.0 & 3.9\\
                   & 75 &1063 & 785 & 0.73 & -0.7 & 2.6 & -0.1 & 1.1 & 5.1\\
\hline
         10$^{20}$ & 30 & 948 & 625 & 0.65 &  2.8 & 3.0 & -4.1 & 6.6 & 9.1\\
                   & 45 &1022 & 651 & 0.63 &  0.4 & 2.1 & -0.1 & 2.7 & 4.2\\
                   & 60 &1062 & 808 & 0.76 & -0.5 & 2.3 &  0.3 & 2.0 & 3.5\\
                   & 75 &1127 &1010 & 0.89 & -1.0 & 1.6 &  0.0 & 0.9 & 4.1\\
\hline
3$\times$10$^{20}$ & 30 &1101 &1094 & 0.99 &  4.7 & 2.1 &  0.0 & 3.8 & 6.2\\
                   & 45 &1119 &1114 & 0.99 & -0.9 & 1.5 &  0.0 & 1.3 & 2.3\\
                   & 60 &1128 &1111 & 0.98 & -1.0 & 1.0 &  0.0 & 0.7 & 2.1\\
                   & 75 &1136 &1119 & 0.98 & -0.5 & 0.8 &  0.0 & 0.4 & 1.4\\
\hline
\hline
\end{tabular}
\label{tab:angular_reso}
\end{table*}
It is important to underline that a more detailed study of the bias should be 
performed. The reduction of
the bias would be capable of improving the overall performance of $\gamma_{68\%}$ as it
includes both statistical and systematic uncertainties. The reconstruction
performance will improve also by reducing the GTU to 1 $\mu$s.

\subsection{ESAF estimate of stereo energy resolution}

\begin{figure*}[tbp]
\centering
\includegraphics[width=0.98\columnwidth]{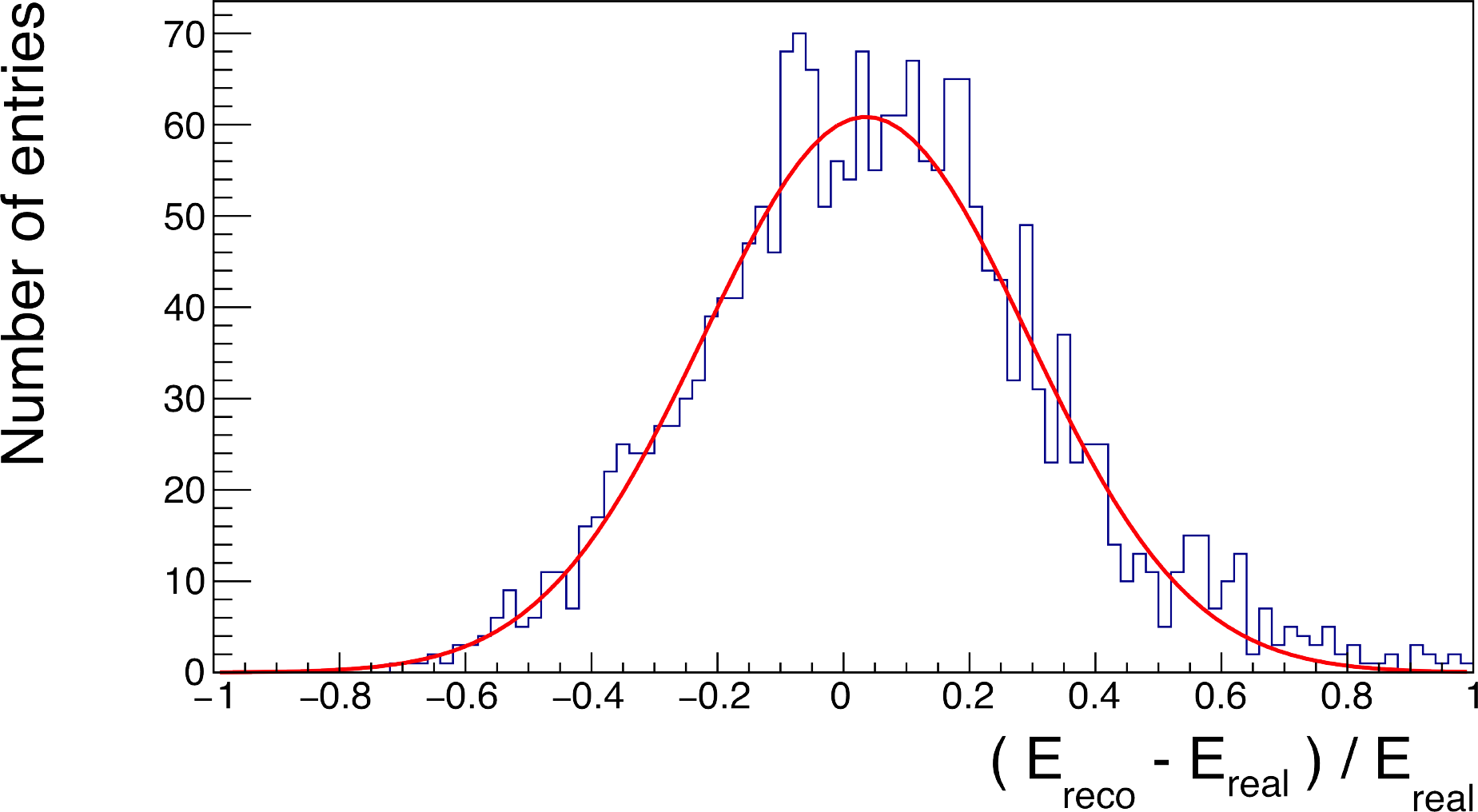}
\includegraphics[width=0.98\columnwidth]{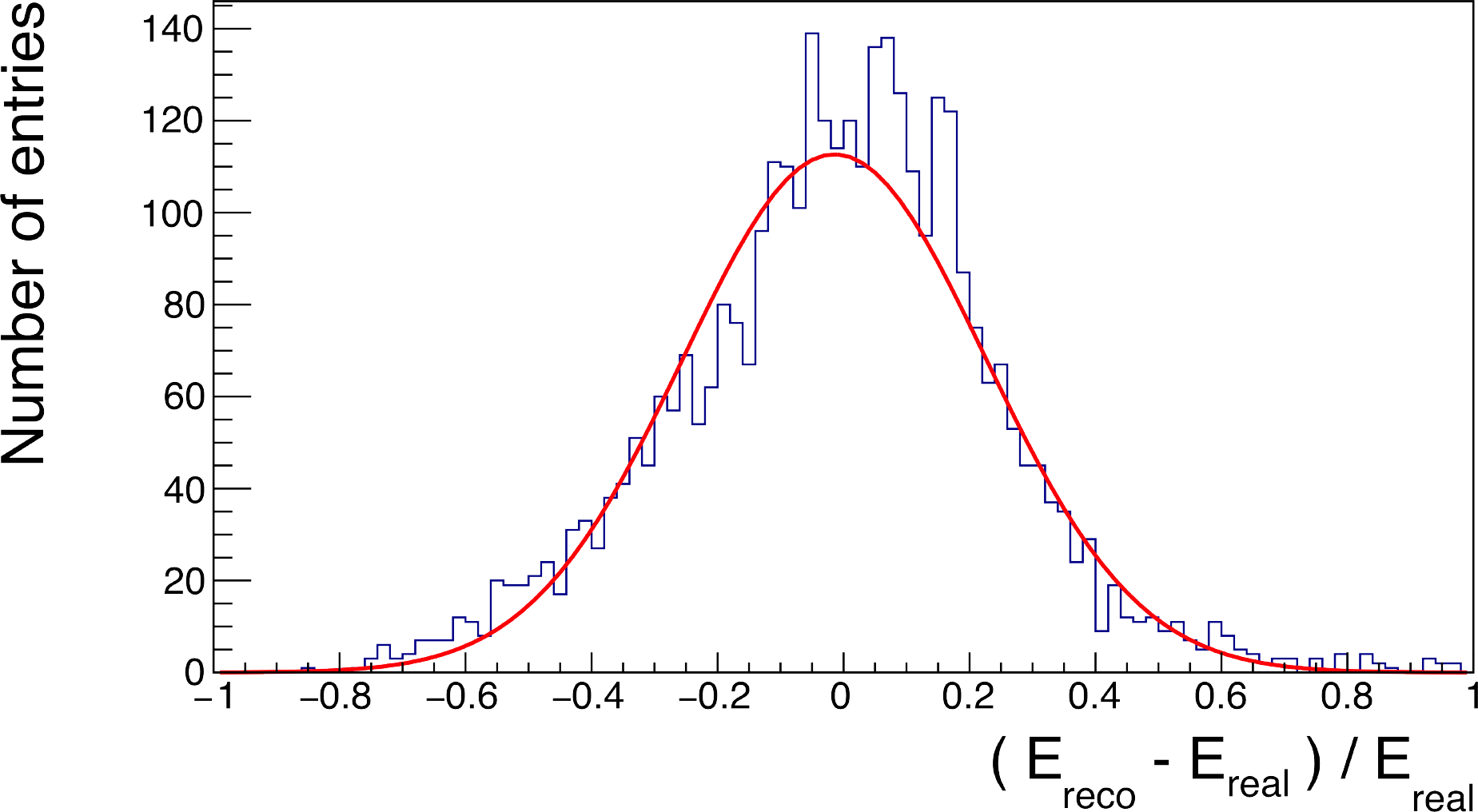}
    \caption{ESAF simulated energy resolution assuming 1$^\circ$ zenith and azimuth angular resolution. Left: 50 EeV results, right: 100 EeV results. \label{ESAF1degERes}}
\end{figure*}

The ESAF simulation was used to estimate the stereo reconstructed energy resolution via imposing an angular resolution defined by Gaussian distributions with $\sigma = 1^\circ$ for the zenith and azimuth angles. Isotropic UHECRs were generated from $0^\circ - 90^\circ$ zenith angles at two energies, 50 EeV and 100 EeV, assuming nadir pointing of a single POEMMA telescope. The events were then reconstructed assuming $1^\circ$ resolutions.  The results for the energy reconstruction are shown in Fig.~\ref{ESAF1degERes}. For 50 EeV, the energy resolution is 26\% with a $+3.5\%$ bias, while for 100 EeV the energy resolution is 24\% with a $-1.5\%$ bias.  Since the two POEMMA telescopes provide independent measurements of each EAS, the combined resolution is obtained by dividing by $\sqrt{2}$, yielding 18\% at 50 EeV and 17.0\% at 100 EeV.

\subsection{Summary}

A first estimation of POEMMA performance in terms of trigger exposure and 
quality of event
reconstruction has been assessed using the ESAF code for the clear atmosphere 
case. The same trigger algorithms and
reconstruction procedures developed for JEM-EUSO have been applied to POEMMA. A parametrized
optics has been adopted.
Results of the EAS reconstruction refer to the nadir configuration, while the trigger exposure
has been obtained also for tilted configurations.
It is important to underline that a 2.5$\mu$s GTU has been used in this study. 
POEMMA is expected to adopt a shorter GTU of 1$\mu$s, therefore, the quality 
in trigger 
reconstruction and in the trigger efficiency will improve. 
Therefore, the results obtained in this study have to be considered as conservative ones.
It is worth mentioning finally that $\sim88\%$ of the triggered events were 
successfully
reconstructed for the energy parameter above $log(E/eV)=19.6$. This is important because it
indicates that the exposure curve provides already a reliable information on 
the number of events that could be retained for further analysis.

The angular reconstruction still needs improvements in terms of reconstruction 
efficiency.
Most probably a too severe threshold on the number of points has been applied. 
Nevertheless,
because the energy reconstruction provides also an angular reconstruction, in 
case of a failing
of the {\it PWISE algorithm}, the information of the angular resolution could 
be retained from the chain employed for the energy reconstruction.

\section{Methodology for Anisotropy Searches Cross-Correlating with Astrophysical Catalogs}
\label{appendix:d}

\begin{figure*}[tb]
\includegraphics[trim = 40mm 38mm 17mm 40mm, clip, width=0.49\textwidth]{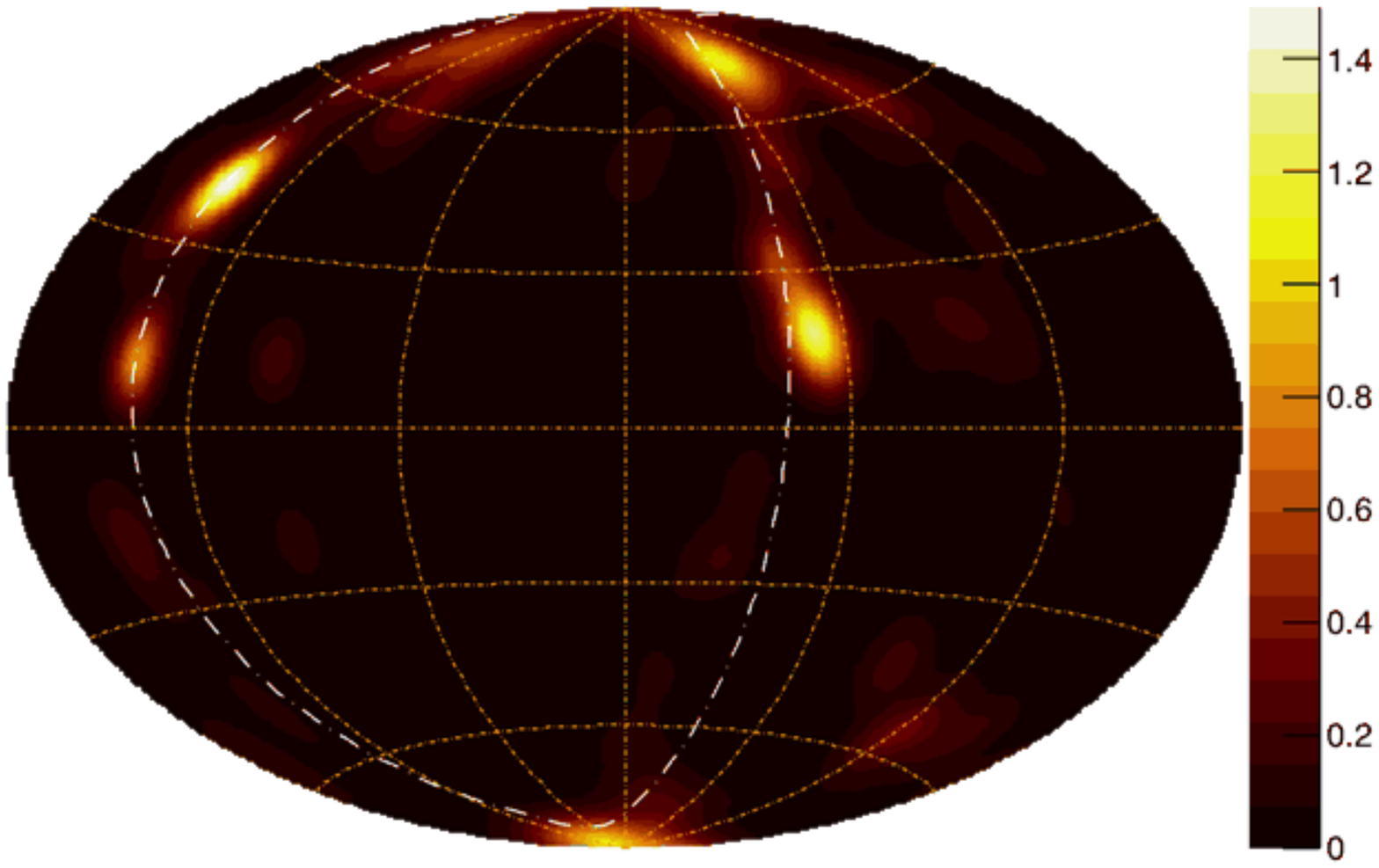}\includegraphics[trim = 40mm 38mm 17mm 40mm, clip, width=0.49\textwidth]{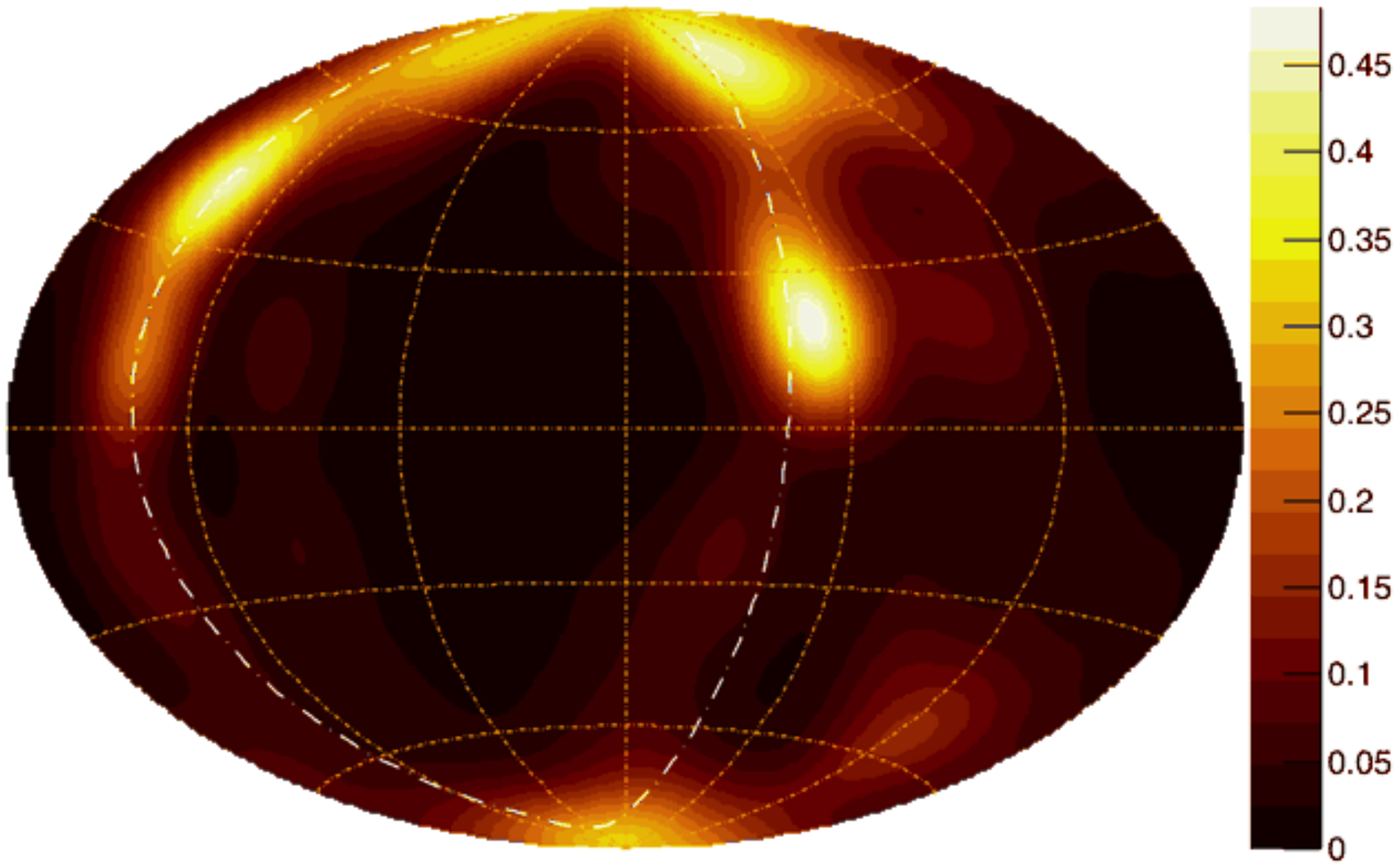}
\includegraphics[trim = 40mm 38mm 17mm 40mm, clip, width=0.49\textwidth]{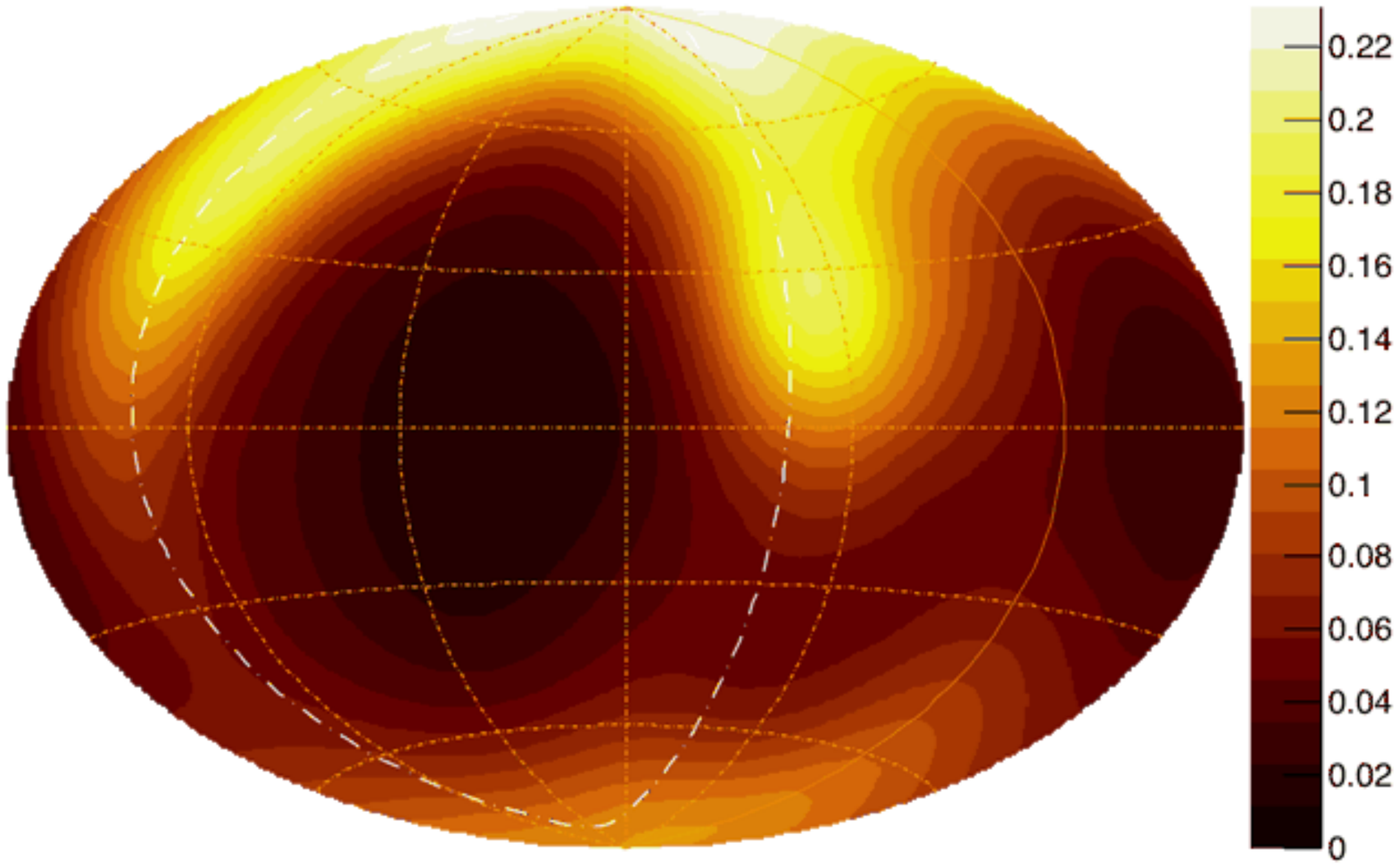}
\includegraphics[trim = 40mm 38mm 17mm 40mm, clip, width=0.49\textwidth]{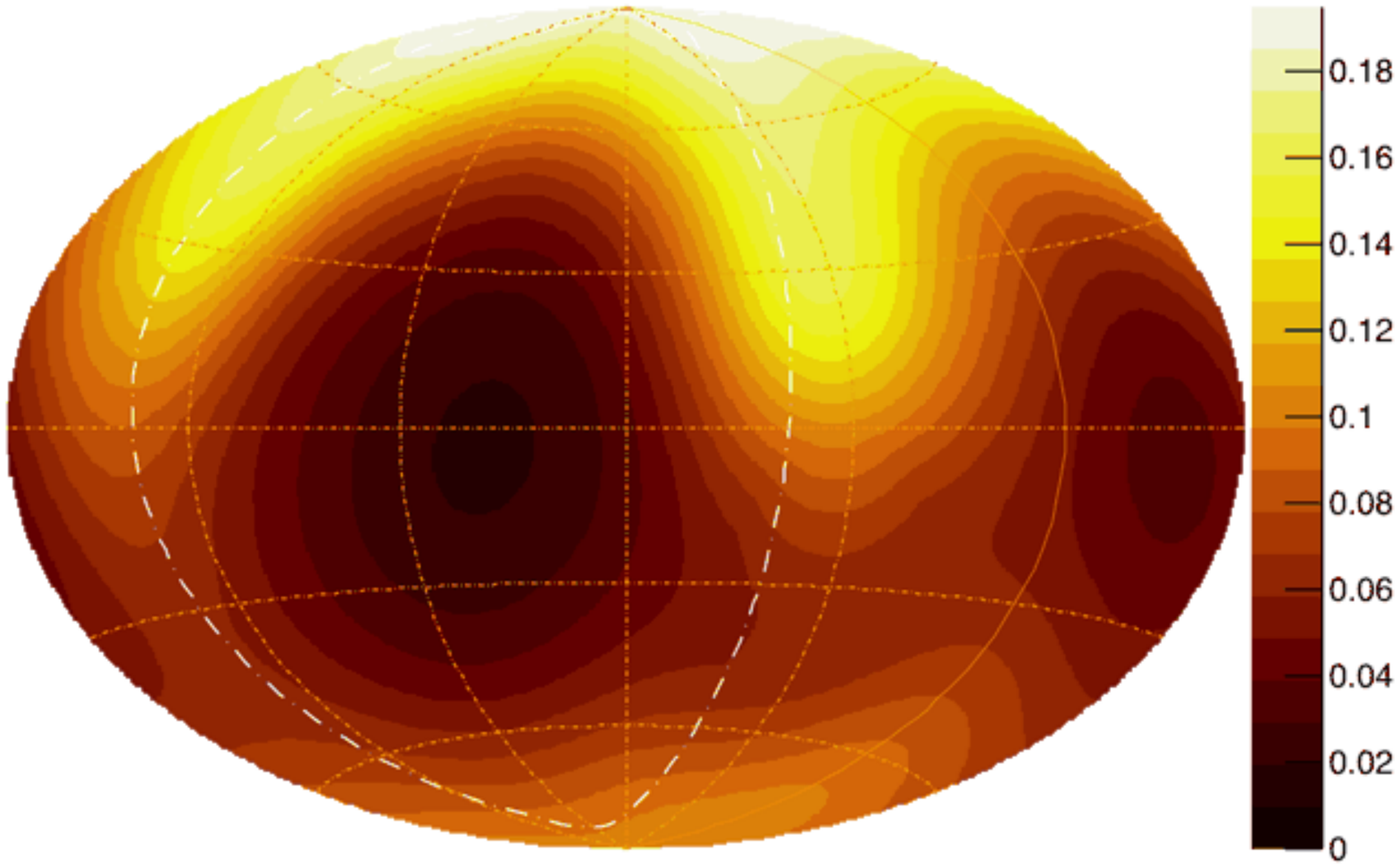}
\caption{Normalized source sky maps for nearby galaxies from the 2MRS catalog~\cite{2012ApJS..199...26H} weighted by K-band flux corrected for Galactic extinction, the attenuation factor accounting for energy losses incurred by UHECRs through propagation, and the exposure of POEMMA. The maps have been smoothed using a von Miser-Fisher distribution with concentration parameters corresponding to angular spreads of $\Theta = 5^{\circ}$ (top left), $10^{\circ}$ (top right), $20^{\circ}$ (bottom left), and $25^{\circ}$ (bottom right). The color scales indicate $\mathcal{F}_{\rm src}$ as a function of position on the sky. The white dot-dashed line indicates the Supergalactic Plane. \label{fig:2mrs_skymaps}}
\end{figure*}

\begin{figure*}[tb]
\includegraphics[trim = 40mm 38mm 17mm 40mm, clip, width=0.49\textwidth]{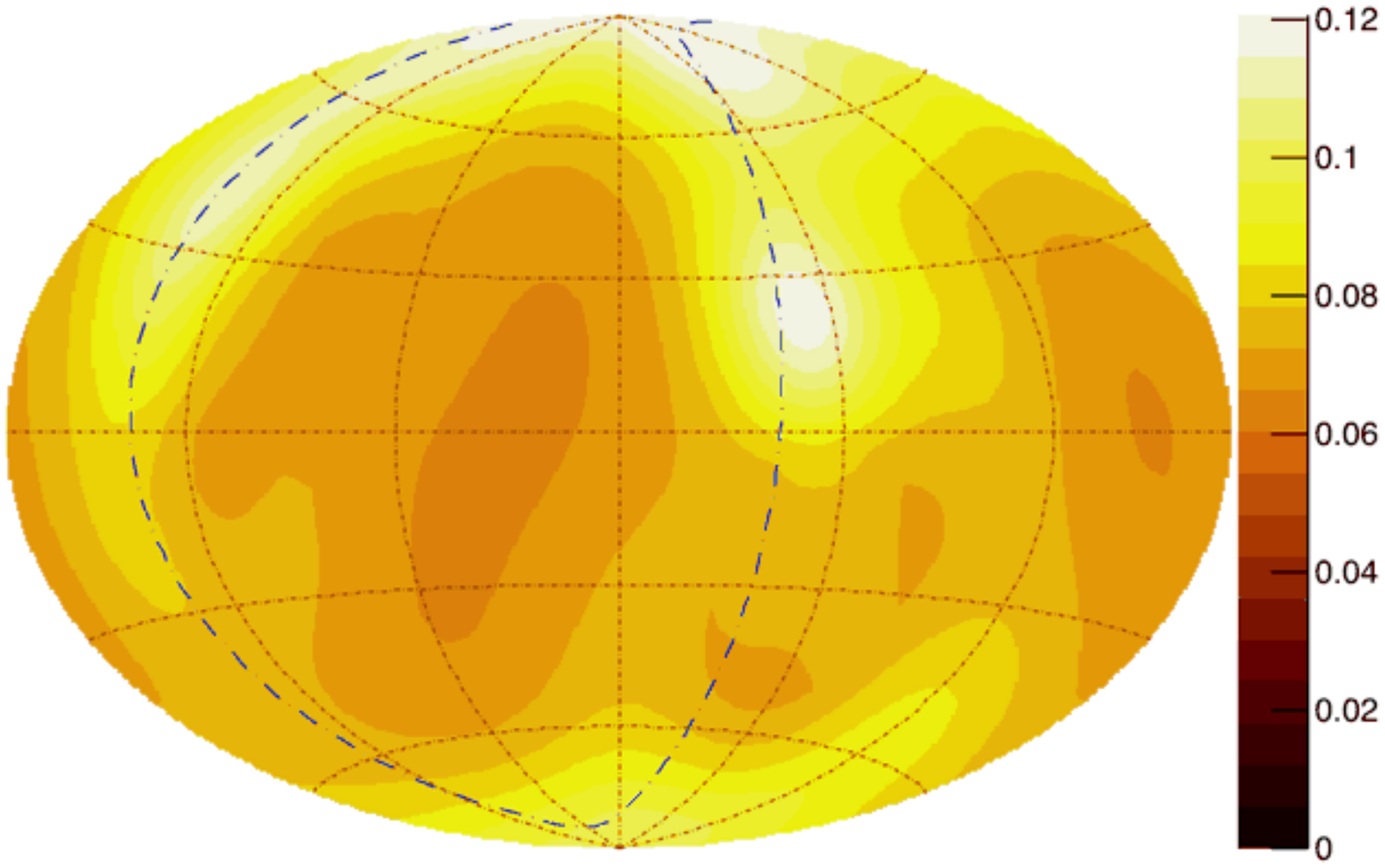}\includegraphics[trim = 40mm 38mm 17mm 40mm, clip, width=0.49\textwidth]{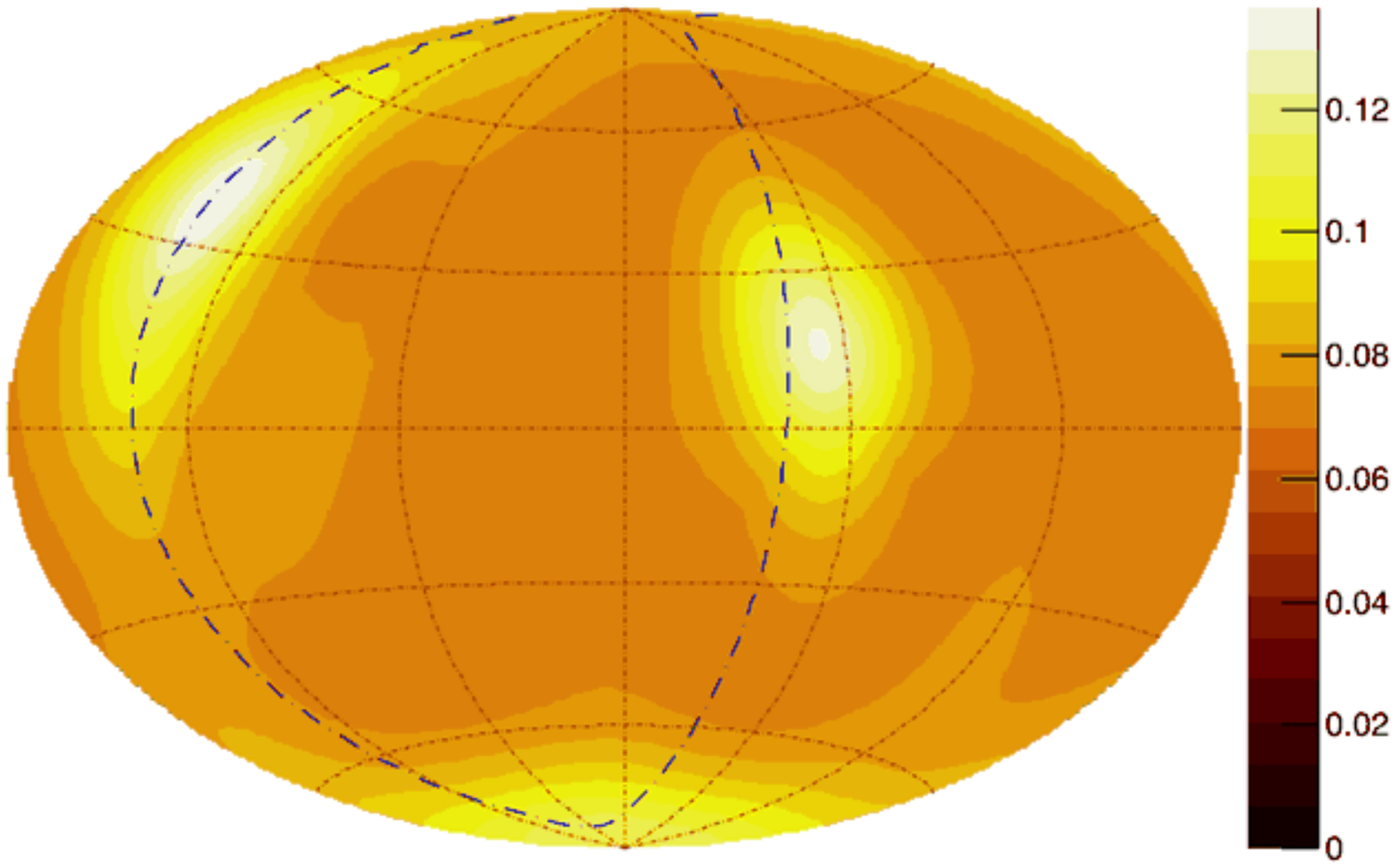}
\includegraphics[trim = 40mm 38mm 17mm 40mm, clip, width=0.49\textwidth]{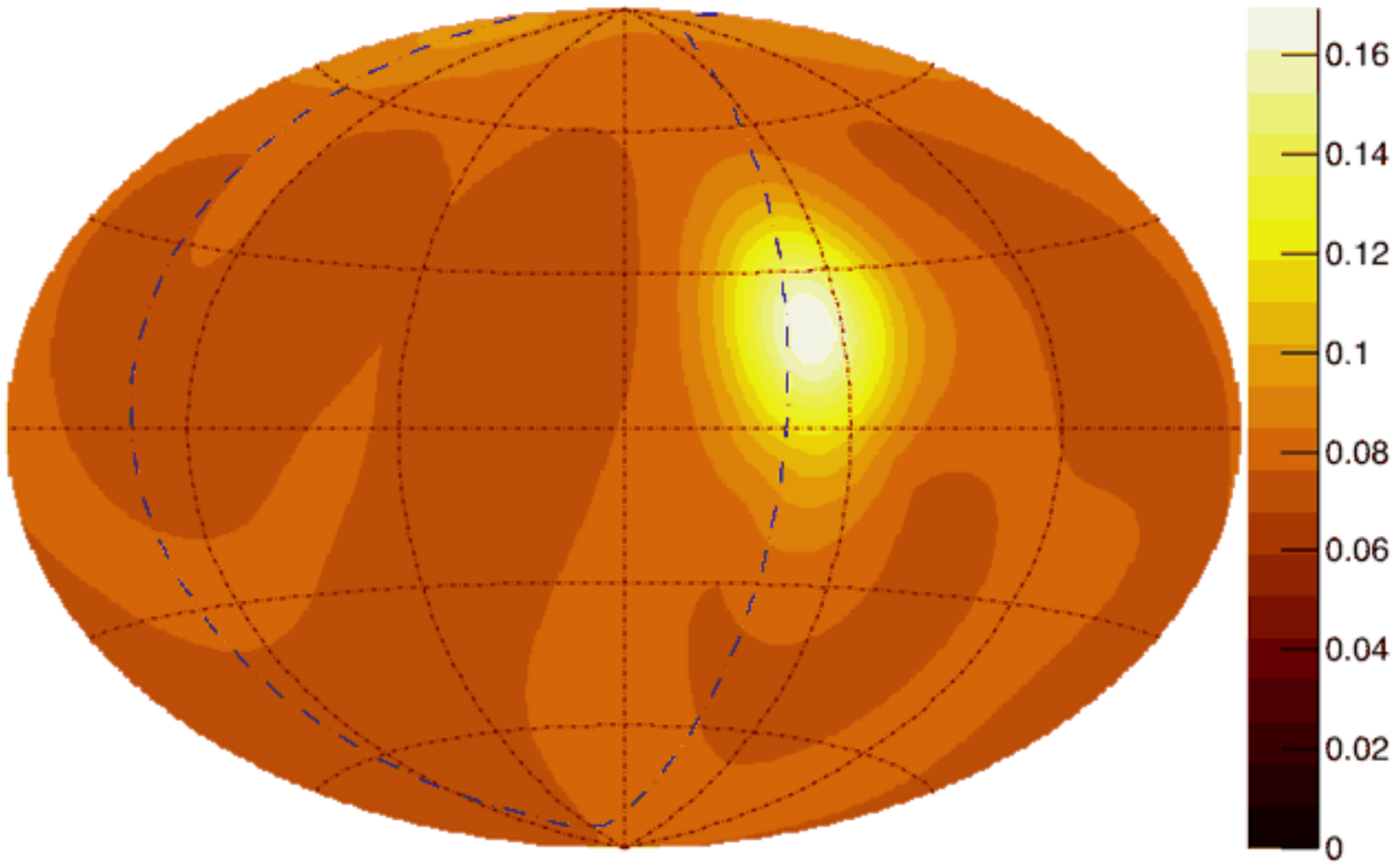}
\caption{Probability density sky maps corresponding to the select astrophysical scenarios with parameters selected based on the best-fit parameters from Ref.~\cite{Aab:2019ogu}. \textit{Top left:} 2MRS galaxies with $f_{\rm sig} = 19$\% and $\Theta = 15^{\circ}$. \textit{Top right:} Starburst galaxies with $f_{\rm sig} = 11$\% and $\Theta = 15^{\circ}$. \textit{Bottom:} \textit{Swift}-BAT AGNs with $f_{\rm sig} = 8$\% and $\Theta = 15^{\circ}$. The color scales indicate $\mathcal{F}_{\rm sky}$ as a function of position on the sky. The blue dot-dashed line indicates the Supergalactic Plane. \label{fig:alt_hypothesis}}
\end{figure*}

In order to determine the reach of POEMMA in detecting anisotropy through cross-correlating with astrophysical catalogs, we perform a detailed statistical study simulating POEMMA datasets assuming either the starburst or AGN hypothesis and comparing them against the null hypothesis of isotropy. In so doing, we follow the likelihood ratio approach of Abbasi et al.~[\citenum{Abbasi:2018tqo}; see also, \citenum{2010APh....34..314A,Aab:2018chp,Anchordoqui:2018qom}]. For each constructed mock dataset (see Sec.~\ref{subsubsec:tsanalysis}), we express the astrophysical hypotheses as sky models in which the flux of UHECRs consists of either of a purely isotropic distribution or an isotropic component with an anisotropic component arising from astrophysical sources:
\begin{equation}
    \mathcal{F}_{\rm sky}\left(\hat{n}\right) = \frac{\omega\left(\hat{n}\right)}{\mathcal{C}}\left[\left(1 - f_{\rm sig}\right)\frac{1}{4\pi} + f_{\rm sig}\mathcal{F}_{\rm src}\left(\hat{n}\right)\right]\,,
\end{equation}
where $\mathcal{F}_{\rm sky}\left(\hat{n}\right)$ is the probability density sky map, $\hat{n}$ is the unit vector for a given location on the sky, $\omega\left(\hat{n}\right)$ is the exposure in the direction of $\hat{n}$, $f_{\rm sig}$ is the signal fraction (the fraction of UHECRs originating from the sources), $\mathcal{F}_{\rm src}$ is the normalized source sky map for the flux of UHECRs originating from the sources, and $\mathcal{C}$ is a normalization factor to ensure that $\int \mathcal{F}_{\rm sky}\left(\hat{n}\right) d\Omega$. We construct $\mathcal{F}_{\rm src}$ as the weighted sum of the UHECR flux from sources of the source class under consideration, assuming that the flux of UHECRs from each individual source is proportional to its electromagnetic flux. Each source is weighted by a von Mises-Fisher distribution\footnote{The equivalent of a Gaussian distribution on the surface of a sphere. For the 2-sphere, it is given by $\mathcal{G}\left(\hat{n},\hat{s};\kappa\right)=\kappa \exp\left(\hat{n}\cdot\hat{s}\right)/\left(4\pi\sinh{\kappa}\right)$, where $\kappa = \Theta^{-2}$ is the concentration parameter.} with angular spread $\Theta$ and an attenuation factor that accounts for UHECR energy losses through propagation. We then construct normalized source sky maps from sources in one of the flux-limited catalogs used for this study (see Sec.~\ref{subsubsec:tsanalysis}). In Fig.~\ref{fig:2mrs_skymaps}, we provide examples of normalized source sky maps ($\mathcal{F}_{\rm src}\left(\hat{n}\right)$) constructed from nearby galaxies in the 2MRS catalog for various values for the concentration parameter. In Fig.~\ref{fig:alt_hypothesis}, we provide examples of probability density sky maps ($\mathcal{F}_{\rm sky}\left(\hat{n}\right)$) corresponding to select astrophysical scenarios in which the flux of UHECRs consists of an isotropic component and an anisotropic component as defined above. Parameters for the displayed scenarios are chosen to coincide with the best-fit search parameters reported by Ref.~\cite{Aab:2019ogu} for the given astrophysical source catalogs.

The null and alternative hypotheses are tested using mock datasets of a given number of events drawn from the astrophysical source sky maps assuming a given signal fraction and angular spread. For each mock dataset, we calculate the TS as a function of the search parameters. The TS value is computed as the likelihood ratio for the modeled sky maps for the null and alternative hypotheses:
\begin{equation}
    {\rm TS} = 2\ln\left(\frac{L\left(\mathcal{F}_{\rm sky}\right)}{L\left(\mathcal{F}_{\rm iso}\right)}\right)\,,
\end{equation}
where $\mathcal{F}_{\rm iso} = \omega\left(\hat{n}\right)/4\pi$ and the likelihood is given by
\begin{equation}
    L\left(\mathcal{F}\right) = \frac{1}{\mathcal{N}}\prod_{i}\mathcal{F}\left(\hat{n}_{i}\right)\,.
\end{equation}

\begin{table*}
\caption{TS values for various astrophysical scenarios.}
\centering
\begin{tabular}{cc}
\hline
    \begin{tabular}{|C{1.0in}|C{0.5in}C{0.5in}C{0.5in}C{0.5in}|}
    \toprule
    \midrule
    \hline
     Catalog & $f_{\rm sig}$ & $\Theta$ & TS & $\sigma$\\
     \hline
    \multirow{10}{1.0in}{\centering SBG} & $5$\% & $5^{\circ}$ & 33.6 & 5.4 \\
    & $5$\% & $10^{\circ}$ & 12.8 & 3.1 \\
    & $5$\% & $15^{\circ}$ & 6.2 & 2.0 \\
    & $5$\% & $20^{\circ}$ & 3.4 & 1.3 \\
    & $5$\% & $25^{\circ}$ & 1.7 & 0.8 \\
    \cline{2-5}
    & $10$\% & $5^{\circ}$ & 115.4 & 10.5 \\
    & $10$\% & $10^{\circ}$ & 48.1 & 6.6 \\
    & $10$\% & $15^{\circ}$ & 24.7 & 4.6 \\
    & $10$\% & $20^{\circ}$ & 13.0 & 3.2 \\
    & $10$\% & $25^{\circ}$ & 7.0 & 2.2 \\
    \hline
    \multirow{10}{1.0in}{\centering 2MRS} & $5$\% & $5^{\circ}$ & 7.7 & 2.3 \\
    & $5$\% & $10^{\circ}$ & 3.7 & 1.4 \\
    & $5$\% & $15^{\circ}$ & 2.4 & 1.0 \\
    & $5$\% & $20^{\circ}$ & 1.5 & 0.7 \\
    & $5$\% & $25^{\circ}$ & 0.9 & 0.5 \\
    \cline{2-5}
    & $10$\% & $5^{\circ}$ & 29.8 & 5.1 \\
    & $10$\% & $10^{\circ}$ & 14.2 & 3.4 \\
    & $10$\% & $15^{\circ}$ & 8.7 & 2.5 \\
    & $10$\% & $20^{\circ}$ & 5.9 & 1.9 \\
    & $10$\% & $25^{\circ}$ & 3.9 & 1.5 \\
    \hline
    \multirow{10}{1.0in}{\centering \textit{Swift}-BAT AGN} & $5$\% & $5^{\circ}$ & 39.9 & 6.0 \\
    & $5$\% & $10^{\circ}$ & 19.1 & 4.0 \\
    & $5$\% & $15^{\circ}$ & 10.4 & 2.8 \\
    & $5$\% & $20^{\circ}$ & 6.7 & 2.1 \\
    & $5$\% & $25^{\circ}$ & 4.2 & 1.5 \\
    \cline{2-5}
    & $10$\% & $5^{\circ}$ & 129.9 & 11.2 \\
    & $10$\% & $10^{\circ}$ & 67.2 & 7.9 \\
    & $10$\% & $15^{\circ}$ & 39.6 & 6.0 \\
    & $10$\% & $20^{\circ}$ & 24.9 & 4.6 \\
    & $10$\% & $25^{\circ}$ & 16.1 & 3.6 \\
    \hline
    \midrule
    \bottomrule
    \end{tabular}
     & 
    \begin{tabular}{|C{0.5in}C{0.5in}C{0.5in}C{0.5in}|}
    \toprule
    \midrule
    \hline
    $f_{\rm sig}$ & $\Theta$ & TS & $\sigma$\\
    \hline
    $15$\% & $5^{\circ}$ & 228.1 & 14.9 \\
    $15$\% & $10^{\circ}$ & 101.9 & 9.8 \\
    $15$\% & $15^{\circ}$ & 54.2 & 7.1 \\
    $15$\% & $20^{\circ}$ & 29.5 & 5.1 \\
    $15$\% & $25^{\circ}$ & 16.1 & 3.6 \\
    \hline
    $20$\% & $5^{\circ}$ & 369.8 & 19.1 \\
    $20$\% & $10^{\circ}$ & 170.9 & 12.9 \\
    $20$\% & $15^{\circ}$ & 92.9 & 9.4 \\
    $20$\% & $20^{\circ}$ & 51.4 & 6.9 \\
    $20$\% & $25^{\circ}$ & 27.9 & 4.9 \\
    \hline 
    $15$\% & $5^{\circ}$ & 63.2 & 7.7 \\
    $15$\% & $10^{\circ}$ & 31.4 & 5.2 \\
    $15$\% & $15^{\circ}$ & 20.0 & 4.1 \\
    $15$\% & $20^{\circ}$ & 13.6 & 3.3 \\
    $15$\% & $25^{\circ}$ & 8.9 & 2.5 \\
    \hline
    $20$\% & $5^{\circ}$ & 105.7 & 10.0 \\
    $20$\% & $10^{\circ}$ & 55.2 & 7.1 \\
    $20$\% & $15^{\circ}$ & 35.2 & 5.6 \\
    $20$\% & $20^{\circ}$ & 23.5 & 4.5 \\
    $20$\% & $25^{\circ}$ & 15.1 & 3.5 \\
    \hline
    $15$\% & $5^{\circ}$ & 248.5 & 15.6 \\
    $15$\% & $10^{\circ}$ & 137.7 & 11.5 \\
    $15$\% & $15^{\circ}$ & 82.4 & 8.8 \\
    $15$\% & $20^{\circ}$ & 53.2 & 7.0 \\
    $15$\% & $25^{\circ}$ & 36.4 & 5.7 \\
    \hline
    $20$\% & $5^{\circ}$ & 392.3 & 19.6 \\
    $20$\% & $10^{\circ}$ & 224.4 & 14.8 \\
    $20$\% & $15^{\circ}$ & 139.3 & 11.6 \\
    $20$\% & $20^{\circ}$ & 91.5 & 9.3 \\
    $20$\% & $25^{\circ}$ & 62.4 & 7.6 \\
    \hline
    \midrule
    \bottomrule 
    \end{tabular}\\
\hline
\end{tabular}{}
\end{table*}

\section{Determination of the $p$-air interaction length}
\label{appendix:b}

To assess POEMMA's ability to measure $\sigma_{p-{\rm air}}$, we use
the method of fitting the shape of the large $X_{\rm max}$ tail of
${\rm d}N/{\rm d}X_{\rm max}$ with an exponential
$\exp(-X_{\rm max}/\Lambda_{\eta})$ \cite{Collaboration:2012wt}, where $\Lambda_{\eta}$ is
proportional to the $p$-air interaction length which is, in turn,
proportional to the inverse of the proton-air cross section. Here, $\eta$ refers to
the fraction of the showers included in the tail, predominantly proton induced, as discussed in Sec.  \ref{sec:sigma}.

The value of $\eta$ in our analysis is translated to a starting value of $\xmax$, $\xmax^{\rm start}$.
The attenuation parameter $\Lambda_\eta$ is determined from the unbinned events, including smearing to account for the detector. With a large enough number of events, one can find the extremum of log-likelihood (see, e.g., Ref. \cite{Ulrich:2015yoo}), but for POEMMA's total $\sim 1,400$ cosmic ray events above 40 EeV, a reasonable approximation is
\begin{equation} 
\Lambda_\eta^{\rm opt} = \sum_{i=1,N_{\rm evts}}(X_{{\rm max},i}-\xmax^{\rm start})/N_{\rm evts}\ .
\label{eq:lopt}
\end{equation}

\begin{figure*}[tb]
\centering
\includegraphics[width=0.49\linewidth]{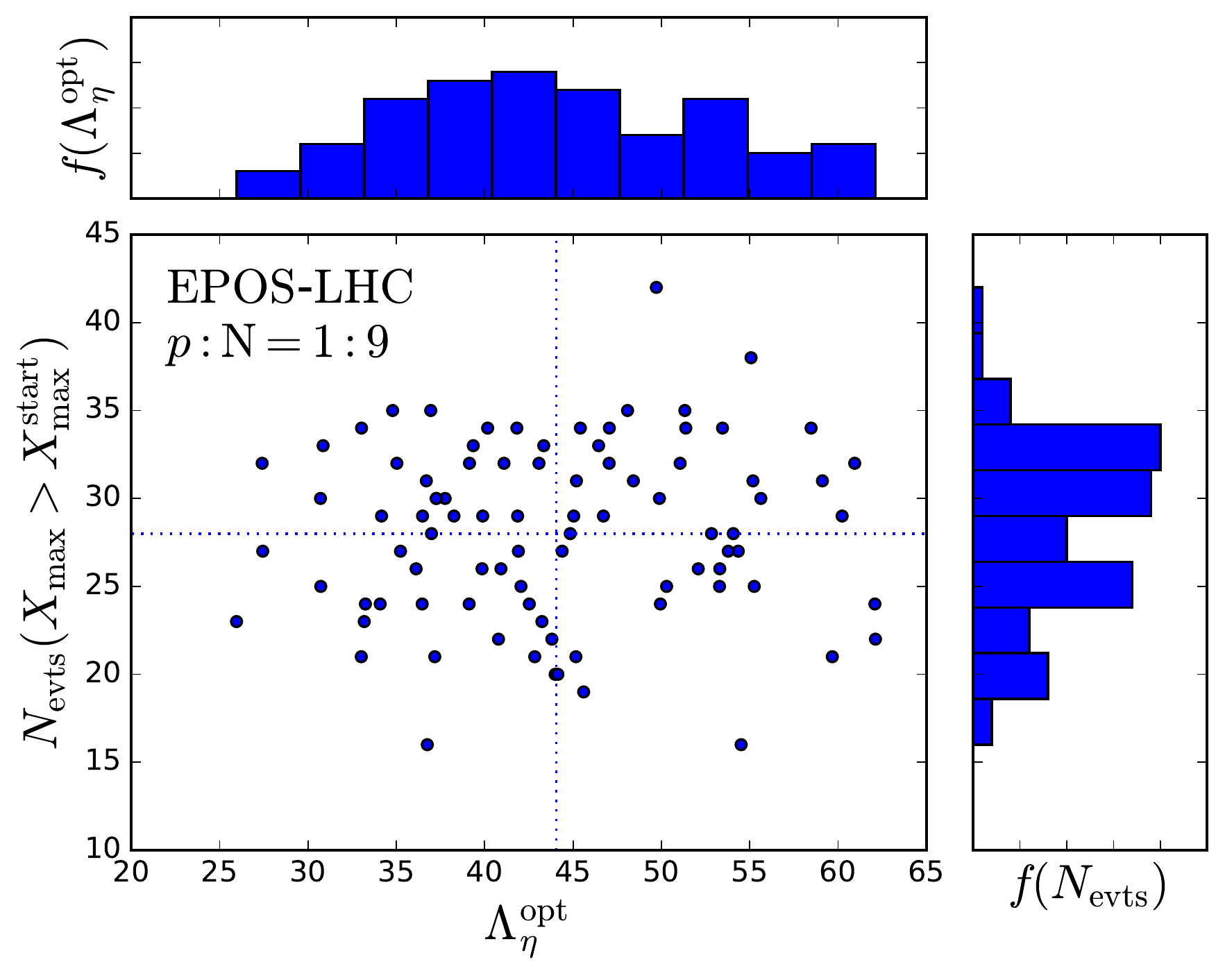}
\includegraphics[width=0.49\linewidth]{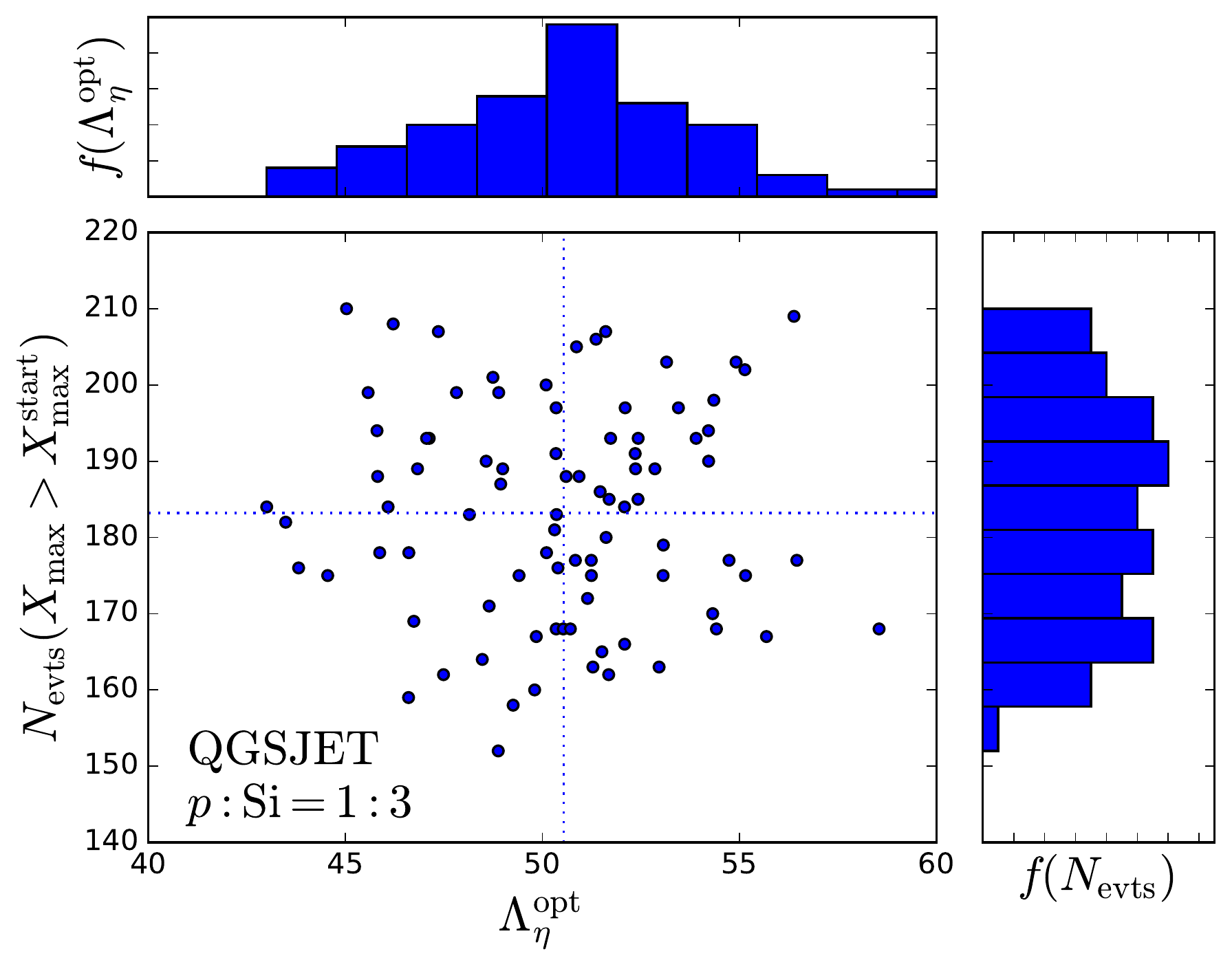}
\caption{Values of $\Lambda_\eta^{\rm opt}$ 
and the number of events for $E_{\rm CR}>40$ EeV
and $\xmax>\xmax^{\rm start}$ for $p$:N = 1:9, $\eta=0.02$ with EPOS-LHC (left) and $p$:Si = 1:3, $\eta=0.13$ with QGSJET (right). The blue histograms show the frequencies of $\Lambda_\eta^{\rm opt}$ and $N_{\rm evts}(\xmax>\xmax^{\rm start})$. The vertical and horizontal dotted lines show the mean values of
$\Lambda_\eta^{\rm opt}$ and $N_{\rm evts}(\xmax>\xmax^{\rm start})$, respectively.}
\label{fig:CRscatter}
\end{figure*}

With a high statistics Monte Carlo, we generate events according to a cosmic ray spectrum above the ankle parameterized by
\begin{eqnarray}
\nonumber
J(E)&=&J_0\Biggl(\frac{E}{E_{\rm ankle}}\Biggr)^{-\gamma_2} \Biggl[1+\Biggl(\frac{E_{\rm ankle}}{E_s}\Biggr)^{\Delta\gamma}\Biggr] \\
&\times &
\Biggl[ 1+\Biggl(\frac{E}{E_s}\Biggr)^{\Delta\gamma}\Biggr]^{-1}\ ,
\end{eqnarray}
where we take the central values of 
$E_{\rm ankle}=5.08$ EeV, $E_s=39$ EeV, 
$\gamma_2=2.53$, $\Delta\gamma=2.5$ as measured 
by the Auger Collaboration and reported in Ref. \cite{Fenu:2017hlc}.
We model the $\xmax$
distribution with the generalized Gumbel form with parameters
determined in Ref. \cite{Arbeletche:2019qma} for the QGSJETII.04 \cite{Ostapchenko:2010vb}
and EPOS-LHC \cite{Pierog:2013ria} shower models.
The cosmic ray energies are smeared according to 
\begin{equation}
\ln(E/E_{norm}) = \ln(E_0/E_{norm})+\Delta
\end{equation}
for $E_{norm}=10^{18}$ eV, with $\Delta$ generated with a Gaussian distribution
with $\sigma_\Delta=0.2$. The $\xmax$ values are 
smeared with a Gaussian distribution with $\sigma_X=35$ g/cm$^2$.  For the QGSJETII.04
(EPOS-LHC)  model,
$\xmax^{\rm start}=896\ (906)$ g/cm$^2$ for the cosmic ray composition of $p$:N = 1:9 and $\eta=0.02$.
We find $\xmax^{\rm start}=834\ (851)$ g/cm$^2$ for $p$:Si = 1:3 with the QGSJETII.04
(EPOS-LHC)  model with $\eta=0.13$.
Examples of $\sim 80-90$ Monte Carlo trials of the $1,400$ events POEMMA will measure with smeared energy above 40 EeV and $\xmax>\xmax^{\rm start}$ are shown in Fig. 
\ref{fig:CRscatter}. 

\begin{figure*}[tb]
\centering
\includegraphics[width=0.49\linewidth]{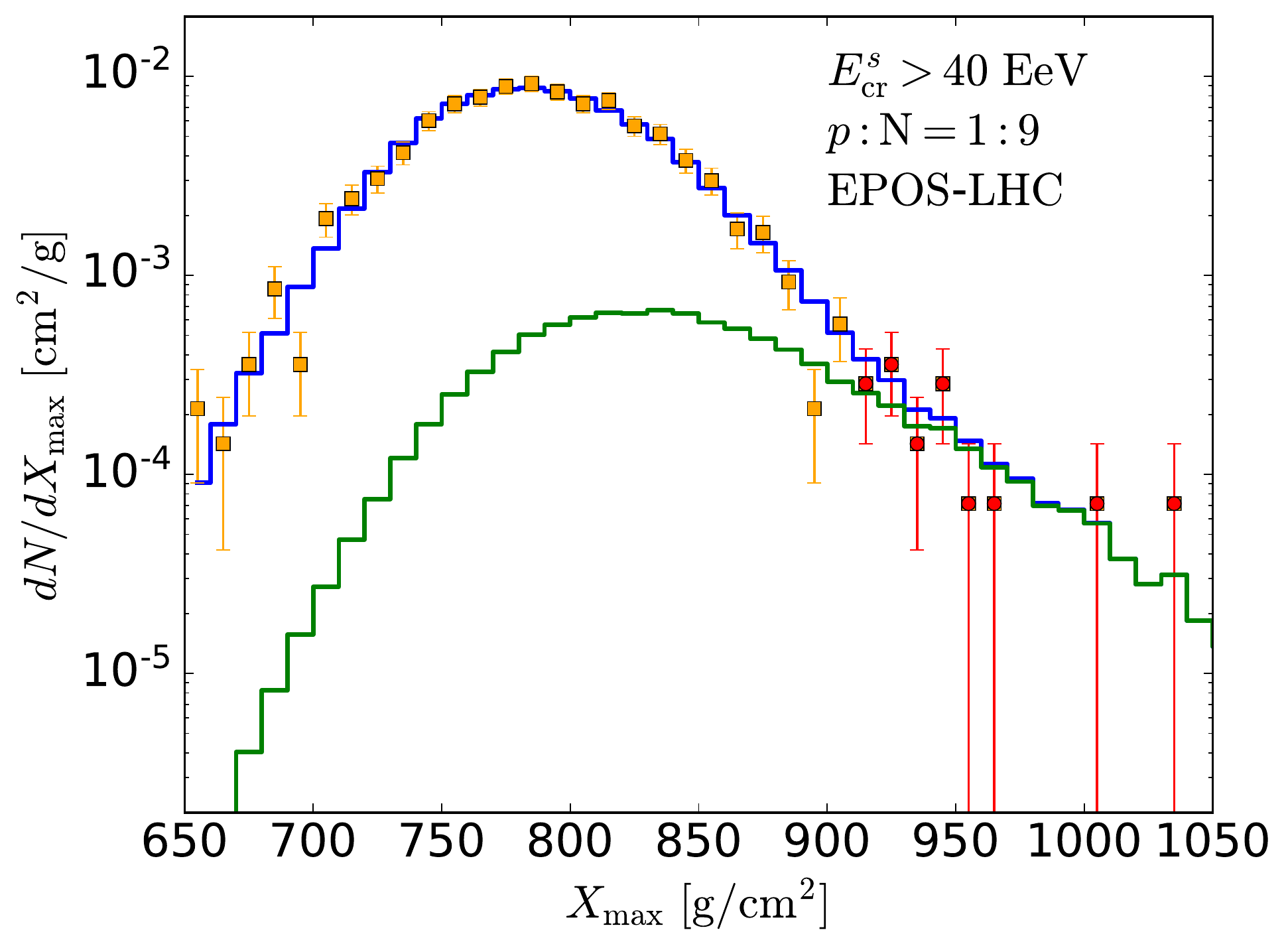}
\includegraphics[width=0.49\linewidth]{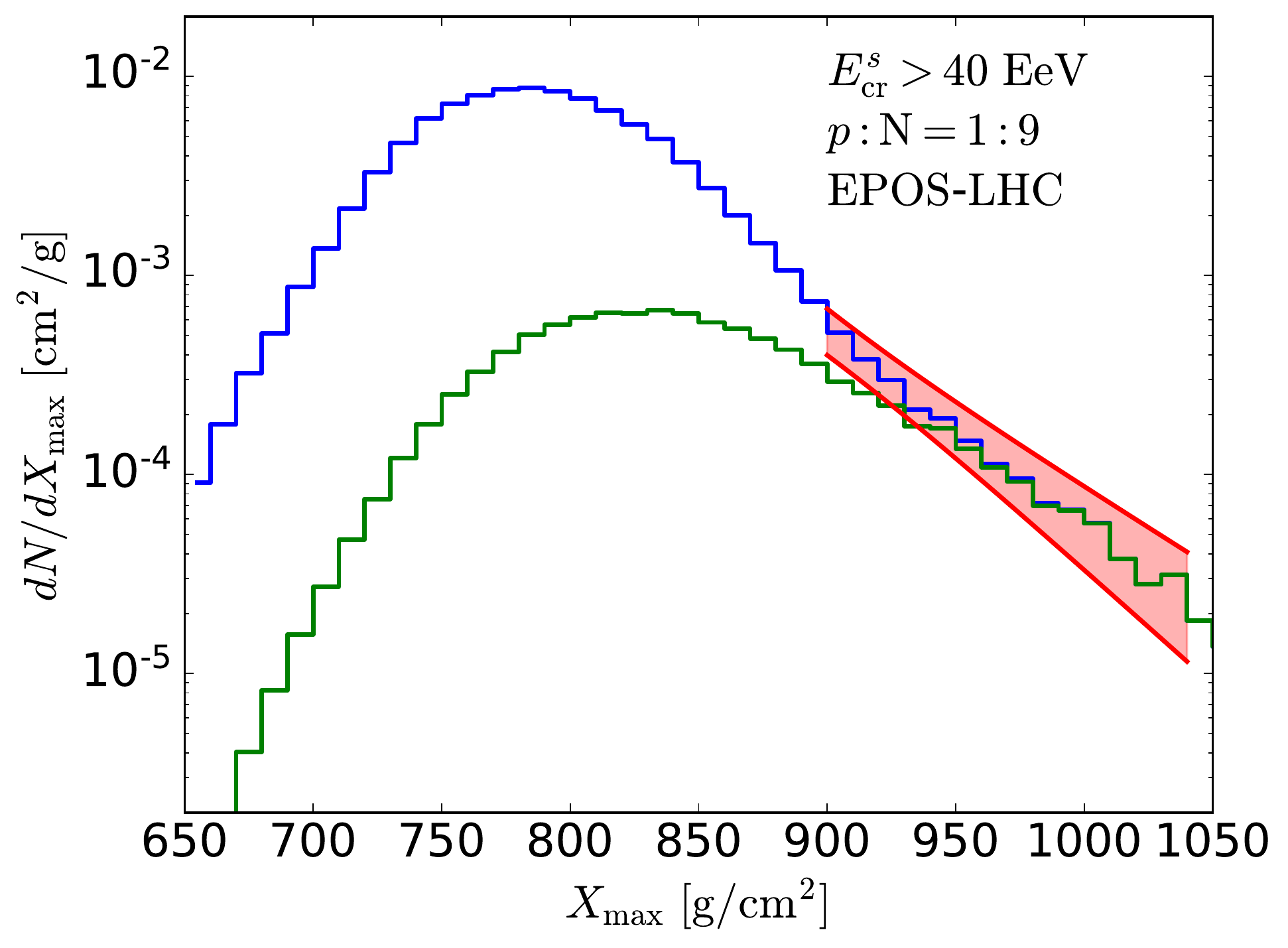}
\includegraphics[width=0.49\linewidth]{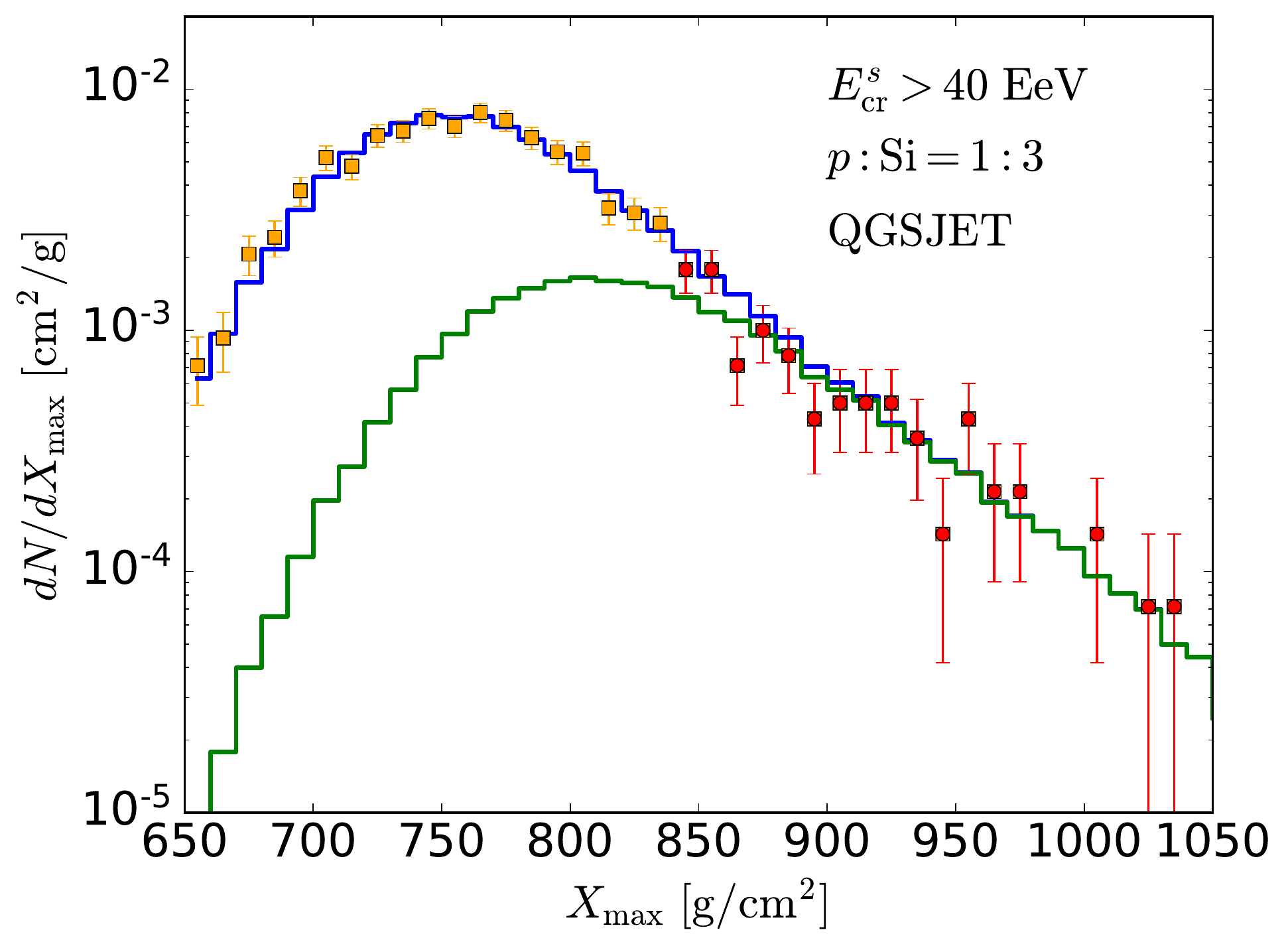}
\includegraphics[width=0.49\linewidth]{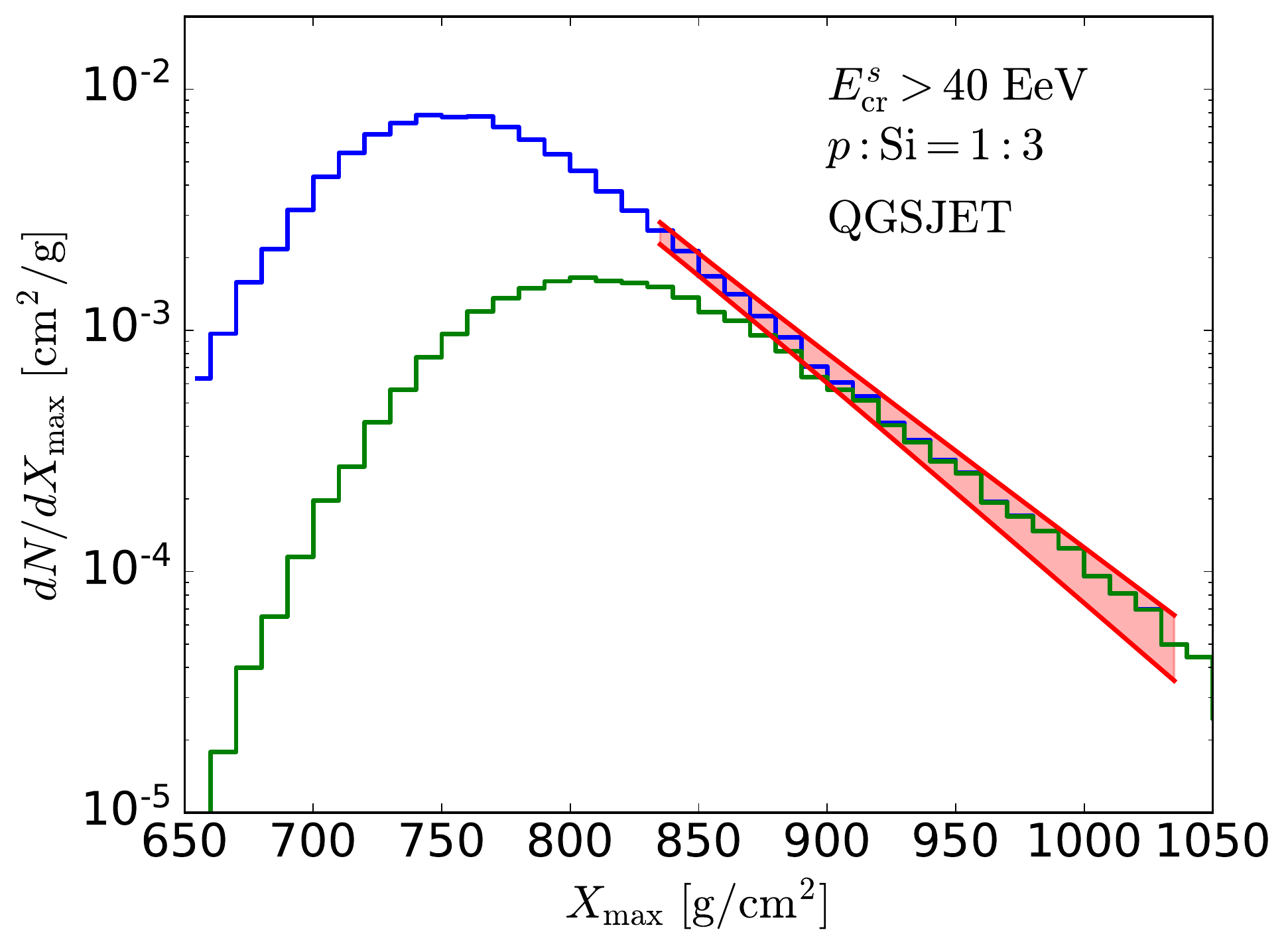}
\caption{The energy and $\xmax$ smeared distribution ${\rm d}N/{\rm d}X_{\rm max}$ for QGS{\sc JET}II.04 (upper panels) and EPOS-LHC (lower panels) from the 
parameterization of Ref. \cite{Arbeletche:2019qma} for $E_{\rm CR}>40$ EeV and $p$:N = 1:9, $\eta=0.02$ (upper) and $p$:Si = 1:3, $\eta=0.13$ (lower). The blue histograms show the full $\xmax$ distributions, while the green histograms show the proton components. The data points, with error bars, show the distribution of events for one sample of $N_{\rm evts}=1,400$ events with error bars according to $\sqrt{N_i}/N_{\rm evts}$, for $N_i$ the number of events in the bin $i$. The red data points show $\xmax$ bins above $\xmax^{\rm start}$.
In the right panels, and shaded red band shows the slope of the tail determined by $\Lambda_\eta^{\rm opt}$ with $1\sigma$ statistical errors on $\Lambda_\eta^{\rm opt}$ and the number of events above $\xmax^{\rm start}$.}
\label{fig:CRdist}
\end{figure*}

Figure \ref{fig:CRdist} shows the smeared, high statistics histograms  of the unit normalized $\xmax$ distributions of all cosmic rays above 40 EeV (blue histograms) and the proton component (green histograms).
The $\xmax$ distributions for one sample of $ N_{\rm evts}= 1,400$ for two composition models
and a hadronic interaction model are shown with the data points in the panels in Fig. \ref{fig:CRdist}. The error bars show $\sqrt{N_{i}}/N_{\rm evts}/\Delta \xmax$ for $\Delta\xmax=10$ g/cm$^2$ for this set of 1,400 events. The red points show the $\xmax$ bins which are above the bin that includes $\xmax^{\rm start}$. In the right panels, the shaded red bands show the range of slopes of the tail determined by $\Lambda_\eta^{\rm opt}$ with $1\sigma$ statistical errors on both  $\Lambda_\eta^{\rm opt}$ and the number of events above $\xmax^{\rm start}$, given 1,400 total events for POEMMA. 
Thus, $\Lambda_\eta^{\rm opt}$ can be converted to $\sigma_{p-{\rm air}}$ with the associated statistical uncertainty shown in Fig. \ref{fig:xs}.

\section{POEMMA's Simulated UHE Neutrino Sensitivity}
\label{appendix:c}

\begin{figure*}
\centering
\includegraphics[width=0.46\textwidth]{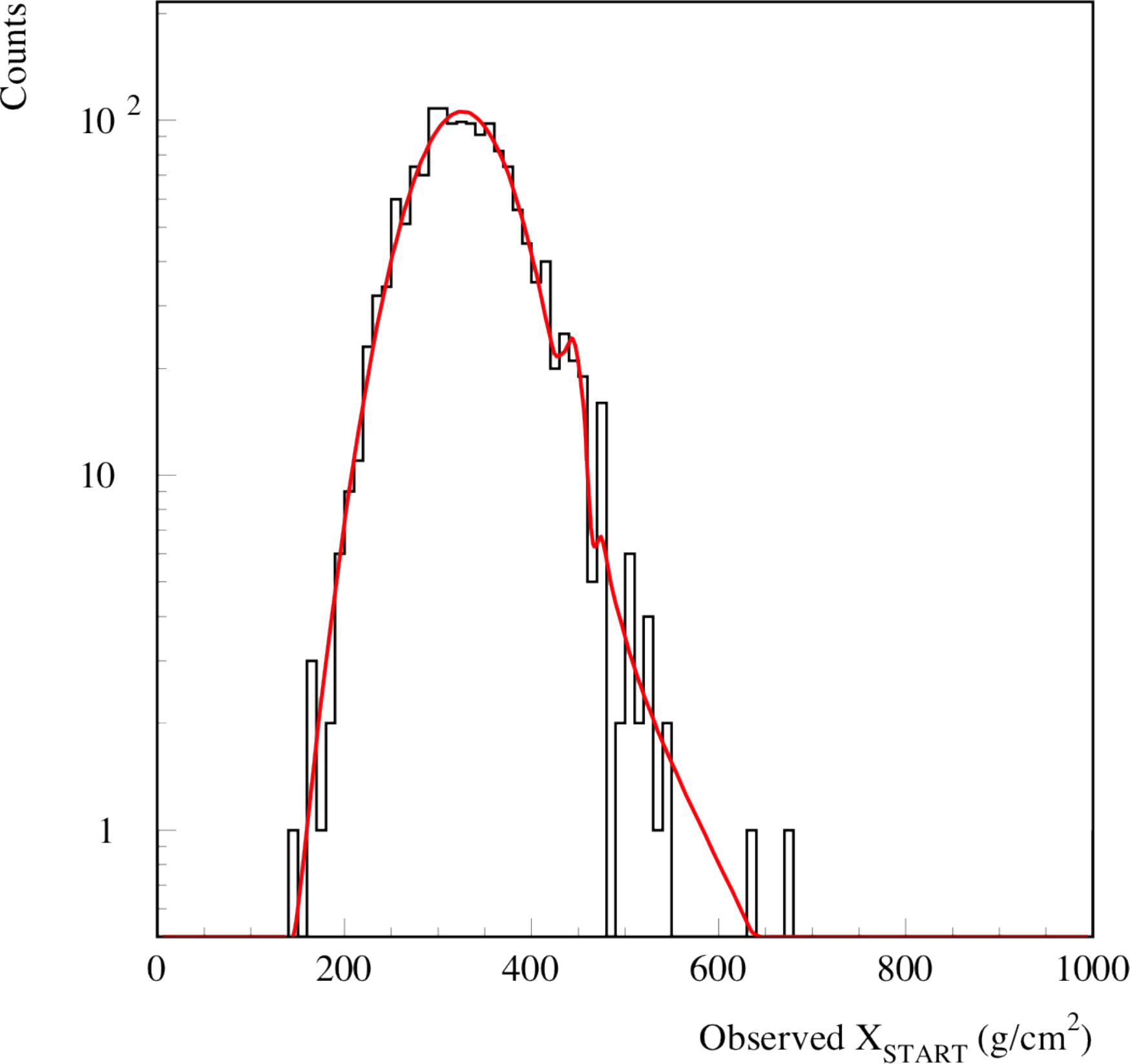}
\includegraphics[width=0.46\textwidth]{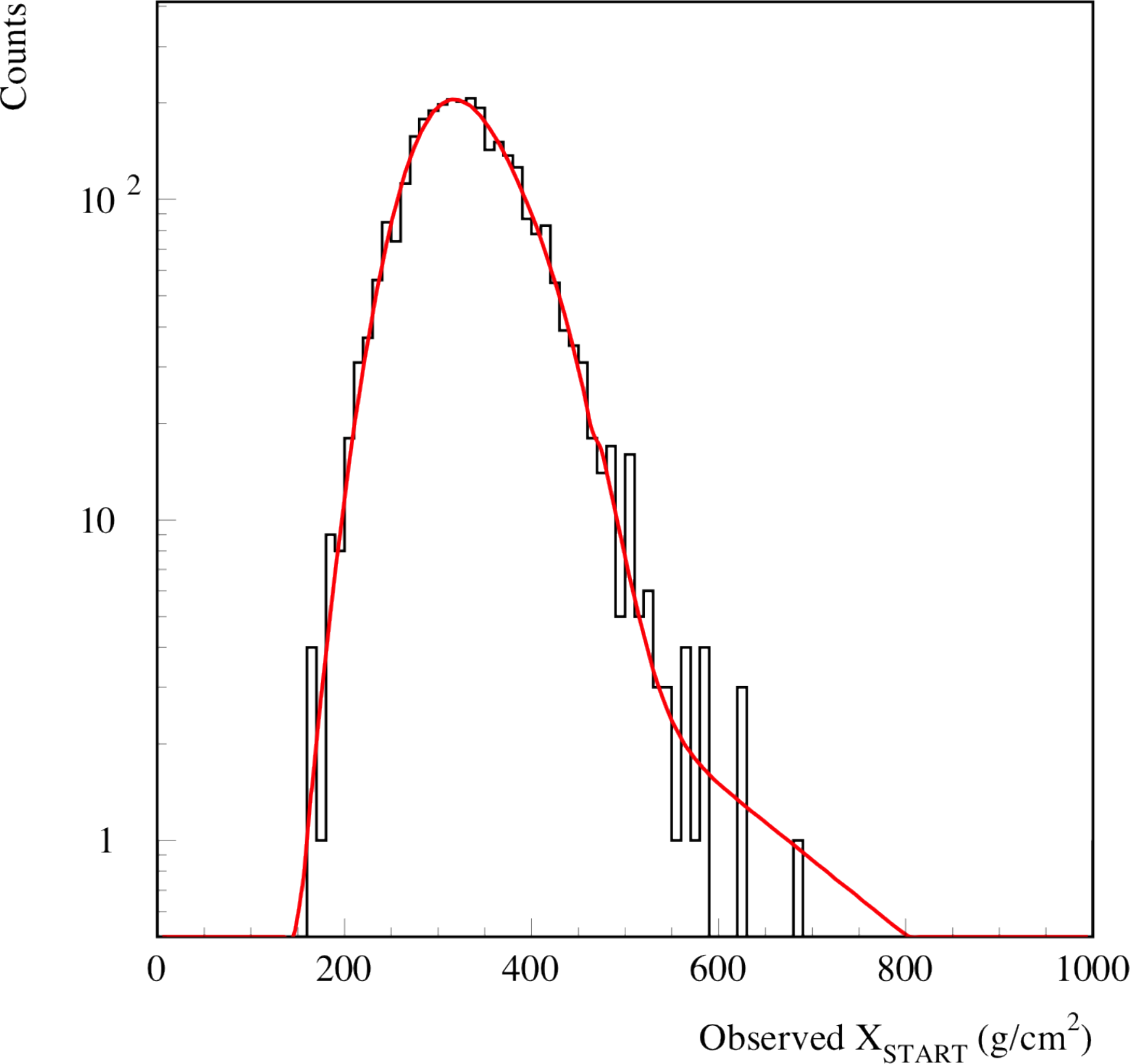}
\includegraphics[width=0.46\textwidth]{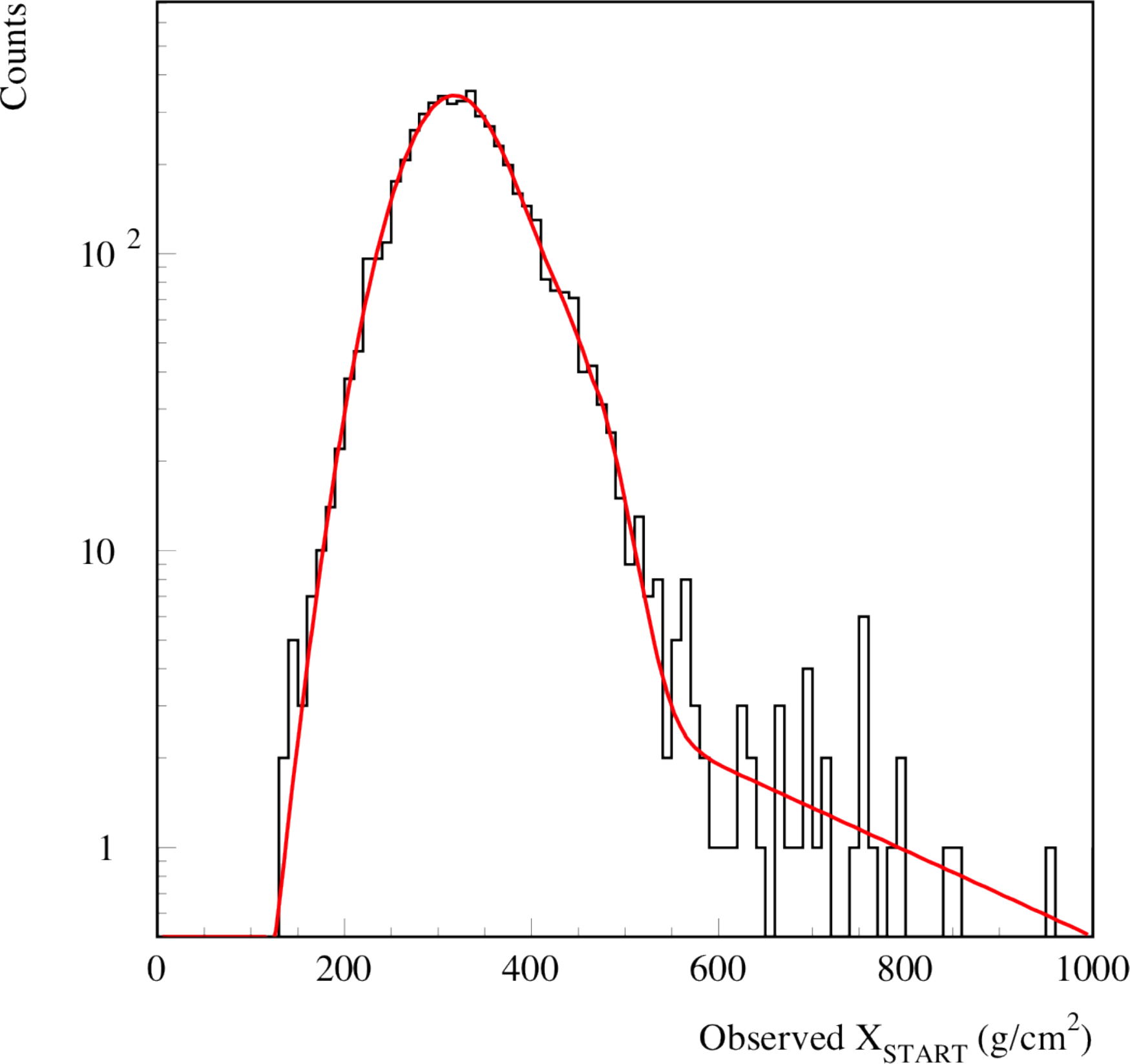}
\includegraphics[width=0.46\textwidth]{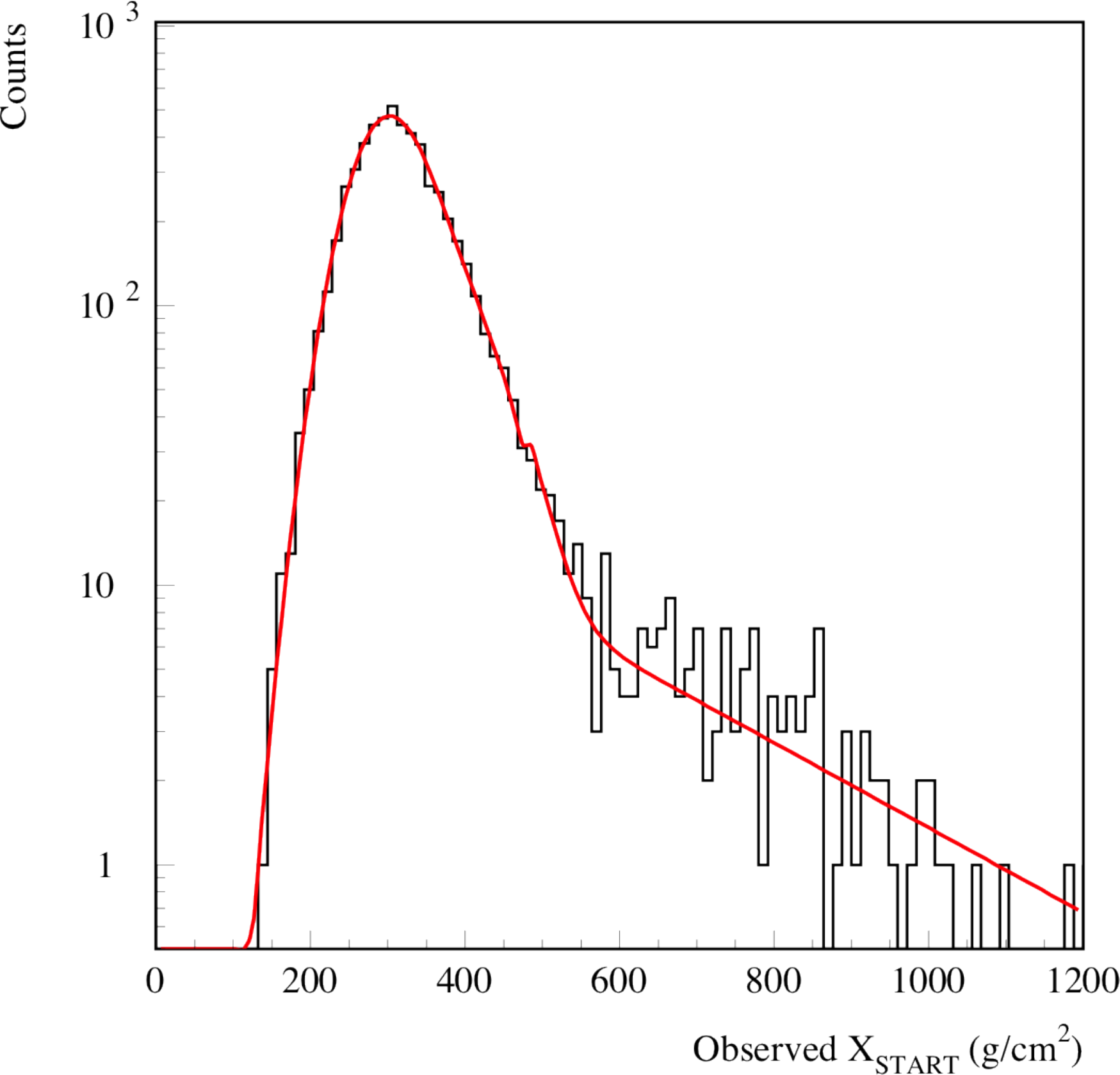}
\caption{The $X_{\rm Start}$ distributions defined by the observed starting point of proton EASs and a parametric fit using two Gaussian and an exponential distribution. {\it Top Left}: 40 EeV (Fit Reduced $\chi^2 = 1.13$); {\it Top Right}: 60 EeV (Fit Reduced $\chi^2 = 1.34$); {\it Bottom Left}: 100 EeV (Fit Reduced $\chi^2 = 0.95$); {\it Bottom Right}: (Fit Reduced $\chi^2 = 0.85$).}
\label{fig:UHECRxstart}
\end{figure*}

In stereo observation mode, POEMMA monitors $\sim 10^{13}$ metric tons of atmosphere. Thus, POEMMA has high sensitivity via air fluorescence EAS observations for UHE neutrino interactions that occur deep in the atmosphere that are well separated from UHECR EASs. The POEMMA stereo simulation package was used to simulate neutrino interactions above 30 EeV assuming an isotropic flux. For neutrino energies above a PeV, the neutrino and antineutrino interaction cross sections are virtually equal, and we just denote these as neutrinos. For neutrino energies above 10 EeV, the y-dependence of the cross section puts, on average, 80\% of the neutrino energy into the daughter lepton and 20\% into a hadronic shower. For charged-current (CC) electron neutrino interactions this leads to an EAS with 100\% of the neutrino energy. The EAS development is simulated as that of a proton, but initiated at the neutrino interaction point. On one hand this is conservative since a pure electromagnetic EAS will have $\sim 20\%$ higher $N_{\rm part}$ at EAS maximum, but on the other hand, the Landau–Pomeranchuk–Migdal effect \cite{Landau:436540,PhysRev.103.1811} will reduce this. Given this, proton EAS profiles are used since a model of the LPM effect is included in the EAS generator. 
The selection of $X_{\rm Start} \ge 2000 ~{\rm g/cm}^2$ is based on the analysis of the $X_{\rm Start}$ distributions for UHECR protons at 40, 60, 100, and 200 EeV. The event selection is based on the stereo selection described earlier, which yields approximately 1$^\circ$ angular resolution which implies a $1.5 \times 10^{-4}$ fractional error on the determination of $X_{\rm Start}$. Fig. \ref{fig:UHECRxstart} shows the distribution of $X_{\rm Start}$ for UHECRS along with parametric fits used to define the quasi-gaussian peaks and exponential tails of the distributions. These parametric fits define probability distributions that can be evaluated to determine the probability for $X_{\rm Start} \ge 2000 ~{\rm g/cm}^2$ and $X_{\rm Start} \ge 1500 ~{\rm g/cm}^2$, the later is used for comparison to the effects on the neutrino aperture for different selection criteria. Table \ref{XstartProbs} provide the UHECR background probabilities as a function of $X_{\rm Start}$ and EAS energies. Note that the assumption that the UHECR flux is pure protons is the most conservative assumption for this background analysis. Table \ref{XstartProbs} also provides the anticipated statistics in representative energy ranges based on the Auger and TA spectra assuming 5 years observation as shown in Fig. \ref{fig:spectrum} for stereo (Nadir) observations.

\begin{table*}[tb]
\caption{UHECR proton background probabilities as a function of energy and $X_{\rm Start}$ as well as the anticipated UHECR statistics in representative energy ranges based on 5 year observation with the Auger and TA measured spectra.}
\centering
\begin{tabular}{c c c c c}
\hline
\hline
~~~~$X_{\rm Start}$~~~~ & ~~~~40 EeV~~~~ & ~~~~60 EeV~~~~  & ~~~~100 EeV~~~~ & ~~~~200 EeV~~~~\\ \hline 
$\ge 1500~{\rm g/cm}^2$ & $4.9 \times 10^{-8}$ & $7.2 \times 10^{-5}$ & $5.8 \times 10^{-4}$ & $9.1 \times 10^{-4}$ \\
$\ge 2000~{\rm g/cm}^2$ & $9.0 \times 10^{-11}$ & $4.8 \times 10^{-6}$ & $1.1 \times 10^{-4}$ & $2.1 \times 10^{-4}$ \\
\hline 
 & & & & \\
Spectrum & ${\rm E}_{\rm prot} <$ 50 EeV~~  & ~~50 EeV $ \le {\rm E}_{\rm prot} <$ 100 EeV~~ & ~~100 EeV $ \le {\rm E}_{\rm prot} <$ 200 EeV~~ & ~~${\rm E}_{\rm prot} \ge$ 200 EeV\\
Auger (5 years) & 3106 & 271 & 66 & 5 \\
TA (5 Years) & 5006 & 925 & 327 & 10 \\
\hline \hline
\end{tabular}
\label{XstartProbs}
\end{table*}

Using the background probabilities and anticipated UHECR event statistics, the mean number of background events can be calculated and evaluated assuming Poisson statistics to calculate the probability of observing $\ge 1$ and $\ge 2$ UHECR background events with POEMMA in 5 years. These are also detailed in Table \ref{BackgroundProbs}.

\begin{table*}[tb]
\caption{UHECR observed proton background probabilities as a function of energy and $X_{\rm Start}$ based on 5 year observation with the Auger and TA measured spectra.}
\centering
\begin{tabular}{c c c c c c}
\hline
\hline
~~~~$X_{\rm Start}$~~~~ & ~~~~40 EeV~~~~ & ~~~~60 EeV~~~~  & ~~~~100 EeV~~~~ & ~~~~200 EeV~~~~ & Sum \\ \hline 
\multicolumn{6}{c}{Auger Spectrum: $N_{\rm Obs} \ge 1$} \\
$\ge 1500~{\rm g/cm}^2$ & ~~~$1.5 \times 10^{-4}$~~~ & ~~~$1.9 \times 10^{-2}$~~~ & ~~~$3.8 \times 10^{-2}$~~~ & ~~~$4.5 \times 10^{-3}$~~~ &  $6.1 \times 10^{-2}$ \\
$\ge 2000~{\rm g/cm}^2$ & $2.8 \times 10^{-7}$ & $1.3 \times 10^{-3}$ & $7.2 \times 10^{-3}$ & $1.0 \times 10^{-3}$  & $9.6 \times 10^{-3}$ \\
\hline
\multicolumn{6}{c}{Auger Spectrum: $N_{\rm Obs} \ge 2$} \\
$\ge 1500~{\rm g/cm}^2$ & ~~~$1.2 \times 10^{-8}$~~~ & ~~~$1.9 \times 10^{-4}$~~~ & ~~~$7.1 \times 10^{-4}$~~~ & ~~~$1.0 \times 10^{-5}$~~~ &  $9.1 \times 10^{-4}$ \\
$\ge 2000~{\rm g/cm}^2$ & $3.9 \times 10^{-14}$ & $8.4 \times 10^{-7}$ & $2.6 \times 10^{-5}$ & $5.3 \times 10^{-7}$  & $2.8 \times 10^{-5}$ \\
\hline
\multicolumn{6}{c}{TA Spectrum: $N_{\rm Obs} \ge 1$} \\
$\ge 1500~{\rm g/cm}^2$ & ~~~$2.5 \times 10^{-4}$~~~ & ~~~$6.4 \times 10^{-2}$~~~ & ~~~$1.7 \times 10^{-1}$~~~ & ~~~$9.0 \times 10^{-3}$~~~ &  $2.5 \times 10^{-1}$ \\
$\ge 2000~{\rm g/cm}^2$ & $4.7 \times 10^{-7}$ & $4.4 \times 10^{-3}$ & $3.5 \times 10^{-2}$ & $2.1 \times 10^{-3}$  & $4.2 \times 10^{-2}$ \\
\hline
\multicolumn{6}{c}{Ta Spectrum: $N_{\rm Obs} \ge 2$} \\
$\ge 1500~{\rm g/cm}^2$ & ~~~$3.0 \times 10^{-8}$~~~ & ~~~$2.1 \times 10^{-3}$~~~ & ~~~$1.6 \times 10^{-2}$~~~ & ~~~$4.1 \times 10^{-5}$~~~ &  $1.8 \times 10^{-2}$ \\
$\ge 2000~{\rm g/cm}^2$ & $1.0 \times 10^{-13}$ & $9.8 \times 10^{-6}$ & $6.3 \times 10^{-4}$ & $2.1 \times 10^{-6}$  & $6.4 \times 10^{-4}$ \\

\hline \hline
\end{tabular}
\label{BackgroundProbs}
\end{table*}

The results show that assuming the POEMMA UHECR statistics based on the measured Auger UHECR spectrum, the probability of getting at lease one UHECR background event in the neutrino sample is 6.1\% for $X_{\rm Start} \ge 1500 ~{\rm g/cm}^2$ while it is $< 1$\% for $X_{\rm Start} \ge 2000 ~{\rm g/cm}^2$. Assuming the POEMMA UHECR statistics based on the measured TA UHECR spectrum, the probability of getting at lease one UHECR background event in the neutrino sample is 25\% for $X_{\rm Start} \ge 1500 ~{\rm g/cm}^2$ while it is 4.2\% for $X_{\rm Start} \ge 2000 ~{\rm g/cm}^2$. This motivates the use of $X_{\rm Start} \ge 2000 ~{\rm g/cm}^2$ for the POEMMA air fluorescence neutrino acceptance.

\begin{figure}[tb]
\centering
\includegraphics[width=0.97\columnwidth]{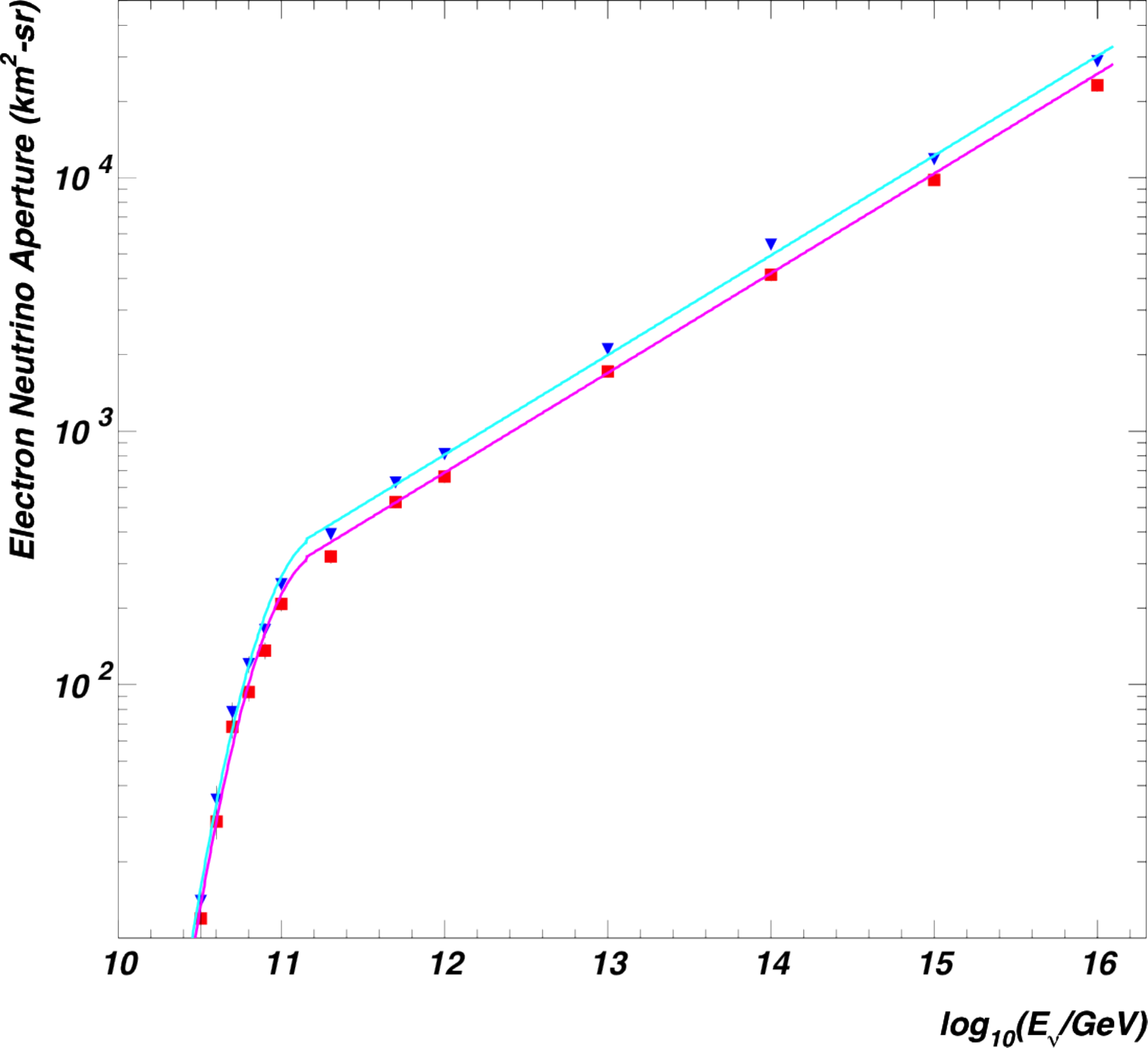}
\caption{Comparison of the instantaneous electron neutrino apertures based on stereo air fluorescence measurements. Upper points and curve are for $X_{\rm Start} \ge 1500 ~{\rm g/cm}^2$ while the lower points and curve are for $X_{\rm Start} \ge 2000 ~{\rm g/cm}^2$. The lower curve is 85\% of the upper curve over the energy band.}
\label{fig:nueAperComp}
\end{figure}

The effects of the $X_{\rm Start}$ selection on the simulated electron neutrino aperture is shown in Fig. \ref{fig:nueAperComp}. A parametric fit is used to describe the aperture based on simulations at specific energies. The comparison of the results for $X_{\rm Start} \ge 2000 ~{\rm g/cm}^2$ show a modest 15\% reduction over the entire energy band as compared to the $X_{\rm Start} \ge 1500 ~{\rm g/cm}^2$ results.  This comparison shows the relative insensitivity of the electron neutrino aperture for modest changes in the observed $X_{\rm Start}$ requirement.

Using the CC electron neutrino aperture, the apertures of the other neutrino flavors for both CC and NC can be obtained.  For the NC, the emergent lepton is a neutrino and thus an EAS with 20\% of the incident neutrino energy is produced but with a lower rate given by the ratio of NC to CC neutrino cross sections.  Effectively this shifts the NC neutrino aperture curve up by a factor of five in neutrino energy, compared to the CC electron neutrino aperture, with a reduction given by $\frac{\sigma_{\rm NC}(E_\nu)}{\sigma_{\rm CC}(E_\nu)}$ for each of the three neutrino flavors. For the CC $\mu_\mu$ and $\nu_\tau$ apertures, we conservatively assume that only 20\% of the neutrino energy is observed, e.g. the EAS generated by the emergent muon or $\tau$-lepton is not observed. While these UHE muons are well above their critical energy, the charged-particle production in these muonic EASs versus electron-initiated EAS is much reduced \cite{Vankov}. Thus it is assumed the air fluorescence signal is below POEMMA's detection threshold. For $\tau$-leptons, the decay length given by $\gamma$c$\tau$ is nearly 5,000 km  at 100 EeV. The conservative approach is to assume the $\tau$-lepton decays outside POEMMA's FoV and only 20\% of the incident neutrino is observed in the EAS.  These assumptions lead to the muon and $\tau$-lepton CC neutrino apertures shifted up by a factor of five in neutrino energy, compared to the CC electron neutrino aperture. The ensemble of these results are presented in the rightmost plot in Fig.~\ref{fig:eenus}.

\bibliographystyle{utphys}
\bibliography{references}
\end{document}